\documentclass[10pt,journal]{IEEEtran} 
 
\ifCLASSOPTIONcompsoc
  \usepackage[nocompress]{cite}
\else
  \usepackage{cite}
\fi

\usepackage{cite}
\usepackage{amsmath,amssymb,amsfonts,amsthm}
\usepackage[ruled]{algorithm2e}
\usepackage{graphicx}
\usepackage{subfigure}
\usepackage{textcomp}
\usepackage{color}
\usepackage{booktabs}
\usepackage{multirow}
\usepackage{threeparttable}
\usepackage[table]{xcolor}
\usepackage[hyphens]{url}
\usepackage{hyperref}
\newtheorem{theorem}{\textbf{Theorem}}

\newcommand{\wjsSec}{\textcolor{black}}
\newcommand{\wjsRR}{\textcolor{black}}
\newcommand{\wjs}{\textcolor{black}}
\usepackage{balance}
\usepackage{pifont}
\usepackage{tikz}
\usepackage[normal]{caption}
\newcommand*\emptycirc[1][0.8ex]{\tikz\draw (0,0) circle (#1);}
\newcommand*\halfcirc[1][0.8ex]{
      \begin{tikzpicture}
       \draw[fill] (0,0)-- (90:#1) arc (90:270:#1) -- cycle ;
       \draw (0,0) circle (#1);
      \end{tikzpicture}}
\newcommand*\fullcirc[1][0.8ex]{\tikz\fill (0,0) circle (#1);}

\begin{document}

\title{pvCNN: Privacy-Preserving and Verifiable Convolutional Neural Network Testing}

\author{Jiasi Weng,
        Jian Weng$^{\ast}$, \IEEEmembership{Member, IEEE},
        Gui Tang,
        Anjia Yang,
        Ming Li,
        Jia-Nan Liu
\IEEEcompsocitemizethanks{\IEEEcompsocthanksitem J.s. Weng, J. Weng, G. Tang, A. Yang and M. Li are with the College of Cyber Security of Jinan University, Guangzhou 510632, and Pazhou Lab, Guangzhou 510335, China. J.-N. Liu is affiliated with the School of Computer Science and Technology, Dongguan University of Technology, Dongguan 523808, and Guangzhou Fongwell Data Limited Company, Guangzhou 511400, and Pazhou Lab, 510335, China.
E-mail: wengjiasi@gmail.com,  cryptjweng@gmail.com, guitang001@gmail.com, anjiayang@gmail.com, limjnu@gmail.com, j.n.liu@foxmail.com. Jian Weng is the corresponding author.} 
}  
%

%\markboth{IEEE Transactions on Information Forensics and Security}%

\maketitle

%model: owner/server
\begin{abstract}
We propose a new approach for privacy-preserving and verifiable convolutional neural network (CNN) testing in a distrustful multi-stakeholder environment.
The approach is aimed to enable that a CNN model \emph{developer} convinces a \emph{user} of the truthful CNN performance over non-public data from \emph{multiple testers}, while respecting model and data privacy. 
%
%The proposed approach (a) supports publicly verifiable and static proofs, (b) preserves privacy, and (c) respects testing in batches.
%Following the efforts of the state-of-the-art relevant work,  
%The proposed approach starts from Fiore \emph{et al.}'s (PKC'20) new zk-SNARK systems for verifiable computation over ring polynomials, enjoying good properties, such as public verifiability, data confidentiality and no need of interaction.
% 
To balance the security and efficiency issues, we appropriately integrate three tools with the CNN testing, including collaborative inference, homomorphic encryption (HE) and zero-knowledge succinct non-interactive argument of knowledge (zk-SNARK).
% .
% 
%On top of this, we make two-fold efforts to reduce proof generation overhead and verification costs.
% 

We start with strategically partitioning a CNN model into a private part kept locally by the model developer, and a public part outsourced to an outside server.
Then, the private part runs over the HE-protected test data sent by a tester, and transmits its outputs to the public part for accomplishing subsequent computations of the CNN testing.
Second, the correctness of the above CNN testing is enforced by generating zk-SNARK based proofs, with an emphasis on optimizing proving overhead for two-dimensional (2-D) convolution operations, since the operations dominate the performance bottleneck during generating proofs.
We specifically present a new quadratic matrix program (QMP)-based arithmetic circuit with \emph{a single multiplication gate} for expressing 2-D convolution operations between multiple filters and inputs in a batch manner.
Third, we aggregate multiple proofs with respect to a same CNN model but different testers' test data (\emph{i.e.}, different statements) into one proof, and ensure that the validity of the aggregated proof implies the validity of the original multiple proofs.
Lastly, our experimental results demonstrate that our QMP-based zk-SNARK performs nearly $13.9\times$ faster than the existing quadratic arithmetic
program (QAP)-based zk-SNARK in proving time, and $17.6\times$ faster in \textsf{Setup} time, for high-dimension matrix multiplication.
Besides, the limitation on handling a bounded number of multiplications of QAP-based zk-SNARK is relieved.\looseness=-1
\end{abstract}

%\begin{IEEEkeywords}
%Blockchain interoperability, privacy-preserving, smart contracts, TEE.
%\end{IEEEkeywords}

\IEEEpeerreviewmaketitle

%\tableofcontents
%-------------------------------------------------------------------------------
\section{INTRODUCTION}~\label{sec:intro}
Convolutional neural networks (CNNs)~\cite{lecun1989backpropagation, das2018convolutional} have been widely applied in various application scenarios, such as healthcare analysis, autonomous vehicle and face recognition.
%.
But real-life reports demonstrate that neural networks often exhibit erroneous decisions, leading to disastrous consequences, \wjs{\emph{e.g.}, self-driving crash due to the failure of identifying 
unexpected driving environments~\cite{google16driving}, and prejudice due to racial biases embedded in face recognition systems~\cite{amazon20face}}.
%
%Some examples include \textsf{(e1)~}self-driving crash due to the failure of identifying 
%unexpected driving environments~\cite{google16driving}, and \textsf{(e2)~}social fighting derived from racial biases embedded in face recognition systems~\cite{amazon20face}.
%For example, an autonomous vehicle recently causes a crash, since it cannot handle .
%
%Another case reveals that several commercial face recognition systems embed gender and racial biases~\cite{amazon20face}.
%Amazon's face recognition performs unfairly on African Americans and White Americans.
%
The reports emphasize that when applying
CNN models in security-critical scenarios, model users 
should be sufficiently cautious of benchmarking the CNN models to obtain the truthful multi-faceted performance, like robustness and fairness, not limited to natural accuracy.
\wjs{For example, when a CNN model is deployed in a self-driving car for identifying camera images, a user should be assured that the CNN model is always highly accurate, even for perturbed images;
for face recognition applications, users want to know that the underlying CNN models can accurately recognize face images without discrimination.}

%independent of some sensitive features in an image like race.

%especially for those from third-party developers, 
% 
%Hence, before CNNs are deployed in security-critical settings, their performance needs to be measured as comprehensively as possible, \emph{e.g.}, accuracy, robustness and fairness.
%

A widely adopted approach to benchmarking a CNN model is black-box testing~\cite{wicker2018feature, aggarwal2021testing}.
Black-box testing enables users to have a black-box access to the CNN model, that is, feeding the model with a batch of test data and merely observing its outputs, \emph{e.g.}, the proportion of correct predictions.
Such approach is suitable for the setting where model users and model developers are not the same entities.
The main reasons are that the parameters of CNN models are intellectual properties and are vulnerable to privacy attacks, such that developers are unwilling to reveal the parameters.
We are also aware that existing excellent white-box testing approaches\cite{ma2018deepgauge} can test CNN models in a fine-granularity fashion using model parameters, but black-box testing is preferred in the above setting where model parameters are inaccessible.

Starting by black-box testing, two essential issues need consideration to probe a CNN model's truthful and multi-faceted performance.
\textbf{(\emph{i}) Black-box testing strongly relies on the test dataset\cite{ma2018deepgauge}, which in turn requires multi-source data support in reality.}
In order to support testing a model's multi-faceted performance, the available test dataset should be as various as possible.
One evidence is that multiple benchmarks are built by embedding dozens of types of noises into ImageNet to measure the robustness of CNN models\cite{hendrycks2018benchmarking}.
Despite the necessity, building such benchmarks consumes many manpower and multi-party efforts, even for large companies, \emph{e.g.}, a recent project named Crowdsourcing Adverse Test Sets for Machine Learning (CATS4ML) launched by Google Research~\cite{cats4ML}. 
% https://ai.googleblog.com/2021/02/uncovering-unknown-unknowns-in-machine.html 
%The CATS4ML project was proposed by a team of researchers from the University of Cambridge, including Nicolas Papernot, Marta Kwiatkowska, and Daniel R. Thomas, among others. They published a research paper titled "Crowdsourcing Adversarial Test Suites for Machine Learning Systems" at the 2018 Conference on Neural Information Processing Systems (NeurIPS), which outlined the concept and goals of the CATS4ML project. Since then, the CATS4ML project has gained traction in the research community, with multiple institutions and organizations participating in the effort to collect diverse and challenging test cases for machine learning models.
%
%
Thus, the off-the-shelf test datasets for supporting multi-faceted CNN testing are likely from multiple sources. 
%\footnote{https://ai.googleblog.com/2021/02/uncovering-unknown-unknowns-in-machine.html}. 
%
%
\textbf{(\emph{ii}) Black-box testing opens a door for untrusted model developers to cheat in testing.}
They might forge untruthful outputs without correctly running the processes of CNN testing on given test datasets.
%
%It is derived from that users, developers and the role of providing test data are mutually distrustful.
% 
Additionally, if developers can in advance learn a given test dataset, they might craft a CNN which typically adapts to the test dataset, which deviates from our initial goal of testing their truthful multi-faceted performance.
From this point, test datasets cannot be public, not merely due to that datasets themselves are privacy-sensitive in many security-critical scenarios.
Motivated by the two issues above, there needs an approach for users to validate the correctness of the black-box CNN testing over multi-source test datasets, in which test datasets are not public and the CNN model is privacy-preserving.

\begin{figure}[h]
\scriptsize
%\vspace{-20pt}
    \centering
    \includegraphics[width=0.6\linewidth]{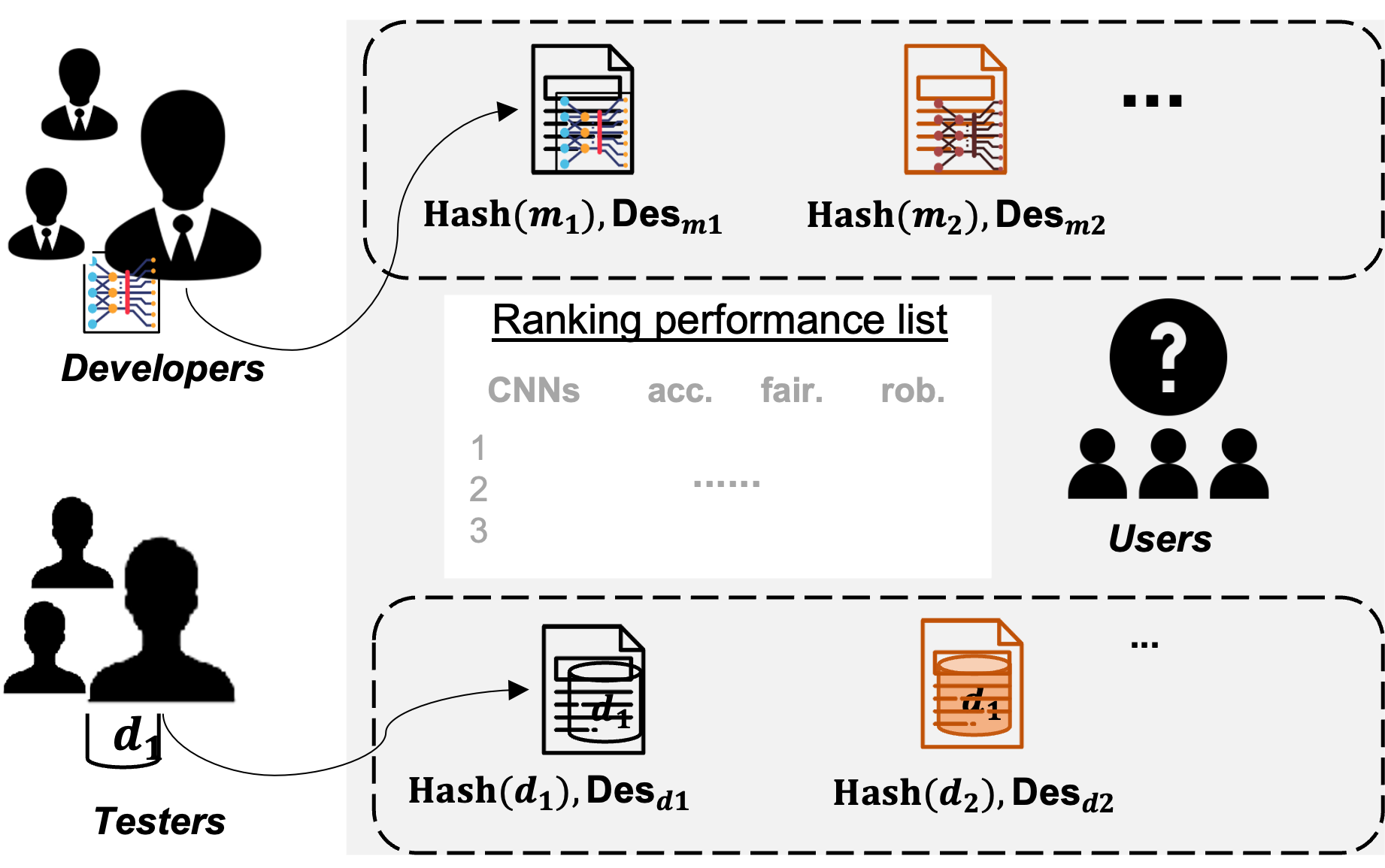}
    \caption{Scenario example ($m_1, m_2$ refer to CNN models, and $d_1, d_2$ refer to test datasets; $\textsf{Des}$ means non-private descriptions for the CNN models or test datasets).}
    \label{fig:intro} %% label for entire figure
\vspace{-10pt} 
\end{figure} 
 
While many \wjs{awesome} efforts have been done to make CNN prediction/testing verifiable~\cite{ghodsi2017safetynets, collantes2020safetpu, lee2020vcnn, fengzen, keuffer2018efficient, zhao2021veriml, madi2020computing, liuzkcnn} using cryptographic proof techniques~\cite{thaler2013time, groth2016size, agrawal2018non, campanelli2019legosnark, fiore2020boosting, mouris2021zilch}, we still need a new design satisfying our scenario requirements.
%\footnote{differ}
We now clarify our scenario requirements, which cannot be fully satisfied by the previous work (see explanation in Section~\ref{sec:work}).
%and explain why previous works are not really satisfactory to the requirements.
%.  
As shown in Fig.~\ref{fig:intro}, a publicly accessible platform (in gray color), \emph{e.g.}, ModelZoo\cite{caffe2019zoo}, Kaggle\cite{kaggle2010data} and \wjsSec{AWS Marketplace~\cite{awsModel}}, can allow third-party \emph{developers} to advertise their CNN models towards users for earning profits. 
For \emph{users} who will pay for a CNN model, they may be understandably concerned about its performance, given previous reports of misleading claims.
\wjsSec{The platform might have a motivation to maintain a good reputation for sustainability by publishing good models.} 
To address the concerns, the platform can announce a crowdsourcing task to test the CNN model, such that \emph{testers} participate in probing the model performance using their test datasets.
\wjsRR{Note that the model parameters and test datasets are not disclosed on the platform, while the model architectures can be public.}
Our design empowers untrusted CNN developers by providing them with a way to prove the truthful performance of their models over multiple test datasets to convince potential users.
The design needs to satisfy three requirements as follows:
 
\noindent(a) \emph{\textbf{Public verifiability and no need of interaction.}}
Since the user is later-coming and not pre-designated, proof generation and verification w.r.t the CNN testing should not share private information between the developer (\emph{i.e.}, prover) and the user (\emph{i.e.}, verifier), and thereby requiring public verifiability.
The proving ability should also be non-interactive, due to that the CNN developer, the user and testers are not always simultaneously online.

\noindent(b) \emph{\textbf{Privacy preservation.}}
During the CNN testing, model parameters should not be exposed to users, since they may be intellectual properties for profits, and the parameters memorize sensitive training data\cite{xu2019verifynet}. Also, the test data should also be protected against the CNN developer, considering the CNN developer might strategically craft the CNN model based on the test data.
Besides, any verifier should fail to learn private information from the final CNN performance and proofs.

\noindent(c) \emph{\textbf{Batch proving and verification.}}
A CNN model usually receives test data in batches, so it is a natural requirement to support batch operations in proof generation.
In terms of verification in our scenario, a later-coming user has to verify multiple proofs w.r.t a single CNN model which is tested with the multi-tester test data.
Hence, enabling the user to verify multiple proofs in a batch manner can be another desirable requirement.\looseness=-1

\subsection{Our Designs} 
\wjs{To satisfy the three requirements, we populate our designs by strategically integrating fully homomorphic encryption (FHE), zero-knowledge Succinct Non-interactive ARgument of Knowledge (zk-SNARK) and an idea of collaborative inference~\cite{he2019model, ryffel2019partially}.
We adopt zk-SNARK-based proof systems, mainly due to that the class of proof technique supports many properties like public delegatability and public verifiability which suit our scenario involving multiple distrustful entities.}
To summarize, our designs include three components: (1) privacy-preserving CNN testing based on an FHE algorithm and collaborative inference, (2) generating zk-SNARK proofs for the above privacy-preserving CNN testing, with an emphasis on proposing a new quadratic matrix programs (QMP)-based zk-SNARK for proving 2-D convolutional relations, and (3) enabling zk-SNARK proof aggregation for verifying multiple proofs in a batch manner.\looseness=-1 

\wjsSec{Firstly, we start by letting a developer locally run the CNN testing process over the FHE-protected test data from every tester.
Our starting point is to enable performing the whole CNN testing processing over ciphertext domain.
While FHE can provide stronger data confidentiality, compared to other secure computing techniques, such as secure multi-party computation and differential privacy~\cite{aono2017privacy}, the straightforward adoption is computation-expensive, nearly 4 to 5 orders slower than computing on unencrypted data.
Moreover, proving overhead on encrypted data (in ring field $\mathcal{R}_q$) and encoded CNN parameters (in ring field $\mathcal{R}_t$) becomes unaffordable, compared to generating proofs over plaintext domain.
We hence find a middle point for balancing privacy and efficiency, by introducing a collaborative inference strategy, and thereby making partial testing process run on ciphertext domain.}
Particularly, the CNN developer can split his CNN model into two parts: one is private, called \textsf{PriorNet}, and another one is less private, called \textsf{LaterNet}.
Then, the developer can locally evaluate \textsf{PriorNet} on the FHE-protected test data, and delegate \textsf{LaterNet} to a computation-powerful server for accomplishing the subsequent computation by feeding it with the plaintext outputs of \textsf{PriorNet}.
Next, due to that the plaintext outputs of \textsf{PriorNet} might be susceptible to model inversion attacks~\cite{he2019model} by the malicious server, \wjs{an off-the-shelf and generic adversarial training strategy~\cite{ryffel2019partially} is able to protect the \textsf{PriorNet}'s outputs against the attacks.} 

The second step is to prove the correctness of the above splitting CNN testing on FHE-encrypted test data (\wjs{\emph{i.e.}, polynomial ring elements}).
We concretely adopt a proving roadmap recently proposed by Fiore \emph{et al.}\cite{fiore2020boosting}, in which \textsf{Step~1} is to prove the simultaneous evaluation of multiple polynomial ring elements in the same randomly selected point (for \textsf{PriorNet}), and \textsf{Step~2} is to prove the satisfiability of quadratic arithmetic program (QAP)-based arithmetic circuits (for \textsf{PriorNet} and \textsf{LaterNet}) whose inputs and outputs are committed. 

But we do not directly apply \textsf{Step~2} into our case, since the direct adoption causes high proving and storage overhead due to a large number of multiplication gates for handling convolutional relations\cite{lee2020vcnn}.
We thus present a new method to reduce the number of multiplication gates during proof generation, for improving the proving time of the convolutional relations.
At first, we represent the 2-D convolution operations between $M$ filters which are $m\times m$ matrices and $Mn^2$ test inputs which are $n\times n$ matrices into \emph{\textbf{a single matrix multiplication (MM)}} computation.
The MM computation is conducted between a $Mn^2\times Mn^2$ filter matrix which is strategically assigned with the $M$ filters and zeros as padding for operation correctness, and another $Mn^2\times Mn^2$ input matrix which is assigned with the $Mn^2$ test inputs.
After that, we start from the QAP-based zk-SNARK and present a new QMP formula\cite{alesiani2021method}.
We then express the above single MM computation as a circuit over $Mn^2\times Mn^2$ matrices \emph{\textbf{by an only one-degree QMP}}.
As a result, the number of multiplication gates always is $1$ and the proving time linearly increases by the matrix dimension, \emph{i.e.}, $Mn^2 \times Mn^2$.
With the effort, we prove the satisfiability of the QMP-based circuits via checking the divisibility in randomly selected point of the set of matrix polynomials of the QMP.
Note that there are heterogeneous operations in each neural network layer during CNN testing.
We express the convolution and full connection operations as QMP-based circuits, while expressing the activation and max pooling functions as QAP-based circuits, and then separately generate specialized proofs for them.
Furthermore, we add \emph{commit-and-prove} (CaP) components based on LegoSNARK~\cite{campanelli2019legosnark} for gluing the separately generated specialized proofs, aiming to prove that the current layer indeed takes as inputs the previous layer's outputs.\looseness=-1

Thirdly, we enable batch verification via proof aggregation, and only a single proof is generated for a CNN model tested by multi-tester inputs.
We aggregate multiple proofs with regard to the same CNN but different test inputs from multiple testers based on Snarkpack\cite{gailly2021snarkpack}.
We note that the aggregation computation and proof validation can be delegated to computation-powerful parties in a competition manner, \emph{e.g.}, decentralized nodes who run a secure consensus protocol to maintain the publicly accessible platform previously described in Fig.~\ref{fig:intro}.
The validation result will be elected according to the consensus protocol, and then uploaded to the public platform and attached to the corresponding CNN model.
To the end, a future user who is interested in the CNN model can enjoy a lightweight verification by merely checking the validation result, via browsing the platform.

\iffalse
As for experiments, we present a proof-of-concept implementation, and evaluate the implementation. 
%
Our QMP-based method for matrix multiplication performs about $17.6\times$ faster than the existing QAP-based method in \textsf{Setup} time, and $13.9\times$ faster in proving time.
%
We can also handle the matrix multiplication in dimensions more than $220\times220$ while the matrix dimension the previous QMP-based method can handle is limited to $220\times220$~\cite{keuffer2018efficient}.
%
However, our proof size is 2 or 3 orders larger than that of the QAP-based zk-SNARK.
%
Despite the demerit, we observe that the order of magnitude becomes smaller with an increasing number of multiplications, which demonstrates our method can be more suitable to handle the computation involving in a large number of multiplications.
\fi

%\subsection{Contribution}
\wjs{In summary, this work makes the following contributions:
\begin{itemize}
    \item We generate zk-SNARK proofs for the CNN testing based on FHE and collaborative inference, which protects both of the model and data privacy and ensures the integrity.
    \item We present a new QMP-based arithmetic circuit to express convolutional relations for efficiency improvement.
    \item
    We aggregate multiple proofs respective to a same CNN model and different testers’ inputs for reducing the entire verification cost.
    \item We give a proof-of-concept implementation, and the experimental results demonstrate that our QMP-based method for matrix multiplication performs about $17.6\times$ faster than the existing QAP-based method in \textsf{Setup} time, and $13.9\times$ faster in proving time. The code is
    available at \href{https://github.com/muclover/pvCNN}{https://github.com/muclover/pvCNN}.
\end{itemize}}

\subsection{Organization}
\wjs{In Section~\ref{sec:work}, we revisit related work. In Section~\ref{sec:preliminaries}, we introduce intrinsic preliminaries. Section~\ref{sec:problem} elaborates the scenario problem this work is concerned about.
Section~\ref{sec:design} presents our solutions with customized proof designs. In Section~\ref{sec:analysis} and Section~\ref{sec: experiments}, we show the security analysis and experimental results, respectively.}

\section{RELATED WORK}~\label{sec:work}
\begin{table}[h]
\vspace{-15pt}
\scriptsize
 \centering
 \caption{Comparison of verifiable CNN prediction/testing schemes. Consider applying a two-dimensional (2-D) convolution to a filter matrix of $m \times m$ and a test data matrix of $n \times n$. The column of Proving Time is for such 2-D convolutions on $M$ filter matrices and $Mn^2$ test data matrices (under plaintext). For simplicity, we note $M>m$ and the channel number of test data is $1$. \wjs{Herein, the requirements include (a) public verifiability and no need of interaction, (b) privacy preservation of data and model and (c) batch proving and verification, respectively. See concrete description in Section~\ref{sec:intro}.}}
 %(We discuss $M\leq m$ in Appendix~\ref{app:res})
 \label{tab:pre} 
 \begin{threeparttable}
 \begin{tabular}{lccc|c}
  \toprule 
  \multirow{2}*{\textbf{Schemes}} &\multicolumn{3}{c|}{\textbf{Requirements}} & \multirow{2}*{\textbf{Proving Time}}\\ 
  \cline{2-4}
  & (a) & (b) & (c) &   \\
  \toprule
  SafetyNets\cite{ghodsi2017safetynets} & \emptycirc & \emptycirc & \emptycirc & $O(M\cdot Mn^2 \cdot(n^2\cdot m^2))$  \\
  Keuffer's\cite{keuffer2018efficient} & \fullcirc &  \halfcirc  & \emptycirc & $O(M\cdot Mn^2 \cdot(n^2\cdot m^2))$   \\
  SafeTPU\cite{collantes2020safetpu} & \emptycirc & \emptycirc & \halfcirc &  $O(M\cdot Mn^2 \cdot(n^2\cdot m^2 \cdot \frac{m^2}{n^2}))$  \\
  Madi's\cite{madi2020computing} &  \emptycirc  & \fullcirc  &  \emptycirc & $O(M\cdot Mn^2 \cdot(n^2\cdot m^2))$  \\
  vCNN\cite{lee2020vcnn} & \fullcirc & \halfcirc & \emptycirc &   $O(M\cdot Mn^2 \cdot(n^2+m^2))$\\
  VeriML\cite{zhao2021veriml} & \fullcirc & \fullcirc &  \emptycirc &  $O(M\cdot Mn^2 \cdot(n^2\cdot m^2))$  \\
  ZEN\cite{fengzen} & \fullcirc  & \halfcirc & \halfcirc & $O(M\cdot Mn^2 \cdot(n^2\cdot m^2 \cdot \frac{1}{\textsf{fac}}))$   \\
  zkCNN\cite{liuzkcnn} &  \emptycirc  &  \halfcirc & \emptycirc  &  $O(M\cdot Mn^2 \cdot(n^2+m^2))$  \\
  Ours &  \fullcirc  &  \fullcirc & \fullcirc  &  $O(Mn^2 \cdot Mn^2)$   \\
  \bottomrule
 \end{tabular}
  \begin{tablenotes}
    \item \emptycirc~denotes the requirement is not satisfied; \halfcirc~denotes the scheme partially supports the requirement; \fullcirc~denotes the scheme fully supports the requirement. \wjs{Besides, \textsf{fac} depends on the real size of filters. ZEN's optimization method may not be effective for small filters. See Table~4 in~\cite{fengzen}.}
  \end{tablenotes}
 \end{threeparttable}
 \vspace{-10pt}
\end{table}

\noindent\textbf{Verifiable CNN Prediction/Testing.} 
Many prior schemes~\cite{ghodsi2017safetynets, collantes2020safetpu, lee2020vcnn, fengzen, keuffer2018efficient, zhao2021veriml, madi2020computing, liuzkcnn} fail to fully satisfy our requirements, as summarized in TABLE~\ref{tab:pre}.
The previous schemes can be roughly classified into two major groups, according to their underlying cryptographic proof techniques, such as Sum-Check like protocols~\cite{thaler2013time, zhang2020doubly, goldwasser2015delegating} and Groth16 zk-SNARK-based systems~\cite{groth2016size, agrawal2018non, campanelli2019legosnark, fiore2020boosting, mouris2021zilch}.
Based on the underlying techniques, some schemes clearly elaborate how they support privately verifiable and interactive proofs\cite{ghodsi2017safetynets, collantes2020safetpu, madi2020computing}; some schemes support partial privacy protection\cite{madi2020computing, lee2020vcnn,fengzen}.
%also, most schemes do not require a batch manner\cite{ghodsi2017safetynets, keuffer2018efficient, madi2020computing,lee2020vcnn,zhao2021veriml, liuzkcnn}.
%
A few schemes~\cite{niu2020toward,zhang2020zero} do not consider CNN.
\wjs{In addition, other promising directions are designing an efficient and memory-scalable proof protocol to support complicated neural network inference~\cite{Weng21Mystique}, and using clever verification methods to convert any semi-honest secure inference into a malicious secure one~\cite{dong2023fusion}, when entities are allowed to be simultaneously online.}\looseness=-1
%
%
%Notably, zkCNN\cite{liuzkcnn} is the latest scheme based on Sum-Check like protocols, while the state-of-the-art schemes built on zk-SNARK are vCNN\cite{lee2020vcnn} and ZEN\cite{fengzen}.

We next differentiate this paper from previous schemes in terms of our requirements.
\wjs{
\emph{\textbf{Regarding requirement (a).}} 
We adopt a class of zk-SNARK proof systems that provide public delegatability and public verifiability, which is suitable to our scenario. 
We depart from previous schemes like SafetyNets, SafeTPU and Madi et al.’s work, since they generally need not the properties. 
Compared to the line of Groth16 zk-SNARK-based schemes~\cite{lee2020vcnn,fengzen,keuffer2018efficient,zhao2021veriml}, we make new contributions considering the following requirements (b) and (c).}
\wjs{
\emph{\textbf{Regarding requirement (b).}} 
We protect data and model privacy by enabling CNN inference on encrypted test data while ensuring the correctness of the result.
Specifically, we protect the test data from the entity who runs the CNN inference, preserve the model privacy from the entity who provides the test data, and prevent any verifier who checks the result correctness from learning private information either of the test data or of the model. 
Most existing work protects the privacy either of model or of test data depending on which one is treated as the witnesses on the side of a prover. 
For example, Keuffer’s~\cite{keuffer2018efficient}, vCNN~\cite{lee2020vcnn} and zkCNN~\cite{liuzkcnn} treat model parameters as the witnesses to be proven, while in ZEN~\cite{fengzen}, test data is witness and model parameters should be shared between a prover and a verifier. 
Although Madi’s~\cite{madi2020computing} provides privacy protection on the both sides, by leveraging homomorphic encryption and homomorphic message authenticator, the work requires a designated verifier. 
As requirement (a) mentioned, our work considers a non-designated verifier, since each user as a verifier is not designated in advance.}
\wjs{
\emph{\textbf{Regarding requirement (c).}} 
We improve the proving time for handling 2-D convolutional operations on batches of filters and test data, as well as aggregate multiple proofs for reducing verification cost. 
In terms of proving time, Keuffer’s and VeriML directly apply the Groth16 zk-SNARK~\cite{groth2016size} to the 2-D convolutional operations between each filter and each input, resulting in $O(n^2\cdot m^2)$ proving time. 
For such convolutional operations, vCNN reduces the proving time to $O(n^2+m^2)$ by firstly transforming the original convolutional representations of a sum of products into a product of sums representations, and then employing the quadratic polynomial program (QPP) in the original QAP-based zk-SNARK. 
ZEN also makes effort to reduce the proving time by reducing the number of constraints in the underlying zk-SNARK system based on a new stranded encoding method. 
Essentially, ZEN’s encoding method can be complementary to other zk-SNARK-based applications, including ours, when encoding the original numbers into finite field elements. 
\emph{Despite the effort, they do not consider the batch case as we previously described. 
When handling convolutional operations between $M$ filters and $Mn^2$ input data, we have $O(Mn^2\cdot Mn^2)$ proving complexity. 
It derives from our two-step effort: first, we strategically represent the $M$ filters into a matrix of $Mn^2\times Mn^2$ and meanwhile reshape the $Mn^2$ inputs into another matrix; second, we employ a new quadratic matrix program (QMP) in the QAP-based zk-SNARK, making proving time depend on the dimension of matrices, \emph{i.e.}, $Mn^2\times Mn^2$.
Besides, the previous work does not consider proof aggregation as our work.}}

\wjs{\noindent\textbf{Secure Outsourced ML Inference.}
Our work delegates partial computation of CNN inference to a strong but untrusted server, which relates to a long line of work about secure outsourced ML inference.
One of most relevant work starts from CryptoNets, enabling CNN inference on FHE-encrypted data.
CryptoNets~\cite{gilad2016cryptonets} inspires many follow-up schemes~\cite{mohassel2017secureml, mishra2020delphi,juvekar2018gazelle,kumar2020cryptflow,kai2019lightweight,dong2023fusion} that aim to improve accuracy, communication or computational efficiency by carefully leveraging cryptographic primitives, other than FHE~\cite{gilad2016cryptonets, hesamifard17cryptoDL}.
For example, Huang \emph{et al.}~\cite{kai2019lightweight} achieve efficient and accurate CNN feature extraction using a secret sharing-based encryption technique, which avoids approximating the ReLU function with a low-degree polynomial that is needed by CryptoNets, and thereby ensures accuracy.
However, the integrity of outsourced computation is out of the consideration of the above schemes.
Similarly targeted at the scenario of privacy-preserving outsourced computation, another line of work studies a group of toolkits, which may pave the way for achieving more complex privacy-preserving ML.
The toolkits support general operations of integer numbers~\cite{liu2016efficient} and rational numbers~\cite{liu2016privacy}, as well as enable large-scale computation~\cite{liu2018privacy}; many ML-based applications usually contain such computation
characteristics.}

\section{PRELIMINARIES}\label{sec:preliminaries}
\subsection{CNN Prediction}\label{subsec:cnn}
We present here a prediction process of a CNN model on a step-by-step basis.
The CNN model basically contains two convolution (conv) layers and three full connection (fc) layers in order; other layers between them include activation (act) and max pooling or average pooling (pool) layers, and the output layer is softmax layer.
\wjs{Notice, complex CNN models generally contain the above layers.}
Specifically, taking as inputs a single-channel test data $\textbf{X}$ which is a $n\times n$ matrix, \emph{e.g.}, gray-scale image, the layer-by-layer operations can be conducted sequentially:  $y_o=f^{o}(f^{fc}(f^{fc}(f^{pool}(f^{act}(f^{conv}(f^{pool}(f^{act}(f^{conv}(\textbf{X}))))))))),$
where $f^{o}$ is the output function which selects out the maximal value among the values outputted by the last $f^{fc}$, determining the prediction output.

We proceed to elaborate the layer-by-layer operations in detail.
Denote a $m\times m$ weight matrix $\textbf{W}^{(k)}$ in the $k_{th}$ layer. \emph{(1)~\textbf{Convolutional Layer~$f^{conv}$.}}
Two dimensional (2-D) convolution operations will be applied to input matrices and weight matrices (also called \emph{filters}). Here, we show a 2-D convolution operation applying to two matrices $\textbf{X}$ and $\textbf{W}^{(1)}$ with a $(n-m+1)\times (n-m+1)$ matrix $\textbf{Y}^{(1)}$ as output:
$$\textbf{Y}^{(1)}[i][j]=\sum_{l^{(1)}_r=0,l^{(1)}_c=0}^{m-1,m-1}\textbf{X}[i+l^{(1)}_r][j+l^{(1)}_c]\times \textbf{W}^{(1)}[l^{(1)}_r][l^{(1)}_c],$$
where $i,j\in [0, n-m+1)$ and the row (resp. column) index of the weight is $l^{(1)}_r$ (resp. $l^{(1)}_c$), and the stride size is $1$.
The subsequent $f^{conv}$ are done similarly, applying to the previous layer's output matrices and the weight matrices in the current layer.
\emph{(2)~\textbf{Activation Layer~$f^{act}$.}}
There are two widely used activation functions, applying to each element of the output matrices of $f^{conv}$ in layer $l$, in order to catch non-linear relationships.
Concretely, the two functions are $f^{act}(\textbf{Y}^{(k)}[i][j])=max(\textbf{Y}^{(k)}[i][j],0)$ named the ReLU function, and $f^{act}(\textbf{Y}^{(k)}[i][j])=\frac{1}{1+e^{-\textbf{Y}^{(k)}[i][j]}}$ named the Sigmoid function.
\emph{(3)~\textbf{Pooling Layer~$f^{pool}$.}}
Average pooling and max pooling are two common pooling functions for reducing the dimension of the output matrices in certain layer $k$.
They are applied to each region covered by a $m\times m$ filter, within each output matrix of the previous $f^{act}$ layer.
Specifically, an average pooling function is done by $(\textbf{Y}^{(k)}[0][0], ... , \textbf{Y}^{(k)}[m-1][m-1])/m^2$, and a max pooling function is by \textsf{max}($\textbf{Y}^{(k)}[0][0], ... , \textbf{Y}^{(k)}[m-1][m-1]$).
\emph{(4)~\textbf{Full Connection Layer~$f^{fc}$.}}
In this layer, each output matrix of the previous layer multiplies by each weight matrix, and then add with a bias in the current layer.
We will omit the bias, for simplicity. \emph{(5)~}After executing the mentioned sequential operations,  $f^o$ outputs a prediction label $l_{test}$ for $\textbf{X}$. 

\subsection{Non-Interactive Zero-Knowledge Arguments}
Non-interactive zero-knowledge (NIZK) arguments allow a prover to convince any verifier of the validity of a statement without revealing other information.
Groth~\cite{groth2016size} proposed the most efficient zk-SNARK scheme, with small constant size and low verification time. %(named \textsf{Groth16} later).
%, which is widely applied and implemented in the scenarios of verifiable computation, such as Filecoin and Zcash.
%
Our work leverages a CaP \textsf{Groth16} variant to prove arithmetic circuit satisfiability with committed inputs, parameters and outputs.
% 
%We highlight the usability of the CaP functionality for our scenario in Appendix~\ref{app:cap}. 
% 
We here recall some essential preliminaries. 

\noindent\textbf{Bilinear Groups.}
We review Type-3 bilinear group ($p$, $\mathtt{G}_1$, $\mathtt{G}_2$, $\mathtt{G}_T$, $e$), where $p$ is a prime, and $\mathtt{G}_1, \mathtt{G}_2$ are cyclic groups of prime order $p$.
Note that $g \in \mathtt{G}_1$, $h \in \mathtt{G}_2$ are the generators.
Then, $e:\mathtt{G}_1 \times \mathtt{G}_2 \rightarrow \mathtt{G}_T$ is a bilinear map, that is, $e(g^a, h^b) = e(g, h)^{ab}$, where $a, b \in \mathtt{Z}_p, e(g, h)^{ab} \in \mathtt{G}_T$.

\noindent\textbf{Arithmetic Circuits.}
Arithmetic circuits are the widely adopted computational models for expressing the computation of polynomials over finite fields $\mathtt{F}_p$.
An arithmetic circuit is formed with a directed acyclic graph, where each vertex of fan-in two called \emph{gate} is labelled by an operation $+$ or $\times$, and each edge called \emph{wire} is labelled by an operand in the finite field.
%
%We review the arithmetic circuit definition as follows:
%
%
%\emph{Definition.}~\emph{Given an arithmetic circuit 
%$\phi_{\textsf{AC}}: \mathtt{F}_p^{n_{in}}\rightarrow \mathtt{F}_p^{n_{ot}}$, taking as inputs $n_{in}$ variables or constants from the finite filed $\mathtt{F}_p$, and executing the operations sum or product via its vertices, which returns $n_{ot}$ values in $\mathtt{F}_p$ as outputs. Herein, $p$ is an appropriate modulus.} 
%
%Based on the above definition, we note that the arithmetic circuit's input wires and output wires are labeled by given $n_{in}$ inputs and $n_{ot}$ outputs correspondingly; its gates can be labeled by a set of operations $\{+, \times\}$, usually named as addition gates and multiplication gates.
%
%so as to express the polynomials on the input wires.
%
Besides, there are usually two measures for the complexity of a circuit, such as its size and depth.
%and the maximal degree of the computing polynomials
%
Specifically, the number of wires of the circuit determines its size; the longest path from inputs to the outputs determine its depth.

%Arbitrary arithmetic circuits can be naturally formulated as a set of quadratic arithmetic programs (QAP)~\cite{gennaro2013quadratic}.
%
%We next introduce the definition of QAP (more details see Appendix~\ref{app:ari} and the reference~\cite{gennaro2013quadratic}).
%
%The complementary description about arithmetic constraint systems for QAP refer to Appendix~\ref{app:ari}. More details to build QAP for arithmetic circuits can be founded in~\cite{gennaro2013quadratic}.

\noindent\textbf{Quadratic Arithmetic Program (QAP)~\cite{gennaro2012quadratic}} For an arithmetic circuit with $n_{in}$ input variables, $n_{ot}$ output variables in $\mathtt{F}_p$ ($p$ is a large prime) and $n_*$ multiplication gates, a QAP is consisting of three sets of polynomials $\{l_i(X), r_i(X), o_i(X)\}_{i=0}^m$ and a $n_*$-degree target polynomial $t(X)$. 
The QAP holds and $(a_0,...,a_{n_{in}},a_{m-n_{ot}+1},...,a_m)\in \mathtt{F}_p^{n_{in}+n_{ot}}$ is a valid assignment for the input/output variables of the arithmetic circuit \emph{iff} there is $(a_{n_{in}+1},...,a_{m-n_{ot}})\in \mathtt{F}_p^{m-n_{in}-n_{ot}}$ such that 
$\sum_{i=0}^ma_il_i(X))\cdot \sum_{i=0}^ma_ir_i(X)\equiv \sum_{i=0}^ma_io_i(X)\quad \text{mod}\quad t(X).$
Herein, the size of the QAP is $m$ and the degree is $n_*$. 
%
%In addition, considering two QAP, a new QAP can be built via composition. 
% 
%Its size is no more than the sum of the two QAP' sizes, and its degree is the sum of the degrees of the two QAP.
%

\noindent\textbf{zk-SNARK.}
\textsf{Groth16}~\cite{groth2016size} is a QAP-based zk-SNARK scheme, enabling proving the satisifiability of QAP via conducting a divisibility check between polynomials, which is equivalent to prove the satisifiability of the corresponding circuits.
We figure out some major notations which can pave the way for presenting \textsf{Groth16} and our design later.
Concretely, we can specify the mentioned assignment $(a_0,...,a_{n_{in}},a_{m-n_{ot}+1},...,a_m)\in \mathtt{F}_p^{n_{in}+n_{ot}}$ as \textit{\textbf{statement}} \textsf{st} to be proven, and specify ($a_{n_{in}+1},...,a_{m-n_{ot}}$)~
$\in \mathtt{F}_p^{m-n_{in}-n_{ot}}$ as \textit{\textbf{witness}} \textsf{wt} only known by a prover.
Then, we define the polynomial time computable binary \textit{\textbf{relation}} $\mathtt{R}$ that comprises the pairs of (\textsf{st}, \textsf{wt}) satisfying the denoted QAP 
$\{\{l_i(X), r_i(X),$ $o_i(X)\}_{i\in[0,m]}, t(X)\}$.
The degree of $\{l_i(X), r_i(X), o_i(X)\}$ is lower than that of $t(X)$.
If $\mathtt{R}$ holds on (\textsf{st}, \textsf{wt}), $\mathtt{R}=1$; otherwise, $\mathtt{R}=0$.
Formally, the relation is denoted as 
$\mathtt{R}=\{(\textsf{st}, \textsf{wt})\mid \textsf{st}:=(a_0,...,a_{n_{in}},a_{m-n_{ot}+1},...,a_m),\textsf{wt}:=(a_{n_{in}+1},...,a_{m-n_{ot}}).$
Associated to the relation $\mathtt{R}$, we denote that \textit{\textbf{language}} $\mathtt{L}$ contains the statements that the corresponding witnesses exist in $\mathtt{R}$, that is, $\mathtt{L}=\{\textsf{st}\mid \exists \textsf{wt}~\text{s.t.}~\mathtt{R}(\textsf{st}, \textsf{wt})=1\}$.

Based on the above notations, we are proceeding to review the definition of a zk-SNARK scheme for the relation $\mathtt{R}$.
Specifically, a zk-SNARK scheme is consisting of a quadruple of probabilistic polynomial time (PPT) algorithms (\textsf{Setup}, \textsf{Prove}, \textsf{Verify}, \textsf{Sim}), which satisfies three properties:

\vspace{3pt}
\noindent$\underline{\textsf{Setup}(1^{\lambda},\mathtt{R})\rightarrow(\textsf{crs}, \textsf{td})}$: the \textsf{Setup} algorithm takes a security parameter $\lambda$ and a relation $\mathtt{R}$ as inputs, which returns a common reference string $\textsf{crs}$ and a simulation trapdoor $\textsf{td}$.

\vspace{3pt}
\noindent$\underline{\textsf{Prove}(\textsf{crs},\textsf{st},\textsf{wt})\rightarrow \pi}$: the \textsf{Prove} algorithm takes \textsf{crs}, statement \textsf{st} and witness \textsf{wt} as inputs, and outputs an argument $\pi$.\looseness=-1

\vspace{3pt}
\noindent$\underline{\textsf{Verify}(\mathtt{R},\textsf{crs},\textsf{st},\pi)\rightarrow 0/1}$: the \textsf{Verify} algorithm takes the relation $\mathtt{R}$, common reference string $\textsf{crs}$, statement $\textsf{st}$ and argument $\pi$ as inputs, and returns $0$ as \textsf{Reject} or $1$ as \textsf{Accept}.

\vspace{3pt}
\noindent$\underline{\textsf{Sim}(\mathtt{R},\textsf{td},\textsf{st}) \rightarrow \pi^{'}}$: the \textsf{Sim} algorithm takes the relation $\mathtt{R}$, simulation trapdoor $\textsf{td}$ and statement $\textsf{st}$ as inputs, and returns an argument $\pi^{'}$.

%Formally, a zero-knowledge proof system satisfies the following three properties.

%
\noindent \emph{$\cdot$~\textbf{Completeness.}} Given a security parameter $\lambda$, $\forall (\textsf{st},\textsf{wt})\in \mathtt{R}$, a honest prover can convince a honest verifier the validity of a correctly generated proof with an overwhelming probability, that is, 
%\begin{align*}
$\textsf{Pr}\{ \textsf{Setup}(1^{\lambda},\mathtt{R})\rightarrow(\textsf{crs}, \textsf{td}), \textsf{Prove}(\textsf{crs},\textsf{st},\textsf{wt}) \rightarrow \pi|\\
\textsf{Verify}(\mathtt{R},\textsf{crs},\textsf{st},\pi)\rightarrow 1\}=1-\text{negl}(1^{\lambda}).$
%\end{align*}

\noindent\emph{$\cdot$~\textbf{Computational soundness.}}
For every computationally bounded adversary $\mathcal{A}$, if an invalid proof is successfully verified by a honest verifier, there exists a PPT extractor $\mathcal{E}$ who can extract $\textsf{wt}$ with a non-negligible probability, that is, 
%\begin{align*} 
    $\textsf{Pr}\{\exists (\textsf{st},\textsf{wt})\notin \mathtt{R},\mathcal{A}(\textsf{crs},\textsf{st},\textsf{wt}) \rightarrow \pi | \textsf{Verify}(\mathtt{R},\textsf{crs},\textsf{st},\pi)\rightarrow 1\\
     \wedge \mathcal{E}(\textsf{crs})\rightarrow \textsf{wt}\}=1-\text{negl}(1^{\lambda}).$
%\end{align*}

\noindent \emph{$\cdot$~\textbf{Zero knowledge.}}
For every computationally bounded distinguisher $\mathcal{D}$, there exists the \textsf{Sim} algorithm such that $\mathcal{D}$ successfully distinguishes a honestly generated proof from a simulation proof with a negligible probability, that is,  
%\begin{align*}
    $\textsf{Pr}\{\textsf{Setup}(1^{\lambda},\mathtt{R})\rightarrow(\textsf{crs}, \textsf{td}), \textsf{Prove}(\textsf{crs},\textsf{st},\textsf{wt})\rightarrow \pi,\\
    \textsf{Sim}(\mathtt{R},\textsf{td},\textsf{st}) \rightarrow \pi^{'}|\mathcal{D}(\pi, \pi^{'})=1\}\leq \text{negl}(1^{\lambda}).$
%\end{align*}

\wjsRR{\noindent\textbf{Commit-and-prove zk-SNARK.}
We particularly emphasize the capability of CaP zk-SNARK that is useful for our scenario.
The formal definition can be found in Definition~3.1 of~\cite{campanelli2019legosnark}.
Essentially, the CaP capability allows a prover to convince a verifier of "$C_{y^1}$ commits to $y^1$ such that $y^1=F_1(x)$ and meanwhile $y^2=F_2(y^1)$".}

\wjsRR{Intuitively, we suppose $C_m$ committed to an intermediate model $m$ of a computation and meanwhile the model $m$ is taken as input for completing subsequent computation, that is $m=F(x) \wedge y=G(m)$.
We next explain how such capability matches the features of our scenario: 
(\emph{a}) \emph{\textbf{Compression}}. Before proving the correctness of a CNN testing, the CNN developer (resp. testers) store commitments to the CNN model (resp. test inputs).
Herein, commitment is a lightweight method to compress large-size CNN models and test inputs for saving storage overhead.
In terms of security, both the models and test inputs are sealed as well.
(\emph{b}) \emph{\textbf{Flexibility}}. Committing to the CNN model in advance provides the developer with certain flexibility that enables proving later-defined statements on the previously committed CNN model over unexpected test inputs.
(\emph{c}) \emph{\textbf{Interoperability}}. The proofs corresponding to the process of CNN testing are generated separately due to that a CNN model is partitioned into separate parts, so commitments to the CNN model will provide the interoperability between the separately generated proofs (see Fig.~\ref{fig:relation_up} and Fig.~\ref{fig:relation_down}).}

%
% a model architecture often consists of heterogeneous computation, \emph{i.e.}, linear and non-linear computation, 
%
%To the end, this paper will resort to zk-SNARK with commit-and-prove capability for generating correctness proofs of CNN testing over multi-tester test data.\looseness=-1

\section{Problem Statement}~\label{sec:problem}
%We firstly introduce a basic workflow of our CNN testing scenarios, and then figure out threat assumptions as well as our security goals.   
%In this section, we elaborate a basic workflow between manually distrustful stakeholders, showing how a CNN developer convinces future users of the truth of multi-shot CNN prediction over the test data from multiple testers.
%
%We subsequently present our threat assumptions and security goals.
%This section firstly introduces a process of CNN collaborative inference, and subsequently presents the scenario workflow, and lastly clarifies some necessary threat assumptions.
% 
\vspace{-30pt}

\subsection{Collaborative Inference}\label{sec:splitting}
We take the CNN model described in Section~\ref{subsec:cnn} for image classification as an example, and partition it into two parts.
%
%Recall that we elaborate the prediction process of \textsf{LeNet-5} in Section~\ref{subsec:cnn}.
%
We choose the first pooling layer as a split point, such that the model can be partitioned into a part named \textsf{PriorNet} $F_{\textbf{M}_1}=f^{pool_1}(f^{act_1}(f^{conv_1}(\cdot)))$ and another part named \textsf{LaterNet} $F_{\textbf{M}_2}=f^{o}(f^{fc_4}(f^{fc_3}(f^{pool_2}(f^{act_2}(f^{conv_2}(\cdot))))))$.
\wjsSec{Suppose that the chosen split point is sufficiently optimized~\cite{he2019model} or the model is pre-processed~\cite{ryffel2019partially} for privacy protection, which makes the outputs of the \textsf{PriorNet} not reveal private information of its input data.}
\wjsSec{The choice of the split point depends on the model architecture. A general principle is that the deeper layer a split point locates at, the less privacy will be leaked.}
Besides, we note that \textsf{PriorNet} $\textbf{M}_1$ will be locally evaluated and \textsf{LaterNet} $\textbf{M}_2$ will be delegated.
\begin{figure}[h]
\vspace{-10pt}
    \centering
    \includegraphics[width=0.6\linewidth]{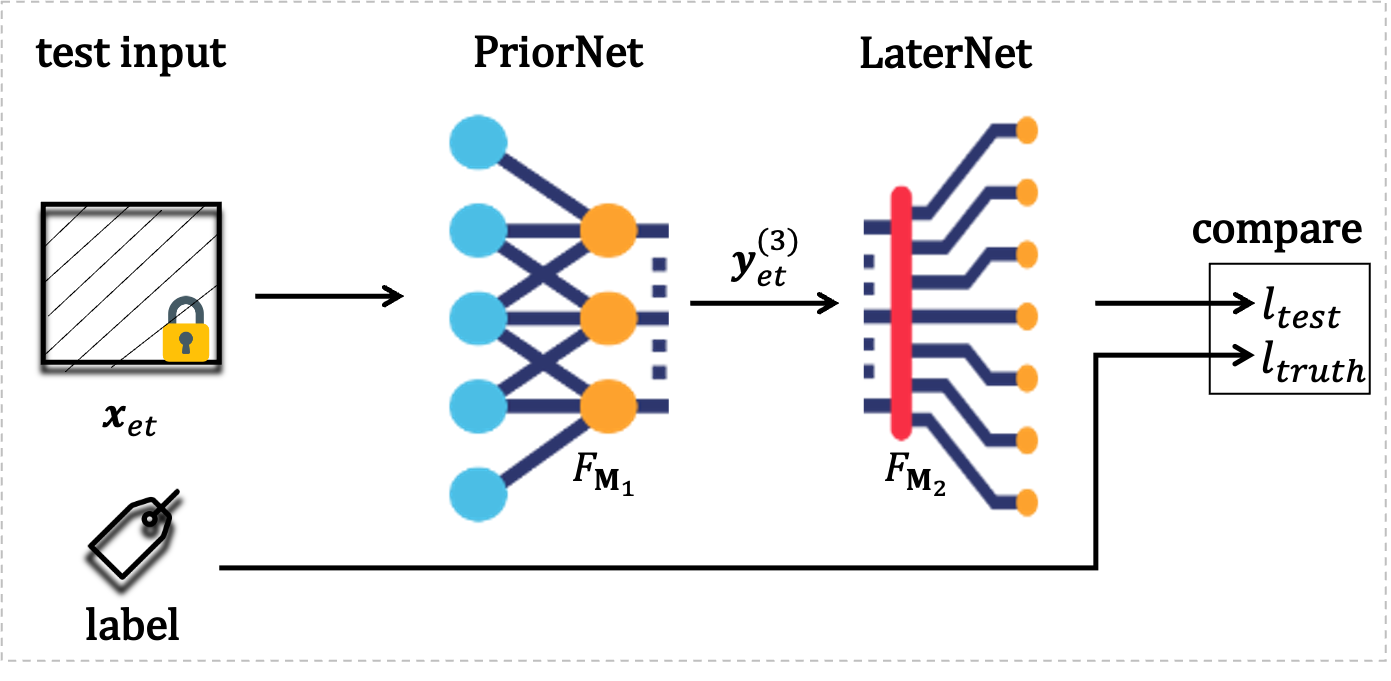}
    \caption{Pipeline of \textsf{PriorNet} and \textsf{LaterNet} inference.}
    \label{fig:split}
    \vspace{-10pt}
\end{figure}
We now elaborate in Fig.~\ref{fig:split} the pipeline of \textsf{PriorNet} and \textsf{LaterNet} inference over a given test input encrypted under a tester's public key of homomorphic encryption.
We note that the corresponding label of the test input is not encrypted.
Concretely, \textsf{PriorNet} is privately performed by the developer. 
It takes as input an encrypted test $\textbf{X}_{et}\in \mathcal{R}_q$, and produces an FHE-encrypted intermediate result $\textbf{Y}^{(3)}_{et}\in \mathcal{R}_q$ as the output.
\wjs{Since the homomorphic encryption cannot efficiently support the operations beyond additions and multiplications~\cite{gilad2016cryptonets, aono2017privacy}, the activation layer within the \textsf{PriorNet} is approximated by a $2$-degree polynomial function.}

Successively, $\textbf{Y}^{(3)}_{et}$'s plaintext $\textbf{Y}^{(3)}$ will be fed into the \textsf{LaterNet} $F_{\textbf{M}_2}$ whose computations are delegated to a service provider.
We can leverage the re-encryption technique to let the service provider obtain the plaintext $\textbf{Y}^{(3)}$.
Since the computations are conducted on plaintext domain, the later activation functions remain unchanged.
Finally, \textsf{LaterNet} outputs the prediction result $l_{\textsf{test}}$.
The service provider proceeds to compare the previously given true label $l_{\textsf{truth}}$ with $l_{\textsf{test}}$, and returns an equality indicator $0$ or $1$.
For a batch of test data, he would count the number of $1$ indicator, and return a proportion of correct prediction.
For the case of using a posterior probability vector as a prediction output, we determine the indicator value according to the distance between the prediction output and the corresponding truth.

\subsection{Basic Workflow} 
\begin{figure}[h]
\vspace{-5pt}
\small
    \centering
    \includegraphics[width=0.8\linewidth]{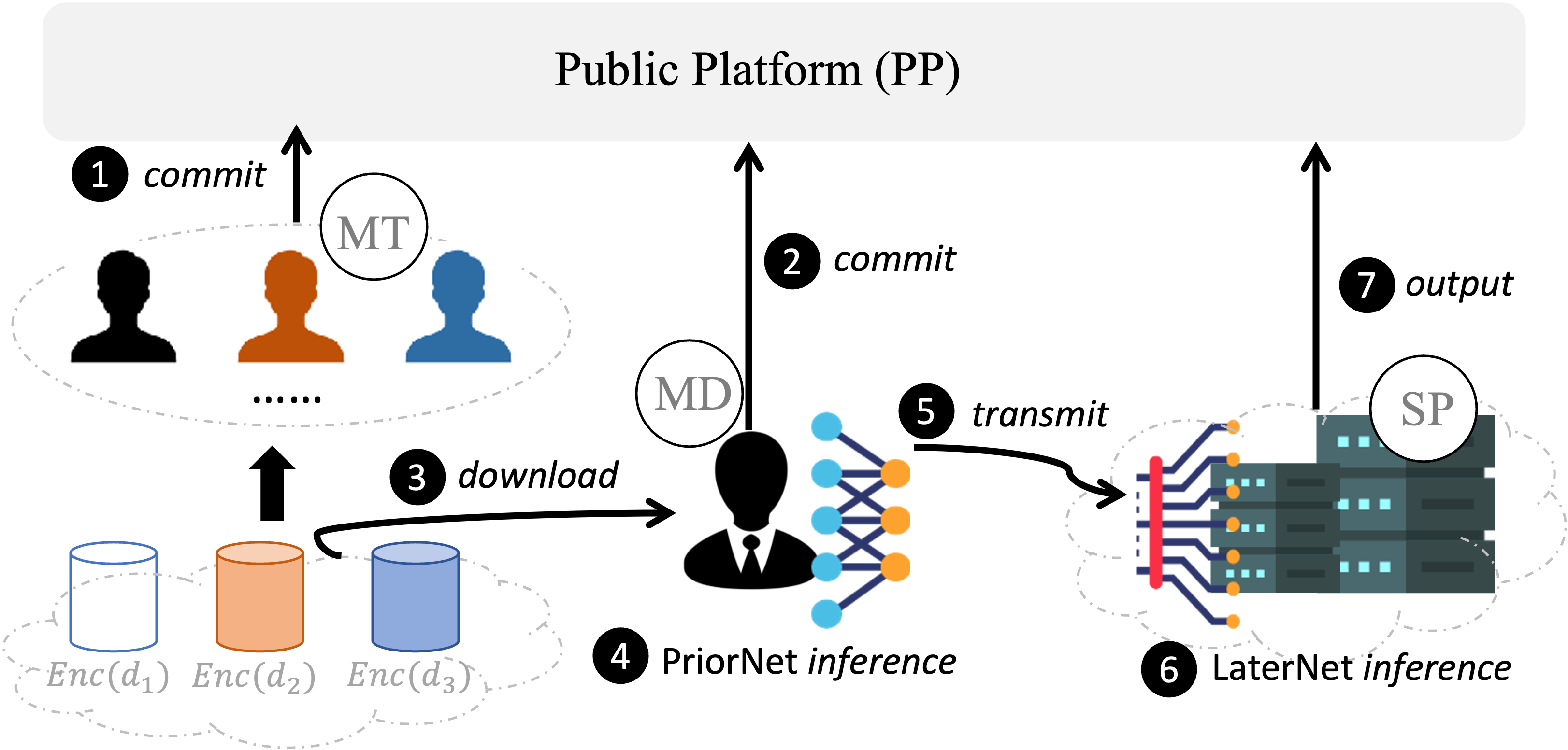}
    \caption{Scenario overview of pvCNN}
    \label{fig:overview}
    \vspace{-10pt}
\end{figure}

We consider four main entities---\emph{public platform (PP)}, \emph{model tester (MT)}, \emph{model developer (MD)} and \emph{service provider (SP)} as demonstrated in Fig.~\ref{fig:overview}.
% and \emph{decentralized nodes (DN)}
% 

Specifically, the \emph{PP} in real world may refer to a publicly accessible and online platform, \wjs{which can be maintained by a set of computational nodes.}
On the platform, an \emph{MD} is allowed to advertise his/her pre-trained CNN model by announcing some usage descriptions and model performance, so as to attract users of interest for earning profits, while an \emph{MT} can provide test data for measuring the model performance.
In order to establish trust among the \emph{MD}, the \emph{MT} and future users in the scenario, the pvCNN wants to enable verifiable announcements on the public platform, that is, any future user is assured of the truthfulness of model performance without learning any information of either \wjsRR{private model parameters} or test data.
We note the \emph{MD} can delegate computation-consuming tasks to the powerful \emph{SP}.
%
\iffalse
We allow model testers to freely participate in the platform by uploading some descriptions about their test data, when they are interested in certain pre-trained CNNs on the platform. They can in advance store test data in an encryption manner on the publicly accessible cloud.
%
Model developers can download the model testers' test data from the cloud to test their CNN models and announce verifiable performance results on the platform, with the aim of advertising their CNN models and the corresponding performance.
%
The service provider offer powerful computational and storage resources for model developers.
\fi
%
 
The basic workflow around the entities is shown below:
~\ding{202}~\emph{MTs} freely participate in the platform and commit to the test data. They also store encrypted test data by using a leveled FHE (L-FHE) scheme~\cite{fansomewhat} on the publicly accessible server.
~\ding{203}~The \emph{MD} splits his CNN model into a private \textsf{PriorNet} which is kept at local device, as well as, a public \textsf{LaterNet} which is sent to the \emph{SP}; 
After that, he generates two commitments to the \textsf{PriorNet} and \textsf{LaterNet}, and sends them to the \emph{PP}.
~\ding{204}~The \emph{MD} downloads the ciphered test data.
~\ding{205}~The \emph{MD} evaluates the plaintext \textsf{PriorNet} on the encrypted test data and generates a proof of executing the inference computation as promised.
~\ding{206}~The \emph{MD} sends the encrypted output of the Step~\ding{205}. Note that the encrypted output can be transformed into the ciphertext under the  \emph{SP}'s public key via re-encryption.
~\ding{207}~The \emph{SP} successively runs \textsf{LaterNet} inference in the clear by taking the plaintext output of \textsf{PriorNet} inference as inputs, and proves that \textsf{LaterNet} is executed correctly with the true inputs and meanwhile \textsf{LaterNet} is consistent to the committed one.
~\ding{208}~The \emph{SP} submits the final test results (\emph{e.g.,} classification correctness rate) to the platform, and distributes proofs to the decentralized nodes for proofs aggregation and validation.

\subsection{Threat Assumptions}\label{subsec:assumption}
\noindent\textbf{Model Developers.}
We consider CNN developers are untrusted; they can provide untruthful prediction results, \emph{e.g.}, outputting meaningless predictions \wjs{or using arbitrary test data or CNN model}. 
They can also be curious about test data during a CNN testing.  
Note that our verifiable computing design does not focus on poisoned or backdoored models\cite{chen2017targeted}, but it greatly relies on multi-tester test data to probe the performance of CNN models in a black-box manner.  
Recently, SecureDL\cite{xu2020secure} and VerIDeep\cite{he2018verideep} resort to sensitive samples to detect model changes, which can be complementary to our work. 
% 
%Considering the splitting process of evaluating a model on test inputs at model owners, sensitive information of test inputs can be learnt by model owners.
%
%Besides, we do not trust the CNNs they provide due to unintended factors, \emph{e.g.,} the inherent limitation on generalization capability~\cite{gong2017adversarial}.
%
%More precisely, the CNNs can inherently mis-predict some inputs when being deployed; such inherent mistakes are hard to acknowledged in advance.

\noindent\textbf{Service Providers.}
They can be computationally powerful entities, and accept outsourced \textsf{LaterNet} from developers.
Despite service providers' powerful capabilities on computation and storage, we do not trust them.
They can try their best to steal private information \wjsSec{with regard to \textsf{PriorNet}} and test inputs. 
They even incorrectly run \textsf{LaterNet} and produce untruthful results due to machine disruption.
Service providers can also collude with a developer to fake correctness proofs, with respect to the independent inference processes based on a splitting CNN, aiming to evade verification by later-coming users.
\wjsSec{But the service providers cannot obtain \textsf{PriorNet} by colluding with the developer, since the developer has no motivation.} 
%
%\wjsRR{Like existing related works~\cite{madi2020computing, zhao2021veriml}, we do not consider hyper-parameter extraction attacks via side-channel analysis~\cite{zhang2021stealing, yan2020cache} from service providers. We let hyper-parameters be public in our setting. In addition, there are defense approaches~\cite{saileshwar2021mirage} against such side-channel analysis.}
%
\wjsRR{Similar to previous studies~\cite{madi2020computing, zhao2021veriml}, we do not account for hyper-parameter extraction attacks through side-channel analysis~\cite{zhang2021stealing, yan2020cache} from service providers in our research. Instead, we assume hyper-parameters to be publicly accessible in our setting. Moreover, there are defensive methods available~\cite{saileshwar2021mirage} to mitigate such side-channel analysis attacks~\looseness=-1.}

\noindent\textbf{Model Testers.}
Testers are organized via a crowdsourcing manner.
We consider that the testers are willing to contribute their test inputs for measuring the quality of the CNNs.
A tester's test inputs concretely contain a batch of data on an input-label basis.
We assume the original inputs need protected but their labels can be public (refer to encrypted inputs and plaintext labels in Section \ref{sec:splitting}).
We also assume each label is consistent with the corresponding input's true label, considering currently popular crowdsourcing-empowered label platforms.\looseness=-1

\noindent\wjs{\textbf{Public Platform and Future Users.}
We assume both the public platform and future users are honest.
Users have access to all of data (not including model parameters or test data) uploaded to the platform and the data cannot be tampered.}
Finally, we assume that all entities communicate via a secure authenticated channel.\looseness=-1
%

\iffalse
We assume that the computational nodes follow a correct consensus protocol and no more than $1/3$ nodes are corrupted.
%
Some permissioned blockchains, \emph{e.g.}, Hyperledger Fabric and Corda.
\fi

%\wjs{Security Goals}

\iffalse
\subsection{Security Goals}\label{sec:goal}
%Here we summarize three security goals of pvCNN and they will be analysed in Section~\ref{sec:analysis}.

\noindent\textbf{Correctness.}
In a CNN prediction, its \textsf{PriorNet} should run correctly over the test data correctly encrypted by testers, and meanwhile, its \textsf{LaterNet} should run correctly by taking the correct intermediate output of the \textsf{PriorNet}, and eventually return correct inference results.

\noindent\textbf{Security.}
%
For any computationally-bounded adversary, he/she cannot produce incorrect inference results of a CNN prediction that pass the validation.

\noindent\textbf{Privacy.}
For any computationally-bounded adversary,
test data cannot be revealed except the tester himself, and a CNN's \textsf{PriorNet} can also not be exposed except the developer himself.
\fi
\section{CONCRETE DESIGN}~\label{sec:design}
\wjs{This section will introduce how to generate publicly verifiable zk-SNARK proofs for CNN testing based on L-FHE and collaborative inference.
We will present high-level ideas in the subsection~\ref{sec:overview}.
We then illustrate the optimization for 2-D convolutional relations in the subsection~\ref{sec:mmrelation}.
In the subsequent subsection~\ref{sec:proof}, we generate zk-SNARK proofs for a whole CNN testing process in a divide-and-conquer fashion, and aggregate multiple proofs in the subsection~\ref{sec:aggregation}.}

\subsection{High-level Ideas}\label{sec:overview}
On top of the scenario of testing \textsf{PriorNet} and \textsf{LaterNet} with encrypted test data from testers, we want to generate zk-SNARK proofs for convincing any future user that (1) the model developer correctly runs \textsf{PriorNet} on the ciphered test data and returns the ciphered intermediate result as output; and 
(2) the service provider exactly takes the intermediate result as input and correctly accomplishes the successive computations of \textsf{LaterNet}, which produces a correct result.
\wjs{Meanwhile, any future user does not learn any information about the model parameters, the intermediate result and the test data.}

\begin{figure}[ht]
    \centering
    \includegraphics[width=0.8\linewidth]{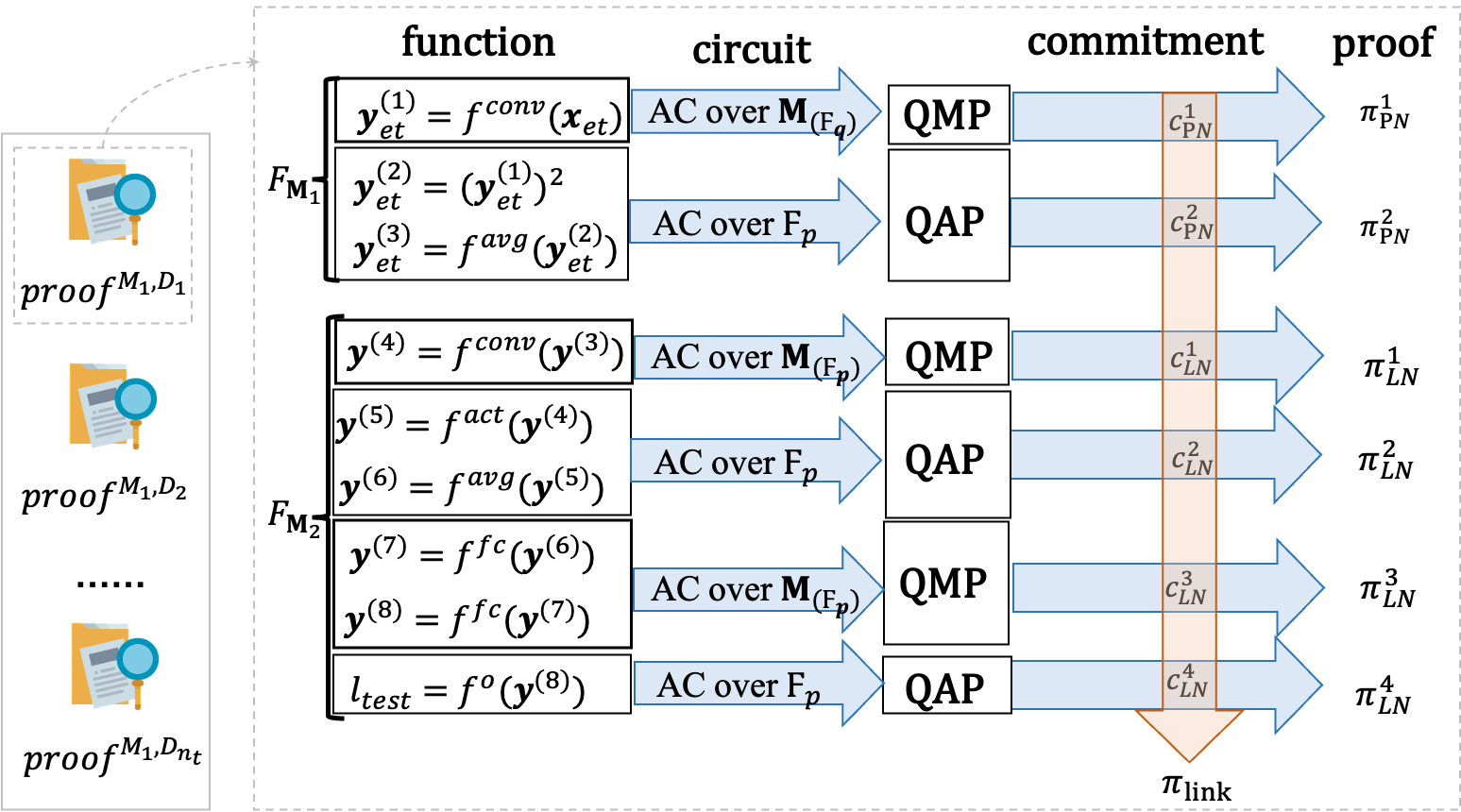}
    \caption{Overview of proof generation. \wjs{See the column of function, we use a 2-degree polynomial to approximate the ReLU function in the \textsf{PriorNet} $F_{\textbf{M}_1}$.}}
    \label{fig:proof}
    \vspace{-12pt}
\end{figure}
Most importantly, we make two efforts to improve proving and verification costs, by reducing the number of multiplication gates and compressing multiple proofs, respectively.
For ease of understanding, we firstly introduce a roadmap of proof generation, and then particularly emphasize our two efforts during proof generation.
As shown in Fig.~\ref{fig:proof}, the roadmap of generating QAP-based zkSNARK proofs is: a) compiling the \emph{functions} of the splitting CNN testing into circuits, b) expressing the \emph{circuits} as a set of QAPs, c) constructing the zk-SNARK proofs for the QAPs, and d) generating \emph{commitments} for linking together the zk-SNARK proofs which are constructed separately.
Based on the roadmap, our main efforts include:
\noindent\textbf{(A)} \textbf{Optimizing convolutional relations for b) (see Section~\ref{sec:mmrelation}).} We allow matrices in constructing the QAP, leading to quadratic matrix programs (QMP)~\cite{alesiani2021method}, which is inspired by vCNN\cite{lee2020vcnn}'s efforts of applying polynomials into QAP (QPPs, quadratic polynomial programs~\cite{KosbaPPSST14TRUESET}).
Our QMP can optimize convolutional relations, considering a convolution layer with multiple filters and inputs can be strategically represented by a single matrix multiplication operation.
The operation of matrix multiplication then is complied into the arithmetic circuits in QMP with only a single multiplication gate.
As a consequence, the number of multiplication gates dominating proving costs is greatly reduced, compared to the previous case of using the original expression of QAP.
\noindent\textbf{(B)} \textbf{Aggregating proofs for c) (see Section~\ref{sec:aggregation}).} We enable aggregating multiple proofs for different statements respective to test datasets over the same QAP/QMP for the same CNN model.
The single proof after aggregation then is validated by a secure committee and the validity result based on the majority voting will be submitted and recorded on our public platform.  
In such way, any future user only needs to check the validation result to determine whether or not the statements over the CNN are valid.\looseness=-1
%

%Next, we shed light on our design consisting in three components, such as (\emph{i}) optimizing matrix multiplication relation, (\emph{ii}) gluing proofs for separate CNN prediction, and (\emph{iii}) aggregating multiple proofs. 
% 
\vspace{2pt}
%\noindent\textbf{\emph{(i)~Optimizing matrix multiplication relations}}

\subsection{Optimizing Convolutional Relations}~\label{sec:mmrelation}
\begin{figure}[t]
    \centering
    \includegraphics[width=1\linewidth]{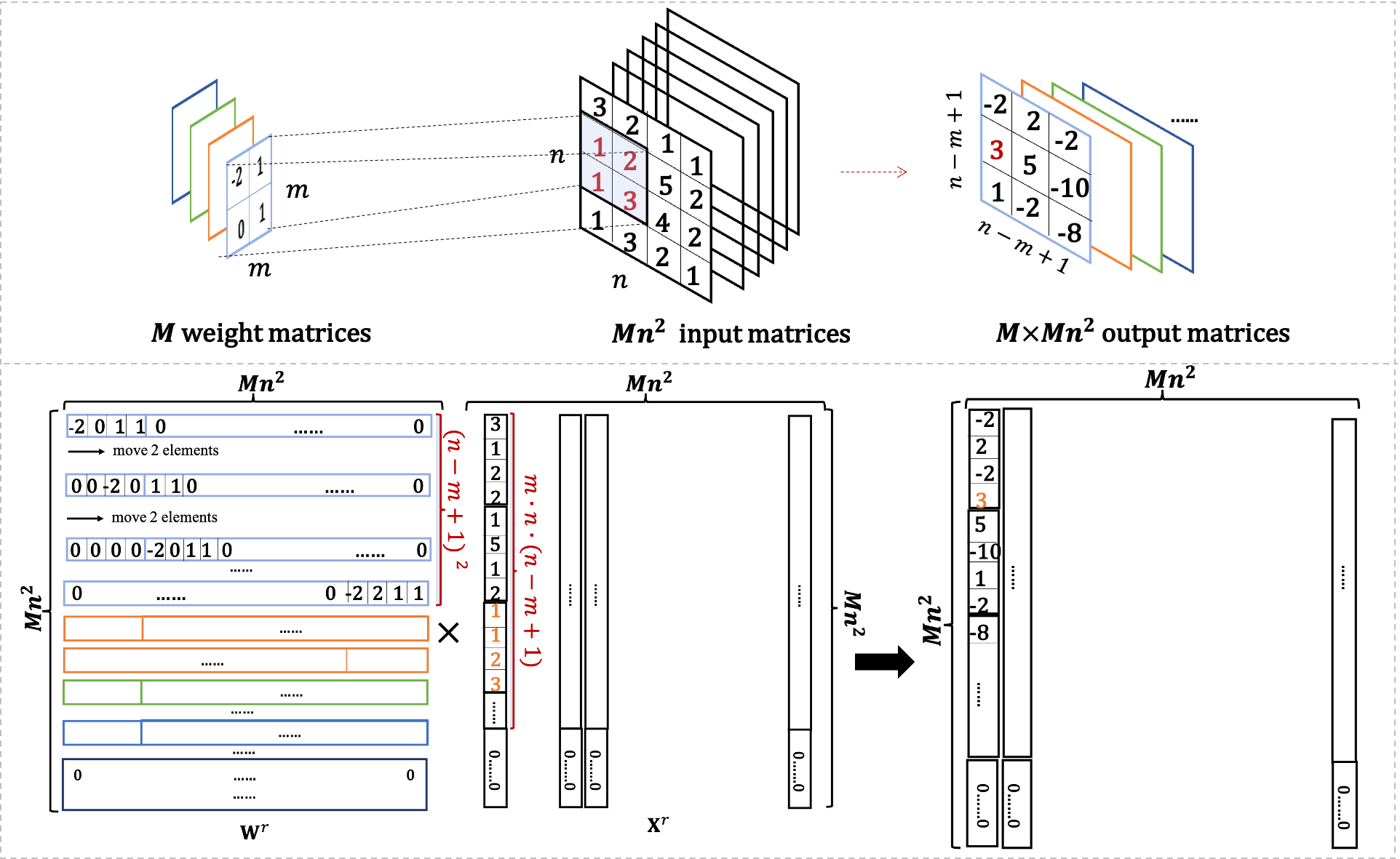}
    \caption{Matrix multiplication for convolution operations. \wjs{We transform the convolution operations of $M$ weight matrices and $Mn^2$ input matrices (demonstrated on the top) into the matrix multiplication operation between two square matrices of $Mn^2\times Mn^2$ (demonstrated on the bottom).}}
    \label{fig:conv}
    \vspace{-15pt}
\end{figure}

\vspace{-5pt}
%
%We will express 2-D convolution operations between input matrices and weight matrices (filters) with the matrix multiplication computation. 
%
%We then optimize the matrix multiplication relation using the expression of arithmetic circuits over matrices.
% 
%Note that the idea can be similarly applied to the full connection operations.
As demonstrated in Fig.~\ref{fig:conv}, we consider 2-D convolution operations between $Mn^2$ amount of $n\times n$ input matrices and $M$ amount of $m\times m$ convolution filters (herein, $M>n$), which produces $M^2n^2$ filtered matrices of $(n-m+1)\times (n-m+1)$, by setting the stride $1$.
Specifically, by computing the inner products between each input matrix and each filter matrix,~$\forall i, j \in [0,n-m]$ and 
$k\in [0, M-1]$, it can be expressed as:
$\textbf{Y}^k[i][j]=
\sum_{l_r=0,l_c=0}^{m-1,m-1}\textbf{X}[i+l_r][j+l_c]\times \textbf{W}^k[l_r][l_c].$
Obviously, the multiplication complexity of such inner product computation is $O(n^2\cdot m^2)$.
Note that there are $M\cdot Mn^2$ amount of such set of $\textbf{Y}$, with respect to $Mn^2$ inputs and $M$ filters.
Thus, $O(M\cdot Mn^2\cdot (n^2\cdot m^2))$ multiplication gates in QAP-based circuits are needed for handling $Mn^2$ inputs and $M$ filters.  
Since the proving time depends on the number of multiplication gates, a proof generation process for 2-D convolution operations becomes impractical.\looseness=-1

\wjs{Now, we seek for improving proving efficiency by reducing the multiplication complexity to $O(Mn^2\cdot Mn^2)$.}
The very first step is to represent the inner products between the $Mn^2$ inputs and $M$ filters with \emph{a single matrix multiplication}.
Concretely, we show a concrete example of the representation in Fig.~\ref{fig:conv}.
First, given $Mn^2$ amount of $n\times n$ input matrices $\{\textbf{X}\}$, they are reshaped into a square matrix $\textbf{X}^{r}$ of $Mn^2\times Mn^2$.
Second, another square matrix $\textbf{W}^{r}$ of the same dimension can be constructed by packing the $M$ filter matrices in a moving and zero padding manner.
As a result, the multiplication between the two square matrices produces a square matrix $\textbf{Y}^r$ of $Mn^2\times Mn^2$, which can also be regarded as a reshaped matrix from the original $M\cdot Mn^2$ amount of output matrices.
Based on such representation, the two square matrices $\textbf{W}^r$ and $\textbf{X}^r$  of the matrix multiplication can later be the left and right inputs of an arithmetic circuit with only one multiplication gate.
We proceed to express the above circuit with our new QMP by using matrices in QAP, which follows the similar principle of using polynomials in QAP~\cite{KosbaPPSST14TRUESET}.

\vspace{3pt}
\textbf{QMP Definition~\cite{alesiani2021method}.}
For an arithmetic circuit with $n_{in}$ input variables, $n_{ot}$ output variables in $\textbf{M}_{(\mathtt{F}_p)}^{s\times s}$ and $n_*$ multiplication gates, 
a QMP is consisting of three sets of polynomials $\{L_i(x), R_i(x), O_i(x)\}_{i=0}^m$ with coefficients in $\textbf{M}_{(\mathtt{F}_p)}^{s\times s}$ and a $n_*$-degree target polynomial $t(x) \in \mathtt{F}_p[x]$. $\mathtt{F}_p$ is a large finite field, \emph{e.g.}, $2^{254}$.
The QMP computes the arithmetic circuit and $(\textbf{A}_0,...,\textbf{A}_{n_{in}},\textbf{A}_{m-n_{ot}+1},...,\textbf{A}_m)\in  \textbf{M}_{(\mathtt{F}_p)}^{s\times s}$ is valid assignment for the $n_{in}+n_{ot}$ input/output variables \emph{iff} there exists $m-n_{in}-n_{ot}$ coefficients $(\textbf{A}_{n_{in}+1},...,\textbf{A}_{m-n_{ot}})\in \textbf{M}_{(\mathtt{F}_p)}^{s\times s}$ for arbitrary $\textbf{X}\in  \textbf{M}_{(\mathtt{F}_p)}^{s\times s}$ such that $t(x)$ divides
$p(x, \textbf{X})=\textsf{tr}\{\textbf{X}^{T}\sum_{i=0}^m\textbf{A}_iL_i(x))\cdot \sum_{i=0}^m\textbf{A}_iR_i(x)\}-\textsf{tr}\{\textbf{X}^{T}\sum_{i=0}^m\textbf{A}_iO_i(x)\}.$
Herein, \textsf{tr} means the trace of a square matrix, and the degree of the QMP is $deg(t(x))$. 
We note that each wire of the arithmetic circuit is labeled by a square matrix $\textbf{A}_i \in \textbf{M}_{(\mathtt{F}_p)}^{s\times s}$. 
Suppose that the arithmetic circuit contains $n_*$ multiplication gates, and then $t(x)=\prod_{r=0}^{n_*-1}t(x-x_r)$.
With regard to a multiplication gate $x_r$, its left input wire is labeled by $\sum_{i=0}^m\textbf{A}_i L_i(x_r))$, right input wire is labeled by $\sum_{i=0}^m\textbf{A}_i R_i(x_r))$, and the output wire is $\sum_{i=0}^m\textbf{A}_i O_i(x_r))$.
\wjs{We note that the multiplication gate is constrained by $\textsf{tr}\{\textbf{X}^{T}\sum_{i=0}^m\textbf{A}_iL_i(x_r))\cdot \sum_{i=0}^m\textbf{A}_iR_i(x_r)\}=\textsf{tr}\{\textbf{X}^{T}\sum_{i=0}^m\textbf{A}_iO_i(x_r)\}.$}
\begin{figure*}[t]
\fbox{\begin{minipage}[h]{1\linewidth}
   
\scriptsize

\vspace{1pt}

\hspace{160pt}\textbf{\textsf{Part~1:}~Layer-wise prediction correctness for \textsf{\textbf{PriorNet}}}.

\vspace{2pt}
\hspace{-3pt}
\textsf{Step 1:}\textcolor[rgb]{0.1,0.1,0.1}{~\textsf{to prove that the prover knows the openings of the commitments in} $\textsf{st}_{\textsf{PN}}^{S1}$}

\vspace{1pt}
\hspace{5pt}
$\mathtt{R}_{\textsf{PN}}^{S1}:=
\big\{
(\textsf{st}_{\textsf{PN}}^{S1}=(C_{\textsf{L}_1},C_{\textbf{X}}, 
%C_{\textbf{Y}^{(1)}},
%\textbf{W}_{ed}^{r(1)},\textbf{X}_{et}^{r(1)},\textbf{Y}_{et}^{r(1)}
C_{\textsf{PN}}^{inter},
%\textbf{Y}_{et}^{r(2)}
%\textbf{Y}_{et}^{r(3)}
\underline{C_{\textsf{L}_1,k},
%C_{\textbf{X},k},
C_{\textsf{L}_2,k},
%\textbf{Y}_{et,k}^{r(2)}
C_{\textsf{L}_3,k}}, k, \textcolor[rgb]{0.68,0.09,0.13}{ck^{\textsf{S1}}, ck_{\textsf{PN}}}),
%\textbf{Y}_{et,k}^{r(3)}
%
\textsf{wt}_{\textsf{PN}}^{S1}=(\textbf{W}_{ed}^{r(1)}, \textbf{X}_{et}^{r(1)}, \textbf{Y}_{et}^{r(1)}, \textbf{Y}_{et}^{r(2)}, \textbf{Y}_{et}^{r(3)}, \underline{\textbf{W}_{ed,k}^{r(1)}, \textbf{X}_{et,k}^{r(1)},\textbf{Y}_{et,k}^{r(1)}, \textbf{Y}_{et,k}^{r(2)}, \textbf{Y}_{et,k}^{r(3)}}, \mathbf{r}_{\textsf{PN}})\mid$

\hspace{50pt}
$
\textbf{W}_{ed,k}^{r(1)}=\textbf{W}_{ed}^{r(1)}(k)
\wedge \textbf{X}_{et,k}^{r(1)}=\textbf{X}_{et}^{r(1)}(k)
\wedge \textbf{Y}_{et,k}^{r(1)}=\textbf{Y}_{et}^{r(1)}(k)
\wedge \textbf{Y}_{et,k}^{r(2)}=\textbf{Y}_{et}^{r(2)}(k)
\wedge \textbf{Y}_{et,k}^{r(3)}=\textbf{Y}_{et}^{r(3)}(k)$

\hspace{50pt}
$\wedge 
C_{\textsf{L}_1}=\textsf{MPoly.Com}(\textbf{W}_{ed}^{r(1)}, \textcolor[rgb]{0.68,0.09,0.13}{ck^{\textsf{S1}}},\mathbf{r}_{\textsf{PN}})
\wedge 
C_{\textbf{X}}=\textsf{MPoly.Com}( \textbf{X}_{et}^{r(1)}, \textcolor[rgb]{0.68,0.09,0.13}{ck^{\textsf{S1}}},\mathbf{r}_{\textsf{PN}}) 
\wedge C_{\textsf{PN}}^{inter}=\textsf{MPoly.Com}(
\textbf{Y}_{et}^{r(1)}, \textbf{Y}_{et}^{r(2)},\textbf{Y}_{et}^{r(3)},
\textcolor[rgb]{0.68,0.09,0.13}{ck_{\textsf{PN}}},\mathbf{r}_{\textsf{PN}})$

\hspace{50pt}
$\wedge \underline{C_{\textsf{L}_1,k}=\textsf{MPoly.Com}(\textbf{W}_{ed,k}^{r(1)}, \textbf{X}_{et,k}^{r(1)}, \textbf{Y}_{et,k}^{r(1)}, \textcolor[rgb]{0.68,0.09,0.13}{ck_{\textsf{PN}}},\mathbf{r}_{\textsf{PN}})}$
$\wedge \underline{C_{\textsf{L}_2,k}=\textsf{MPoly.Com}(\textbf{Y}_{et,k}^{r(2)}, \textcolor[rgb]{0.68,0.09,0.13}{ck_{\textsf{PN}}},\mathbf{r}_{\textsf{PN}})}
\wedge \underline{C_{\textsf{L}_3,k}=\textsf{MPoly.Com}(\textbf{Y}_{et,k}^{r(3)},\textcolor[rgb]{0.68,0.09,0.13}{ck_{\textsf{PN}}},\mathbf{r}_{\textsf{PN}})}\big\}.$

\hspace{50pt}
\textcolor[rgb]{0.1,0.1,0.8}{//$ck_{\textsf{PN}}$ is the commitment key used before the evaluation while $ck^{\textsf{S1}}$ is used during the prediction}

\hspace{50pt}
\textcolor[rgb]{0.1,0.1,0.8}{//$C_{\textbf{X}}$ commits to $\textbf{X}_{et}^{r(1)}$; $C_{\textsf{L}_1}$ commits to $\textbf{W}_{ed}^{r(1)}$}

\vspace{1pt}
\hspace{-3pt}
\textsf{Step 2:}\textcolor[rgb]{0.1,0.1,0.1}{~\textsf{to prove that the equations with respect to} $\textsf{L}_1,\textsf{L}_2,\textsf{L}_3$ \textsf{hold}}

\vspace{2pt}
\hspace{5pt}
$\mathtt{R}_{\textsf{PN}}^{S2}[\textsf{L}_1]:=\big\{
(\textsf{st}_{\textsf{PN}}^{S2}=(
C_{\textsf{L}_1,k}^{'}, k, \textcolor[rgb]{0.68,0.09,0.13}{ck_{\textsf{L}_1}}), \textsf{wt}_{\textsf{PN}}^{S2}=(\textbf{W}_{ed,k}^{r(1)}, \textbf{X}_{et,k}^{r(1)}, \textbf{Y}_{et,k}^{r(1)}, \textbf{r}_{\textsf{PN}}))\mid
\textbf{Y}_{et,k}^{r(1)}=\textbf{W}_{ed,k}^{r(1)}\odot \textbf{X}_{et,k}^{r(1)}
\wedge~C_{\textsf{L}_1,k}^{'}=\textsf{MPoly.Com}(\textbf{W}_{ed,k}^{r(1)}), \textbf{X}_{et,k}^{r(1)}, \textbf{Y}_{et,k}^{r(1)}, \textcolor[rgb]{0.68,0.09,0.13}{ck_{\textsf{L}_1}},\textbf{r}_{\textsf{PN}})\big\}.$

\hspace{10pt}
\textcolor[rgb]{0.1,0.1,0.8}{//$ck_{\textsf{L}_1}$ is built depending on the relation; $C_{\textsf{L}_1,k}^{'}$ commits to the consistent $\textbf{W}_{ed,k}^{r(1)}), \textbf{X}_{et,k}^{r(1)}, \textbf{Y}_{et,k}^{r(1)}$ with the above $C_{\textsf{L}_1,k}$}

%\vspace{5pt}
%\hspace{3pt}
%$\textsf{L}_2$ and $\textsf{L}_3$:

\hspace{5pt}
$\mathtt{R}_{\textsf{PN}}^{S2}[\textsf{L}_2,\textsf{L}_3]:=\big\{
(\textsf{st}_{\textsf{PN}}^{S2}=(C_{\textbf{L}_2,k}^{'},C_{\textbf{L}_2}[\textbf{Y}_{et,k}^{r(1)}],C_{\textbf{L}_3,k}^{'},C_{\textbf{L}_3}[\textbf{Y}_{et,k}^{r(2)}],k,\textcolor[rgb]{0.68,0.09,0.13}{ck_{\textsf{L}_2}},\textcolor[rgb]{0.68,0.09,0.13}{ck_{\textsf{L}_3}}), \textsf{wt}_{\textsf{PN}}^{S2}=(\textbf{Y}_{et,k}^{r(2)}, \textbf{Y}_{et,k}^{r(3)}, \mathbf{r}_{\textsf{PN}})\mid$

\hspace{50pt}
$\{\textbf{Y}^{r(2)}_{et,k}[i][j]=(\textbf{Y}^{r(1)}_{et,k}[i][j])^2+\textbf{Y}^{r(1)}_{et,k}[i][j]\}_{i=0,j=0}^{Mn^2-1,Mn^2-1}
\wedge  \textbf{Y}^{r(3)}_{et,k}=f^{\text{avg}}(\textbf{Y}^{r(2)}_{et,k})
\wedge~C_{\textbf{L}_2,k}^{'}=\textsf{MPoly.Com}(\textbf{Y}_{et,k}^{r(2)},\textcolor[rgb]{0.68,0.09,0.13}{ck_{\textsf{L}_2}},\mathbf{r}_{\textsf{PN}})$
\hspace{50pt}
$\wedge C_{\textbf{L}_2}[\textbf{Y}_{et,k}^{r(1)}]=\textsf{MPoly.Com}(\textbf{Y}_{et,k}^{r(1)},\textcolor[rgb]{0.68,0.09,0.13}{ck_{\textsf{L}_2}},\mathbf{r}_{\textsf{PN}})
\wedge C_{\textbf{L}_3,k}^{'}=\textsf{MPoly.Com}(\textbf{Y}_{et,k}^{r(3)},\textcolor[rgb]{0.68,0.09,0.13}{ck_{\textsf{L}_3}},\mathbf{r}_{\textsf{PN}})
\wedge C_{\textbf{L}_3}[\textbf{Y}_{et,k}^{r(2)}]=\textsf{MPoly.Com}(\textbf{Y}_{et,k}^{r(2)},\textcolor[rgb]{0.68,0.09,0.13}{ck_{\textsf{L}3}},\mathbf{r}_{\textsf{PN}})\big\}.$

\hspace{10pt}
\textcolor[rgb]{0.1,0.1,0.8}{//$C_{\textbf{L}_2,k}^{'}$ (and $C_{\textbf{L}_3,k}^{'}$) commit to the same $\textbf{Y}_{et,k}^{r(2)}$ (resp. $\textbf{Y}_{et,k}^{r(3)}$) with $C_{\textbf{L}_2,k}$ (resp. $C_{\textbf{L}_3,k}$)}

\hspace{10pt}
\textcolor[rgb]{0.1,0.1,0.8}{//$C_{\textbf{L}_2}[\textbf{Y}_{et,k}^{r(1)}]$ should commit to the same $\textbf{Y}_{et,k}^{r(1)}$ that $C_{\textsf{L}_1,k}^{'}$ commits to; similarly, $C_{\textbf{L}_3}[\textbf{Y}_{et,k}^{r(2)}]$ should commit to the same $\textbf{Y}_{et,k}^{r(2)}$ that $C_{\textsf{L}_2,k}^{'}$ commits to}

\end{minipage}}
\caption{Layer-by-layer relation definition for~\textsf{PriorNet}.}\label{fig:relation_up}

\vspace{-10pt}
\end{figure*}

\begin{figure*}[ht]
\fbox{\begin{minipage}[h]{1\linewidth}
   
\scriptsize

\vspace{1pt}

\hspace{160pt}\textbf{\textsf{Part~2:}~Layer-wise prediction correctness for \textsf{\textbf{LaterNet}}}.

%\hspace{3pt}
%$\textsf{L}_4$:

\vspace{5pt}
\hspace{5pt}
$\mathtt{R}_{\textsf{LN}}[\textsf{L}_4]:=\big\{
(\textsf{st}_{\textsf{LN}}=(
C_{\textbf{Y}^{r(3)}},
C_{\textsf{L}_4},
C_{\textbf{Y}^{r(4)}},\textcolor[rgb]{0.68,0.09,0.13}{ck_{\textsf{L}_4}}),
~\textsf{wt}_{\textsf{LN}}=(\textbf{W}^{r(4)},\textbf{Y}^{r(3)}, \textbf{Y}^{r(4)}, \mathbf{r}_{\textsf{LN}}))\mid \textbf{Y}^{r(4)}=\textbf{W}^{r(4)}\times \textbf{Y}^{r(3)}$

\hspace{50pt}
$\wedge~C_{\textbf{Y}^{r(3)}}=\textsf{MPoly.Com}(\textbf{Y}^{r(3)},\textcolor[rgb]{0.68,0.09,0.13}{ck_{\textsf{L}_4}},\mathbf{r}_{\textsf{LN}})
\wedge C_{\textsf{L}_4}=\textsf{MPoly.Com}(\textbf{W}^{r(4)}, \textcolor[rgb]{0.68,0.09,0.13}{ck_{\textsf{L}_4}},\mathbf{r}_{\textsf{LN}})
\wedge C_{\textbf{Y}^{r(4)}}=\textsf{MPoly.Com}(\textbf{Y}^{r(4)},\textcolor[rgb]{0.68,0.09,0.13}{ck_{\textsf{L}_4}},\mathbf{r}_{\textsf{LN}})\big\}.$

\hspace{10pt}
\textcolor[rgb]{0.1,0.1,0.8}{//$C_{\textbf{Y}^{r(3)}}$ should commit to $\textbf{Y}^{r(3)}$ whose encryption is committed in $C_{\textsf{PN}}^{inter}$}

%\vspace{5pt}
%\hspace{3pt}
%$\textsf{L}_5$ and $\textsf{L}_6$:

\vspace{3pt}
\hspace{5pt}
$\mathtt{R}_{\textsf{LN}}[\textsf{L}_5,\textsf{L}_6]:=\big\{
(\textsf{st}_{\textsf{LN}}=(
C_{\textsf{L}_5}[\textbf{Y}^{r(4)}], C_{\textbf{Y}^{r(5)}}, C_{\textsf{L}_6}[\textbf{Y}^{r(5)}], C_{\textbf{Y}^{r(6)}},\textcolor[rgb]{0.68,0.09,0.13}{ck_{\textsf{L}_5}},\textcolor[rgb]{0.68,0.09,0.13}{ck_{\textsf{L}_6}}),
~\textsf{wt}_{\textsf{LN}}=(\textbf{Y}^{r(4)}, \textbf{Y}^{r(5)}, \textbf{Y}^{r(6)},  \mathbf{r}_{\textsf{LN}}))\mid$

\hspace{10pt}
$\{\textbf{Y}^{r(5)}[i][j]=max(\textbf{Y}^{r(4)}[i][j],0)\}_{i=0,j=0}^{Mn^2-1,Mn^2-1}
\wedge \textbf{Y}^{r(6)}=f^{\text{avg}}(\textbf{Y}^{r(5)})$

\hspace{10pt}
$\wedge~C_{\textsf{L}_5}[\textbf{Y}^{r(4)}]=\textsf{MPoly.Com}(\textbf{Y}^{r(4)}, \textcolor[rgb]{0.68,0.09,0.13}{ck_{\textsf{L}_5}},\mathbf{r}_{\textsf{LN}})
\wedge C_{\textbf{Y}^{r(5)}}=\textsf{MPoly.Com}(\textbf{Y}^{r(5)}, \textcolor[rgb]{0.68,0.09,0.13}{ck_{\textsf{L}_6}},\mathbf{r}_{\textsf{LN}})$\textcolor[rgb]{0.1,0.1,0.8}{//$C_{\textsf{L}_5}[\textbf{Y}^{r(4)}]$ should commit to $\textbf{Y}^{r(4)}$ that is committed in $C_{\textbf{Y}^{r(4)}}$ of $\textsf{L}_4$}

\hspace{10pt}
$\wedge C_{\textsf{L}_6}[\textbf{Y}^{r(5)}]=\textsf{MPoly.Com}(\textbf{Y}^{r(5)}, \textcolor[rgb]{0.68,0.09,0.13}{ck_{\textsf{L}_6}},\mathbf{r}_{\textsf{LN}})
\wedge C_{\textbf{Y}^{r(6)}}=\textsf{MPoly.Com}(\textbf{Y}^{r(6)}, \textcolor[rgb]{0.68,0.09,0.13}{ck_{\textsf{L}_6}},\mathbf{r}_{\textsf{LN}})\big\}.$\textcolor[rgb]{0.1,0.1,0.8}{//$C_{\textsf{L}_6}[\textbf{Y}^{r(5)}]$ should commit to $\textbf{Y}^{r(5)}$ that is committed in $C_{\textbf{Y}^{r(5)}}$ of $\textsf{L}_5$}

\vspace{3pt}
\hspace{5pt}
$\mathtt{R}_{\textsf{LN}}[\textsf{L}_7,\textsf{L}_8]:=\big\{
(\textsf{st}_{\textsf{LN}}=(
C_{\textsf{L}_{7,8}},C_{\textsf{L}_{7}}[\textbf{Y}^{r(6)}],C_{\textsf{L}_{8}}[\textbf{Y}^{r(8)}],\textcolor[rgb]{0.68,0.09,0.13}{ck_{\textsf{L}_{7,8}}}), 
~\textsf{wt}_{\textsf{LN}}=(\textbf{W}^{r(7)}, \textbf{W}^{r(8)},\textbf{Y}^{r(6)}, \textbf{Y}^{r(7)}, \textbf{Y}^{r(8)}, \mathbf{r}_{\textsf{LN}}))\mid$
$\textbf{Y}^{r(7)}=\textbf{Y}^{r(6)}\times \textbf{W}^{r(7)}$
%$\wedge \textbf{Y}^{r(8)}=\textbf{Y}^{r(7)}\times \textbf{W}^{r(8)}$

\hspace{10pt}
$\wedge C_{\textsf{L}_{7,8}}=\textsf{MPoly.Com}(\textbf{W}^{r(7)}, \textbf{W}^{r(8)}, \textcolor[rgb]{0.68,0.09,0.13}{ck_{\textsf{L}_{7,8}}},\mathbf{r}_{\textsf{LN}})
\wedge C_{\textsf{L}_{7}}[\textbf{Y}^{r(6)}]=\textsf{MPoly.Com}(\textbf{Y}^{r(6)}, \textcolor[rgb]{0.68,0.09,0.13}{ck_{\textsf{L}_{7,8}}},\mathbf{r}_{\textsf{LN}})
\wedge C_{\textsf{L}_{8}}[\textbf{Y}^{r(8)}]=\textsf{MPoly.Com}(\textbf{Y}^{r(8)}, \textcolor[rgb]{0.68,0.09,0.13}{ck_{\textsf{L}_{7,8}}},\mathbf{r}_{\textsf{LN}})\big\}.$

\hspace{10pt}
\textcolor[rgb]{0.1,0.1,0.8}{//$C_{\textsf{L}_7}[\textbf{Y}^{r(6)}]$ should commit to $\textbf{Y}^{r(6)}$ that is committed in $C_{\textbf{Y}^{r(6)}}$ of $\textsf{L}_6$}

%\vspace{5pt}
%\hspace{3pt}
%$\textsf{L}_o$:

\hspace{5pt}
$\mathtt{R}_{\textsf{LN}}[\textsf{L}_o]:=\big\{
(\textsf{st}_{\textsf{LN}}=(
C_{\textsf{L}_o}[\textbf{Y}^{r(8)}],\textbf{l}_{test}, \textcolor[rgb]{0.68,0.09,0.13}{ck_{\textsf{L}_{o}}}), \textsf{wt}_{\textsf{LN}}=(\textbf{Y}^{r(8)}))\mid
\textbf{l}_{test}[j]=\textsf{argmax}(\textbf{Y}^{r(8)}[j])
\wedge~C_{\textsf{L}_o}[\textbf{Y}^{r(8)}]=\textsf{MPoly.Com}(\textbf{Y}^{r(8)}, \textcolor[rgb]{0.68,0.09,0.13}{ck_{\textsf{L}_{o}}},\mathbf{r}_{\textsf{LN}})\big\}.$

\hspace{10pt}
\textcolor[rgb]{0.1,0.1,0.8}{//$C_{\textsf{L}_o}[\textbf{Y}^{r(8)}]$ should commit to $\textbf{Y}^{r(8)}$ that is committed in $C_{\textbf{Y}^{r(8)}}$ of $\textsf{L}_8$}

\end{minipage}}
\caption{Layer-by-layer relation definition for~\textsf{LaterNet}.}\label{fig:relation_down}

\vspace{-20pt}
\end{figure*} 

\vspace{3pt}
\textbf{QMP for Matrix Multiplication Relations.}
As we introduced before, we can represent convolutional operations as matrix multiplication $\textbf{Y}^{r}=\textbf{W}^{r}\times \textbf{X}^{r}$, where $\textbf{Y}^{r}, \textbf{X}^{r}$ and $\textbf{W}^{r} \in \textbf{M}_{(\mathtt{Z}_p)}^{Mn^2\times Mn^2}$.
It is noteworthy that we can use advanced quantization techniques to transform floating-point numbersm of 8-bit precision into unsigned integer numbers in [$0,255$], which adapts to a large finite field.
Now according to the QMP definition, when the matrix multiplication computation is expressed by the arithmetic circuit over matrices, the corresponding QMP is 
$\{L_{\textbf{W}}(x), R_{\textbf{X}}(x), O_{\textbf{Y}}(x), t(x)\}$ of the arithmetic circuit. Herein, there are three input/output variables, and the degree of the QMP is one.

Similarly, we can apply our above idea into full connection operations, and finally the operations can also be expressed with similarly low-degree QMP.

\subsection{Proof Generation for a Whole CNN Testing}\label{sec:proof}

\wjs{This section elaborates how to generate proofs based on the CaP zk-SNARK for ensuring the result correctness of executing \textsf{PriorNet} and \textsf{LaterNet}.
The proofs convince an arbitrary user that \textsf{PriorNet} is correctly evaluated on the ciphered test inputs, and subsequently, \textsf{LaterNet} is correctly performed by taking the clear outputs of \textsf{PriorNet} as inputs, yielding correct results.
We note that the processes of proof generation for the \textsf{PriorNet} and \textsf{LaterNet} are slightly different.
For \textsf{PriorNet}, the values, such as weights, test inputs and outputs, are represented as polynomial ring elements, and we follow Fiore \emph{et al.}'s work~\cite{fiore2020boosting} to generate proofs for the computation over polynomial rings. 
For \textsf{LaterNet}, we generate zk-SNARK proofs for the computation over scalars.
We will begin with relation definition, and then present concrete realization for generating proofs.
In light of the nature of layer-by-layer computation of model inference, we will use a divide-and-conquer method to generate zk-SNARK proofs for each separate layer, and combine them as a whole via a commit-and-prove methodology.}
%

%
%Step 1 is to prove the simultaneous evaluation of multiple polynomial ring elements in the same randomly selected point (for PriorNet), and Step 2 is to prove the satisfiability of QAP-based arithmetic circuits (for PriorNet and LaterNet) whose inputs and outputs are committed.
%
\wjs{Let us define some notations at the beginning.
We let \textsf{MPoly.Com} be a polynomial commitment scheme from \cite{fiore2020boosting}.
We denote $ck_{\textsf{PN}}$ and $ck_{\textsf{LN}}$ are the commitment keys used for committing to the \textsf{PriorNet} and \textsf{LaterNet};
$\mathbf{r}_{\textsf{PN}}$, $\mathbf{r}_{\textsf{LN}}$ are non-reused commitment randomnesses.
$C_{\textbf{M}_{1,{ed}}}$ means the commitments to the inputs, weights, intermediates of $\textbf{M}_{1}$ while $C_{\textbf{M}_{2}}$ represents the commitments to the weights and intermediates of $\textbf{M}_{2}$.}

\vspace{-2pt}
\textbf{Relation Definition.}
We define the relations $\mathtt{R}_{\textsf{PN}}$ and $\mathtt{R}_{\textsf{LN}}$ with regard to \textsf{PriorNet} and \textsf{LaterNet} execution correctness:
\begin{align*}
\mathtt{R}_{\textsf{PN}}:=\big\{(\textsf{st}_{\textsf{PN}}&=(C_{\textbf{M}_{1,{ed}}}, C_{\textbf{X}_{et}^{r(1)}}, \textbf{Y}^{r(3)}_{et},ck_{\textsf{PN}}),\\ 
\textsf{wt}_{\textsf{PN}}&=(\textbf{M}_{1,{ed}},\textbf{X}_{et},\mathbf{r}_{\textsf{PN}} ))\mid
\textbf{Y}^{r(1)}_{et}=\textbf{W}^{r(1)}_{ed}\odot \textbf{X}^{r(1)}_{et} \\
\wedge \{\textbf{Y}^{r(2)}_{et}[i][j]&=(\textbf{Y}^{r(1)}_{et}[i][j])^2+\textbf{Y}^{r(1)}_{et}[i][j]\}_{i=0,j=0}^{Mn^2-1,Mn^2-1} \\
 \wedge \textbf{Y}^{r(3)}_{et}&=f^{\text{avg}}(\textbf{Y}^{r(2)}_{et}) \\
\wedge C_{\textbf{M}_{1,{ed}}}&=\textsf{MPoly.Com}(\textbf{M}_{1,{ed}}, ck_{\textsf{PN}}, \mathbf{r}_{\textsf{PN}}) \\
\wedge C_{\textbf{X}_{et}}&=\textsf{MPoly.Com}(\textbf{X}_{et}^{r(1)}, ck_{\textsf{PN}}, \mathbf{r}_{\textsf{PN}})\big\},
\end{align*} 
\begin{align*}
\mathtt{R}_{\textsf{LN}}:=\{(\textsf{st}_{\textsf{LN}}
&=(C_{\textbf{Y}^{r(3)}}, C_{\textbf{M}_2}, l_{test}, ck_{\textsf{LN}}),\\ \textsf{wt}_{\textsf{LN}}&=(\textbf{Y}^{r(3)}, \textbf{M}_2, \mathbf{r}_{\textsf{LN}}))\mid
\textbf{Y}^{r(4)}=\textbf{W}^{r(4)}\times \textbf{Y}^{r(3)}\\
\wedge \{\textbf{Y}^{r(5)}[i][j]&=max(\textbf{Y}^{r(4)}[i][j],0)\}_{i=0,j=0}^{Mn^2-1,Mn^2-1}\\
\wedge \textbf{Y}^{r(6)}&=f^{\text{avg}}(\textbf{Y}^{r(5)})\\
\wedge \textbf{Y}^{r(7)}&=\textbf{Y}^{r(6)}\times \textbf{W}^{r(7)}\\
\wedge \textbf{Y}^{r(8)}&=\textbf{Y}^{r(7)}\times \textbf{W}^{r(8)}\\
\wedge ~\textbf{l}_{test}[j]&=\textsf{argmax}(\textbf{Y}^{r(8)}[j])\\
\wedge~C_{\textbf{Y}^{r(3)}}&=\textsf{MPoly.Com}(\textbf{Y}^{r(3)}, ck_{\textsf{LN}}, \mathbf{r}_{\textsf{LN}})\\
\wedge~C_{\textbf{M}_{2}}&=\textsf{MPoly.Com}(\textbf{M}_{2}, ck_{\textsf{LN}}, \mathbf{r}_{\textsf{LN}})\big\}.
\end{align*}

\noindent Besides the relations, we further define layer-by-layer relations for proving the layer-wise computations of the $\mathtt{R}_{\textsf{PN}}$ and $\mathtt{R}_{\textsf{LN}}$ in Fig.~\ref{fig:relation_up} and~\ref{fig:relation_down}.
\wjs{In \textsf{Part~1}, the layer-wise computations of the \textsf{PriorNet} performs in an encryption manner, so we define the relations to be proven with \textsf{Setp~1} and \textsf{Setp~2}, as guided by the work~\cite{fiore2020boosting}.
In \textsf{Part~2}, the \textsf{LaterrNet} runs in a plaintext setting, so we define the corresponding relations similar to the above \textsf{Setp~2} (not needing \textsf{Setp~1}).}
Lastly, in order to combine the layer-wise computations as a whole, we also denote the relations for ensuring that starting from the second layer, each layer's inputs are consistent with the former layer's outputs.\looseness=-1

With the relations, we proceed to generate zk-SNARK proofs accordingly.
At a high level, the computations of the \textsf{PriorNet} and \textsf{LaterNet} are proved correct \emph{iff} both of binary relations $\mathtt{R}_{\textsf{PN}}$ and $\mathtt{R}_{\textsf{LN}}$ return $1$, which means $\mathtt{R}_{\textsf{PN}}$ and $\mathtt{R}_{\textsf{LN}}$ hold on the corresponding pairs of statement and witness $(\textsf{st}, \textsf{wt})$.
For $\mathtt{R}_{\textsf{PN}}$, we generate publicly verifiable proofs that $\textbf{Y}^{r(3)}_{et}$ is computed correctly in an encryption manner, with the trained parameters of \textsf{PriorNet} $\textbf{M}_{1,{ed}}$ and ciphered test inputs $\textbf{X}_{et}^{r(1)}$ through the sequential computations mentioned in the $\mathtt{R}_{\textsf{PN}}$.
In the relation, the witnesses $\textbf{M}_{1,{ed}}$ and $\textbf{X}_{et}^{r(1)}$ are committed using the key $ck_{\textsf{PN}}$, and the generated commitments plus $\textbf{Y}^{r(3)}_{et}$ are treated as the public statement.
For $\mathtt{R}_{\textsf{LN}}$, we generate proofs that \textsf{LaterNet} takes the correct plaintext $\textbf{Y}^{r(3)}$ and returns the correct inference result $l_{test}$ by conducting the sequential operations included in the $\mathtt{R}_{\textsf{LN}}$.
Similarly, the witnesses $\textbf{M}_{2}$ and $\textbf{Y}^{r(3)}$ are committed using the key $ck_{\textsf{LN}}$, and then the commitments and the execution result are public statements.
The processes of proof generation for $\mathtt{R}_{\textsf{PN}}$ and $\mathtt{R}_{\textsf{LN}}$ are conducted separately, since the two parts \textsf{PriorNet} and \textsf{LaterrNet} are executed by different entities.
After completing the proofs for the two parts, we link the proofs together by additionally proving that the output of the \textsf{PriorNet} is exactly equal to the input of the \textsf{LaterNet}.

\begin{figure}[!h]
\fbox{
\begin{minipage}[t]{0.96\linewidth}
%\small
\scriptsize
\hspace{20pt}\textbf{\textsf{Step~1} CaP zk-SNARK Proofs of Knowledge of Commitments}

\vspace{3pt}

\hspace{-3pt}\text{1:}
\underline{~\textsf{MUniEv-$\Pi$.Setup}($\lambda, d_{c}, n_{c})\rightarrow  \textcolor[rgb]{0.68,0.09,0.13}{ck^{S1}}:$}

\vspace{1pt}
\hspace{-3pt}\text{2:}
\textcolor[rgb]{0.1,0.1,0.8}{~//$d_{c}$ refers to the degree of ciphers; $n_{c}$ is the number of ciphers}

\hspace{-3pt}\text{3:}
~~$g,h \overset{\$}{\leftarrow} \mathtt{G}$, $g^{*} \overset{\$}{\leftarrow} \mathcal{G}$,
$\alpha, s, t \overset{\$}{\leftarrow} \mathtt{Z}_p$

\hspace{-3pt}\text{4:}
~~$\hat{g}=g^{\alpha}, \hat{h}=h^{\alpha},
\hat{g^{*}}={g^{*}}^{\alpha}$, $g^{*}_1={g^{*}}^s$, $h_1=h^s$,

\hspace{-3pt}\text{5:}
~~$\{g_{i,j}=g^{s^{i}t^{j}}, \hat{g}_{i,j}=\hat{g}^{s^{i}t^{j}}\}_{i=0,j=0}^{d_c,n_c}$,
\textsf{HASH}:~${\mathtt{G}}^*\times {\mathtt{Z}_q}^*\rightarrow \mathtt{Z}_p$

\hspace{-3pt}\text{6:}
~~$\textcolor[rgb]{0.68,0.09,0.13}{ck^{S1}}:=(g,h,g^{*},\{g_{i,j},\hat{g}_{i,j}\}_{i=0,j=0}^{d_c,n_c},\hat{g},\hat{h},\hat{g^{*}},g^{*}_1,h_1,\textsf{HASH})$
\vspace{2.5pt}

\hspace{-3pt}\text{7:}
\underline{~\textsf{MUniEv-$\Pi$.Prove}($\textcolor[rgb]{0.68,0.09,0.13}{ck^{S1}}, \textbf{W}^{r(1)}_{ed},\textbf{X}^{r(1)}_{et})\rightarrow \pi^{S1}:$} 

\hspace{-3pt}\text{8:}
\textcolor[rgb]{0.1,0.1,0.8}{~//$\textbf{W}^{r(1)}_{ed},\textbf{X}^{r(1)}_{et}$ are square matrices of $l_{\text{new}}=Mn^2$}

\hspace{-3pt}\text{9:}
\textcolor[rgb]{0.1,0.1,0.8}{~//$l_{\text{new}}^2$ is smaller that $n_c$}

\hspace{-3pt}\text{10:}
\textcolor[rgb]{0.1,0.1,0.8}{~//call the commitment scheme \textsf{MPoly.Com} of~\cite{fiore2020boosting}} 

\hspace{-3pt}\text{11:}
~$\textsf{call~MPoly.Com}(\textcolor[rgb]{0.68,0.09,0.13}{ck^{S1}}, \textbf{W}^{r(1)}_{ed})\rightarrow C_{\textsf{W}_1}$:

\hspace{4pt}
$-~r_{\textsf{W}_1} \overset{\$}{\leftarrow} \mathtt{Z}_p$,

\hspace{4pt}
$-~C_{\textsf{W}_1}^1=h^{r_{\textsf{W}_1}} g_{i,j}^{\sum_{i=0,j=0}^{d_c,l_{\text{new}}^2-1}w_{i,j}}$,
$C_{\textsf{W}_1}^2=\hat{h}^{r_{\textsf{W}_1}} \hat{g}_{i,j}^{ \sum_{i=0,j=0}^{d_c,l_{\text{new}}^2-1}w_{i,j}}$,

\vspace{2pt}
\hspace{4pt}
$-~C_{\textsf{W}_1}:=(C_{\textsf{W}_1}^1,C_{\textsf{W}_1}^2)$

\hspace{-3pt}\text{12:}
~$\textsf{call~MPoly.Com}(\textcolor[rgb]{0.68,0.09,0.13}{ck^{S1}}, \textbf{X}^{r(1)}_{et})\rightarrow C_{\textbf{X}}$:

\hspace{4pt}
$-~r_{\textbf{X}} \overset{\$}{\leftarrow} \mathtt{Z}_p$,
$C_{\textbf{X}}^1=h^{r_{\textbf{X}}} g_{i,j}^{\sum_{i=0,j=0}^{d_c,l_{\text{new}}^2-1}\text{x}_{i,j}}$,
$C_{\textbf{X}}^2=\hat{h}^{r_{\textbf{X}}} \hat{g}_{i,j}^{\sum_{i=0,j=0}^{d_c,l_{\text{new}}^2-1}\text{x}_{i,j}}$,

\vspace{2pt}
\hspace{4pt}
$-~C_{\textbf{X}}:=(C_{\textbf{X}}^1,C_{\textbf{X}}^2)$

\hspace{-3pt}\text{13:}
$\textsf{call~MPoly.Com}(\textcolor[rgb]{0.68,0.09,0.13}{ck^{S1}}, \textbf{Y}^{r(1)}_{et})\rightarrow C_{\textbf{Y}}$:

\hspace{4pt}
$-~r_{\textbf{Y}} \overset{\$}{\leftarrow} \mathtt{Z}_p$,
$C_{\textbf{Y}}^1=h^{r_{\textbf{Y}}} g_{i,j}^{\sum_{i=0,j=0}^{d_c,l_{\text{new}}^2-1}\text{y}_{i,j}}$,
$C_{\textbf{Y}}^2=\hat{h}^{r_{\textbf{Y}}} \hat{g}_{i,j}^{\sum_{i=0,j=0}^{d_c,l_{\text{new}}^2-1}\text{y}_{i,j}}$,

\vspace{2pt}
\hspace{4pt}
$-~C_{\textbf{Y}}:=(C_{\textbf{Y}}^1,C_{\textbf{Y}}^2)$

\vspace{2pt}
\hspace{-3pt}\text{14:}
$k=\textsf{HASH}(C_{\textsf{W}_1},C_{\textbf{X}}, \textbf{W}^{r(1)}_{ed}, \textbf{X}^{r(1)}_{et}, \textbf{Y}^{r(1)}_{et})$

\hspace{-3pt}\text{15:}
\textcolor[rgb]{0.1,0.1,0.8}{~//$\textbf{X}^{r(1)}_{et}\in \textbf{M}_{\mathtt{Z}_q[x](x^{d_c}+1)}^{l_{new}^2}$, $\textbf{X}(x,y)=\sum_{i=0,j=0}^{d_c,l_{\text{new}}^2-1}\text{x}_{i,j}x^{i}y^{j}$}

\hspace{-3pt}\text{16:}
\textcolor[rgb]{0.1,0.1,0.8}{~//$\textbf{W}^{r(1)}_{et}\in \textbf{M}_{\mathtt{Z}_q[x](x^{d_c}+1)}^{l_{new}^2}$, $\textbf{W}(x,y)=\sum_{i=0,j=0}^{d_c,l_{\text{new}}^2-1}\text{w}_{i,j}x^{i}y^{j}$}

\hspace{-3pt}\text{17:}
\textcolor[rgb]{0.1,0.1,0.8}{~//$\textbf{Y}^{r(1)}_{et}\in \textbf{M}_{\mathtt{Z}_q[x](x^{d_c}+1)}^{l_{new}^2}$, $\textbf{Y}(x,y)=\sum_{i=0,j=0}^{d_c,l_{\text{new}}^2-1}\text{y}_{i,j}x^{i}y^{j}$}
%

%\hspace{-3pt}\text{20:}
%$(\alpha_1, \alpha_2) \leftarrow \mathtt{Z}_p$ 
\hspace{-3pt}\text{18:}
$\textsf{L}_{1}:=
\textsf{L}_1(x,y)= \textbf{W}(x,y)+ \textbf{X}(x,y)+\textbf{Y}(x,y)=\sum_{i=0,j=0}^{d_c, l_{\text{new}}^2-1}l_{i,j}x^i y^j$

\hspace{-3pt}\text{19:}
$\textsf{call~MPoly.Com}(\textcolor[rgb]{0.68,0.09,0.13}{ck^{S1}}, \textsf{L}_1(s,t))\rightarrow C_{\textsf{L}_1}=(C_{\textsf{L}_1}^1, C_{\textsf{L}_1}^2)$

%\hspace{-3pt}\text{21:}
%$s, t \overset{\$}{\leftarrow} \mathtt{Z}_p$
\hspace{-3pt}\text{20:}
$\textsf{L}_{1,k}:=
\textsf{L}_1(k,y)= \textbf{W}(k,y)+ \textbf{X}(k,y)+\textbf{Y}(k,y)=\sum_{i=0,j=0}^{d_c,l_{\text{new}}^2-1}l_{i,j}k^i y^j$

\hspace{-3pt}\text{21:}
$\textsf{call~MPoly.Com}(\textcolor[rgb]{0.68,0.09,0.13}{ck^{S1}}, \textbf{W}^{r(1)}_{ed,k})\rightarrow C_{\textsf{w},k}=(C_{\textsf{w},k}^1,C_{\textsf{w},k}^2)$

\hspace{-3pt}\text{22:}
$\textsf{call~MPoly.Com}(\textcolor[rgb]{0.68,0.09,0.13}{ck^{S1}}, \textbf{X}^{r(1)}_{ed,k})\rightarrow C_{\textsf{x},k}=(C_{\textsf{x},k}^1,C_{\textsf{x},k}^2)$
%\hspace{1pt}
%$(\alpha_1 \text{w}_{0,0}+\alpha_2 \text{x}_{0,0}+ \text{y}_{0,0})k^0y^0+...+(\alpha_1  \text{w}_{d_c,l_{\text{new}}^2-1}+\alpha_2 \text{x}_{d_c,l_{\text{new}}^2-1}+ \text{y}_{d_c,l_{\text{new}}^2-1})k^{d_c}y^{l_{\text{new}}^2-1}$
%

\hspace{-3pt}\text{23:}
$\textsf{call~MPoly.Com}(\textcolor[rgb]{0.68,0.09,0.13}{ck^{S1}}, \textbf{Y}^{r(1)}_{ed,k})\rightarrow C_{\textsf{y},k}=(C_{\textsf{y},k}^1,C_{\textsf{y},k}^2)$

%\hspace{-3pt}\text{23:}
%$C_{\textsf{L}_1,k}^1
%\hspace{1pt}
%$\textbf{W}(k,y)=\text{w}_{0,0}k^0y^0+...+\text{w}_{d_c,l_{\text{new}}^2}k^{d_c}y^{l_{\text{new}}^2}$\textcolor[rgb]{0.1,0.1,0.8}{~//\textbf{W}^{r(1)}_{ed,k}}
%

%\hspace{-3pt}\text{21:}
%$\textsf{call~MPoly.Com}(\textcolor[rgb]{0.8,0.8,0.1}{ck^{S1}}, \textsf{L}_{1,k})\rightarrow C_{\textsf{L}_1,k}$:
%

%\hspace{-3pt}\text{22:}
%~~~$r_{\textsf{L}_1,k} \overset{\$}{\leftarrow} \mathtt{Z}_p$,

%\hspace{-3pt}\text{23:}
%~~~$C_{\textsf{L}_1,k}^1=h^{r_{\textsf{L}_1,k}}\prod_{j=0}^{l_{\text{new}}^2-1} g_{0,j}^{\text{l}_{j}}$,
%
%$C_{\textsf{L}_1,k}^2=\hat{h}^{r_{\textsf{L}_1,k}}\prod_{j=0}^{l_{\text{new}}^2-1} \hat{g}_{0,j}^{\text{l}_{j}}$,
%
%$C_{\textsf{L}_1,k}:=(C_{\textsf{L}_1,k}^1,C_{\textsf{L}_1,k}^2)$
%

%C_{\textsf{L}_1,k}=\textsf{MPoly.Com}(\textbf{W}_{ed,k}^{r(1)}, \textbf{X}_{et,k}^{r(1)}, \textbf{Y}_{et,k}^{r(1)}

\hspace{-3pt}\text{24:}
$C_{\textsf{L}_1,k}^1=(C_{\textsf{w},k}^1\cdot C_{\textsf{x},k}^1\cdot C_{\textsf{y},k}^1);C_{\textsf{L}_1,k}^2=(C_{\textsf{w},k}^2\cdot C_{\textsf{x},k}^2\cdot C_{\textsf{y},k}^2)$

\vspace{2pt}
\hspace{-3pt}\text{25:}
$\textbf{T}(x,y)=\frac{\textsf{L}_1(x,y)-\textsf{L}_1(k,y)}{(x-k)}$
\vspace{2pt}

\hspace{-3pt}\text{26:}
$\bar{g}=\frac{h_1}{h^k}$, $a,b \overset{\$}{\leftarrow} \mathtt{Z}_p$, 

\hspace{-3pt}\text{27:}
$\textsf{call~MPoly.Com}(\textcolor[rgb]{0.68,0.09,0.13}{ck^{S1}}, \textbf{T}(x,y))\rightarrow C_{\textbf{T}}$:

\hspace{4pt}
$-~r_{\textbf{T}} \overset{\$}{\leftarrow} \mathtt{Z}_p$,
$C_{\textbf{T}}^1=h^{r_{\textbf{T}}} \bar{g}^{\sum_{t=0,i=0,j=0}^{d_c-1,d_c-t,l_{\text{new}}^2-1}l_{i+t,j}k^{i-1}s^it^j}$,

\hspace{12pt}
$C_{\textbf{T}}^1=h^{r_{\textbf{T}}} \hat{\bar{g}}^{\sum_{t=0,i=0,j=0}^{d_c-1,d_c-t,l_{\text{new}}^2-1}l_{i+t,j}k^{i-1}s^it^j}$,

\vspace{2pt}
\hspace{4pt}
$-~C_{\textbf{T}}:=(C_{\textbf{T}}^1,C_{\textbf{T}}^2)$

\hspace{-3pt}\text{28:}
$U=e(h^a\bar{g}^b, g^{*})$, $\mathtt{e}\leftarrow \textsf{HASH}(C_{\textsf{L}_1}, C_{\textsf{L}_1,k},C_{\textbf{T}},U,k)$,

\hspace{-3pt}\text{29:}
$\sigma=a-(r_{\textbf{L}_1,k}-r_{\textbf{L}_1})\cdot \mathtt{e}~\text{mod}~q$,
$\tau=b-r_{\textbf{T}}\cdot \mathtt{e}~\text{mod}~q$,
$\pi^{S1}=(C_{\textbf{T}}, \mathtt{e}, \sigma, \tau)$

\vspace{4pt}
\hspace{-3pt}\text{30:}
\underline{\textsf{MUniEv-$\Pi$.Verify}($\textcolor[rgb]{0.68,0.09,0.13}{ck^{S1}},C_{\textsf{L}_1},C_{\textbf{X}},C_{\textsf{L}_1,k},C_{\textbf{T}},k,\pi^{\textsf{S1}})\rightarrow 0/1$:}

\vspace{2pt}
\hspace{-3pt}\text{31:}
$b_1:=(e(C_{\textsf{L}_1}^1, \hat{g^{*}})==e(C_{\textsf{L}_1}^2,g^{*}))$;
$b_2:=(e(C_{\textbf{X}}^1, \hat{g^{*}})==e(C_{\textbf{X}}^2,g^{*}))$;

\hspace{-3pt}\text{32:}
$b_3:=(e(C_{\textbf{w},k}^1, \hat{g^{*}})==e(C_{\textbf{w},k}^2,g^{*})
\wedge e(C_{\textbf{x},k}^1, \hat{g^{*}})==e(C_{\textbf{x},k}^2,g^{*})$

\hspace{12pt}
$\wedge e(C_{\textbf{y},k}^1, \hat{g^{*}})==e(C_{\textbf{y},k}^2,g^{*}))$;

\hspace{-3pt}\text{33:}
$b_4:=(e(C_{\textbf{T}}^1, \hat{g^{*}})==e(C_{\textbf{T}}^2,g^{*}))$;

\vspace{2pt}
\hspace{-3pt}\text{34:}
$H=e(C_{\textbf{T}}^1, g^{*}_1/{g^{*}}^k)\cdot e(C_{\textbf{L}_1}^1/C_{\textbf{L}_1,k}^1, g^{*})^{-1}$,
$\bar{g}=h_1/h^k$,

\hspace{-3pt}\text{35:}
$U=e(h^{\sigma}\bar{g}^{\tau},g^{*})\cdot H^{\mathtt{e}}$,

\hspace{-3pt}\text{36:}
$b_5:=(\mathtt{e}==\textsf{HASH}(C_{\textbf{L}_1},C_{\textbf{L}_1,k},C_{\textbf{T}},U,k))$;

\hspace{-3pt}\text{37:}
$b=(b_1\wedge b_2\wedge b_3\wedge b_4\wedge b_5)\rightarrow 0/1$.
\end{minipage}
}
\caption{Generating zk-SNARK proofs for relation $\mathtt{R}_{\textsf{PN}}^{S1}$.}
\label{fig:PNS2-1}
\end{figure}

\begin{figure}[h]
\fbox{
\begin{minipage}[t]{0.98\linewidth}
%\small
\scriptsize
\vspace{3pt}
\hspace{50pt}\textbf{\textsf{Step~2} CaP zk-SNARK Proofs over QMP}

\vspace{3pt}

\hspace{-5pt}
\textcolor[rgb]{0.1,0.1,0.8}{~//\textsf{st}=($\textbf{W}^{r(1)}_{ed,k}$, $\textbf{X}^{r(1)}_{et,k}$, $\textbf{Y}^{r(1)}_{et,k}$);~~~$io=m=3$;~QMP=($L(x)$,$R(x)$,$O(x)$,$t(x)$)}
 
%\hspace{-3pt}
%\textcolor[rgb]{0.1,0.1,0.8}{~/}
%QMP=($\textbf{W}^{r(1)}_{ed,k}\cdot L_{\textbf{W}}(x)$, $\textbf{X}^{r(1)}_{et,k}\cdot R_{\textbf{X}}(x)$, $\textbf{Y}^{r(1)}_{et,k}\cdot O_{\textbf{Y}}(x)$, $t(x)$) S
%the degree of the QMP is $1$ 
\hspace{-5pt}
\textcolor[rgb]{0.1,0.1,0.8}{~//$p(x,\textbf{Z})=\textsf{tr}\{\textbf{Z}^T\cdot L(x)\cdot R(x)\}-\textsf{tr}\{\textbf{Z}^T\cdot O(x)\}=h(x,\textbf{Z})t(x)$, $\forall~\textbf{Z}\in \textbf{M}_{(\mathtt{Z}_p)}^{l_{new}^2}$}

\hspace{-3pt}\text{1:}
\underline{\textsf{AC-$\Pi$.Setup}($1^{\lambda}$, QMP)$\rightarrow \textsf{crs}$}:

\hspace{-3pt}\text{2:}
$(\alpha, \beta, \gamma, \delta, \eta, z) \leftarrow \mathtt{Z}_p$,
$\textbf{Z} \leftarrow \textbf{M}_{(\mathtt{Z}_p)}^{l_{new}^2}$, $g, h \leftarrow \mathtt{G}$

\hspace{-3pt}\text{3:}
$\textsf{crs}:=(g, h, g^{\alpha}, g^{\beta}, h^{\beta}, h^{\gamma}, g^{\delta}, h^{\delta}, g^{\frac{\eta}{\delta}}, g^{\frac{\eta}{\gamma}},
\{\textsf{OP}_i=g^{\frac{\beta L_{i}(z)}{\gamma}} \cdot g^{\frac{\alpha R_{i}(z)}{\gamma}}\cdot g^{\frac{O_{i}(z)}{\gamma}}\}_{i=0}^t)$
%
%\textcolor[rgb]{0.1,0.1,0.8}{~//$ck_{\textsf{L}_1}$}

\vspace{3pt}
\hspace{-3pt}\text{4:}
\underline{\textsf{AC-$\Pi$.Prove}(\textsf{crs}, \textsf{st},\textsf{wt})$\rightarrow \pi^{S2}_{\textsf{L}_1}$}:

\hspace{-3pt}\text{5:}
%\textcolor[rgb]{0.1,0.1,0.8}{~//$h(z, Z)=\textsf{tr}\{Z^T\cdot (\textbf{W}^{r(1)}_{ed,k}\cdot L_{\textbf{W}}(z))\cdot (\textbf{X}^{r(1)}_{et,k}\cdot R_{\textbf{X}}(z))\}-\textsf{tr}\{Z^T\cdot (\textbf{Y}^{r(1)}_{et,k}\cdot O_{\textbf{Y}}(z))\}/t(z)$}
$L(x)=\cdot \textbf{W}^{r(1)}_{ed,k}\cdot L_{\textbf{W}}(z)$;
$R(x)=\textbf{X}^{r(1)}_{et,k}\cdot R_{\textbf{X}}(z)$;
$O(x)=\cdot \textbf{Y}^{r(1)}_{et,k}\cdot O_{\textbf{Y}}(z)$

\hspace{-3pt}\text{6:}
$(t,s,v) \overset{\$}{\leftarrow} \mathtt{Z}_p$,
$A:=g^{\alpha}\cdot g^{\textbf{Z}^TL(z)}\cdot g^{\delta t}$, 
$B:=h^{\beta}\cdot h^{R(z)}\cdot h^{\delta s}$,

\hspace{-3pt}\text{7:}
$C:=g^{h(z, \textbf{Z})t(z)/\delta}\cdot A^s \cdot (g^{\beta}\cdot g^{R(z)}\cdot g^{\delta s})^t \cdot g^{-ts\delta} \cdot g^{-v\eta/\delta}$,

\hspace{-3pt}\text{8:}
$D:=g^{\frac{\beta \textbf{Z}^TL(z)}{\gamma}}\cdot g^{\frac{\alpha R(z)}{\gamma}}\cdot g^{\frac{\textbf{Z}^TO(z)}{\gamma}}\cdot g^{v\eta/\gamma}$,

\hspace{-3pt}\text{9:}
$d_1=g^{\frac{\beta L(z)}{\gamma}}\cdot g^{v\eta/\gamma}, d_2=g^{\frac{\alpha R(z)}{\gamma}}\cdot g^{v\eta/\gamma}, d_3=g^{\frac{O(z)}{\gamma}}\cdot g^{v\eta/\gamma}$, 

\hspace{-3pt}\text{10:}
$C_{\textsf{L}_1,k}^{'}:=(d_1,d_2,d_3),$
$~~\pi^{S2}_{\textsf{L}_1}:=(A,B,C,D,C_{\textsf{L}_1,k}^{'})$

\vspace{1pt}
\hspace{-3pt}\text{11:}
\underline{\textsf{AC-$\Pi$.Verify}(\textsf{crs}, $\pi^{S2}_{\textsf{L}_1})\rightarrow 0/1$}:

\hspace{-3pt}\text{12:}
$b:=(\textsf{tr}\{e(A,B)\}==\textsf{tr}\{e(g^{\alpha},h^{\beta})\cdot e(D, h^{\gamma})\cdot e(C,h^{\delta})\})\rightarrow 0/1$.

\end{minipage}
}
\caption{Generating zk-SNARK proofs for relation $\mathtt{R}_{\textsf{PN}}^{S2}[\textsf{L}_1]$.}

\label{fig:PNS2}
\vspace{-10pt}
\end{figure}
 
\vspace{3pt}
\textbf{Proof Generation for $\mathtt{R}_{\textsf{PN}}$.} 
\wjs{As described before, the relation $\mathtt{R}_{\textsf{PN}}$ defines that \textsf{PriorNet} is performed in an encryption manner, by taking the encrypted inputs $\textbf{X}_{et}^{r(1)}$ in ring field $Z_q[x](x^{d_c}+1)$.}
To prove it, we adopt the methodology of efficient verification computation on encrypted data by Fiore \emph{et al.}~\cite{fiore2020boosting}.
We then integrate it with the aforementioned CaP zk-SNARK from \textsf{Groth} scheme~\cite{campanelli2019legosnark} for proving the satisifiability of QMP-based arithmetic circuits that are used to express convolutional operations.

\wjs{Following Fiore \emph{et al.}, there are two steps for completing the proofs.
With respective to the relation $\textsf{R}^{S_1}_{\textsf{PN}}$, \textsf{Step~1} is to prove the knowledge of the commitments to test inputs, trained weights of $\textsf{L}_1$, and intermediate outputs, by proving simultaneous evaluation of multiple ring field polynomials on a same point, as demonstrated in Fig. \ref{fig:PNS2-1}.
%
%In other word, the prover convinces a verifier of the openings of the public commitments with regard to model parameters and test inputs to be evaluated.
%
With respective to the relation $\textsf{R}^{S_2}_{\textsf{PN}}[\textsf{L}_1]$, \textsf{Step~2} shown in Fig. \ref{fig:PNS2} is to prove the arithmetic correctness of the matrix multiplication computation with our idea of using QMP.
The proofs in Fig.~\ref{fig:PNS2} contains the commitments to the QMP, and they are $C_{\textsf{L}_1,k}^{'}=(d_1, d_2, d_3)$.
For another relation $\textsf{R}^{S_2}_{\textsf{PN}}[\textsf{L}_2, L_3]$, the 
computation conducted in the two layers involves a 2-degree polynomial evaluated on $\textbf{Y}^{r(1)}_{et}$ and subsequently an average pooling function.
Hence, we remain using the QAP-based circuit to express them, instead of QMP-based one, and generate proofs by directly leveraging the CaP variant of the \textsf{Groth} scheme~\cite{campanelli2019legosnark}.}

\vspace{3pt}
\textbf{Proof Generation for $\mathtt{R}_{\textsf{LN}}$.} 
We previously describe in the relation $\mathtt{R}_{\textsf{LN}}$ that \textsf{LaterNet} is executed over scalars instead of polynomial rings, due to all of the data here are plaintext.
Hence, \textsf{Step~1} is not needed and we directly adopt the above \textsf{Step~2} to prove the validity of the relations regarding the \textsf{LaterNet}.
Notice, before proof generation, we express matrix multiplication operations in $\mathtt{R}_{\textsf{LN}}[\textsf{L}_4]$ and $\mathtt{R}_{\textsf{LN}}[\textsf{L}_7, \textsf{L}_8]$ using QMP-based arithmetic circuits, while expressing non-linear operations in $\mathtt{R}_{\textsf{LN}}[\textsf{L}_5, \textsf{L}_6]$ and $\mathtt{R}_{\textsf{LN}}[\textsf{L}_o]$ using QAP-based arithmetic circuits.
\begin{figure}[h]
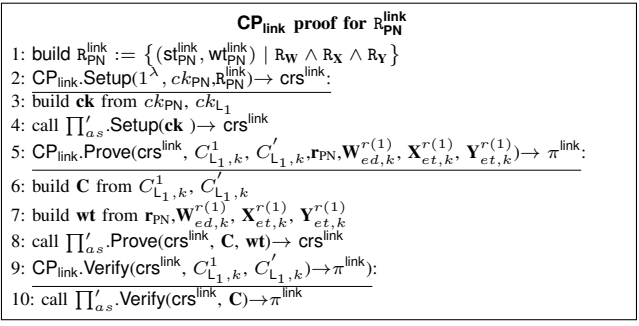

\fbox{
\begin{minipage}[t]{0.9\linewidth}
%\small
\scriptsize
\hspace{80pt} \textbf{$\textsf{CP}_{\textsf{link}}$ proof for $\mathtt{R}_{\textsf{PN}}^{\textsf{link}}$}

\vspace{3pt}

\hspace{-3pt}\text{1:}
\textsf{build}~$\mathtt{R}_{\textsf{PN}}^{\textsf{link}}:=\big\{(\textsf{st}_{\textsf{PN}}^{\textsf{link}}, 
\textsf{wt}_{\textsf{PN}}^{\textsf{link}}) \mid \mathtt{R}_\textbf{W} \wedge \mathtt{R}_\textbf{X} \wedge \mathtt{R}_\textbf{Y}\big\}$

\hspace{-3pt}\text{2:} \underline{$\textsf{CP}_{\textsf{link}}$.\textsf{Setup}($1^{\lambda}, ck_{\textsf{PN}}$,$\mathtt{R}_{\textsf{PN}}^{\textsf{link}}$)$\rightarrow \textsf{crs}^{\textsf{link}}$:}

\hspace{-3pt}\text{3:} 
build $\textbf{ck}$ from $ck_{\textsf{PN}}$, $ck_{\textsf{L}_1}$

\hspace{-3pt}\text{4:} 
call~$\prod_{as}'$.\textsf{Setup}($\textbf{ck}$ )$\rightarrow$ $\textsf{crs}^{\textsf{link}}$

\hspace{-3pt}\text{5:} \underline{$\textsf{CP}_{\textsf{link}}$.\textsf{Prove}($\textsf{crs}^{\textsf{link}}$, $C_{\textsf{L}_1,k}^{1}$, $C_{\textsf{L}_1,k}^{'}$,$\textbf{r}_{\text{PN}}$,$\textbf{W}^{r(1)}_{ed,k}$, $\textbf{X}^{r(1)}_{et,k}$, $\textbf{Y}^{r(1)}_{et,k}$)$\rightarrow \pi^{\textsf{link}}$}:

%$r_{\text{w},k}$,~$r_{\text{x},k}$,~$r_{\text{y},k}$,~$v$
\hspace{-3pt}\text{6:} 
build $\textbf{C}$ from $C_{\textsf{L}_1,k}^{1}$, $C_{\textsf{L}_1,k}^{'}$

\hspace{-3pt}\text{7:} 
build $\textbf{wt}$ from $\textbf{r}_{\text{PN}}$,$\textbf{W}^{r(1)}_{ed,k}$, $\textbf{X}^{r(1)}_{et,k}$, $\textbf{Y}^{r(1)}_{et,k}$

\hspace{-3pt}\text{8:} 
call~$\prod_{as}'$.\textsf{Prove}($\textsf{crs}^{\textsf{link}}$, $\textbf{C}$, $\textbf{wt}$)$\rightarrow$ $\textsf{crs}^{\textsf{link}}$

\hspace{-3pt}\text{9:} \underline{$\textsf{CP}_{\textsf{link}}$.\textsf{Verify}($\textsf{crs}^{\textsf{link}}$, $C_{\textsf{L}_1,k}^{1}$, $C_{\textsf{L}_1,k}^{'}$)$\rightarrow$$\pi^{\textsf{link}}$)}:

\hspace{-3pt}\text{10:} 
call~$\prod_{as}'$.\textsf{Verify}($\textsf{crs}^{\textsf{link}}$, $\textbf{C}$)$\rightarrow$$\pi^{\textsf{link}}$

\end{minipage}
}
\caption{Generating zk-SNARK proofs for relation $\mathtt{R}_{\textsf{PN}}^{\textsf{link}}$.}\label{fig:link}
\vspace{-25pt}
\end{figure}

\vspace{3pt}
\textbf{Proofs Composition.}
\wjs{We now need to combine the previous layer-wise proofs together by following the commit-and-prove methodology.
Concretely, we need to prove two commitments of two successive layer proofs open to the same values.
One of the commitments may commit to the inputs of a current layer while another one may commit to its previous layer's outputs.
Taking two separate proofs $\pi^{\textsf{S1}}$ and $\pi^{\textsf{S2}}_{\textsf{L}_1}$ as an example, we ensure the commitments $C_{\textsf{L}_1,k}^{1}$ and $C_{\textsf{L}_1,k}^{'}$  are opened to the same value, \emph{i.e.},
$\mathtt{R}_{\textsf{PN}}^{\textsf{link}}:=\big\{(\textsf{st}_{\textsf{PN}}^{\textsf{link}}=(C_{\textsf{L}_1,k}^{1}, C_{\textsf{L}_1,k}^{'}, ck_{\textsf{PN}}, ck_{\textsf{L}_1}), 
\textsf{wt}_{\textsf{PN}}^{\textsf{link}}=(\textbf{W}^{r(1)}_{et,k},\textbf{X}^{r(1)}_{et,k},\textbf{Y}^{r(1)}_{et,k},r_{\textsf{PN}},r'_{\textsf{PN}})) \mid$
$C_{\textsf{L}_1,k}^{1}=\textsf{MPoly.Com}(\textbf{W}^{r(1)}_{et,k},\textbf{X}^{r(1)}_{et,k},\textbf{Y}^{r(1)}_{et,k},r_{\textsf{PN}},ck_{\textsf{PN}})
\wedge
C_{\textsf{L}_1,k}^{'}=\textsf{MPoly.Com}(\textbf{W}^{r(1)}_{et,k},\textbf{X}^{r(1)}_{et,k},\textbf{Y}^{r(1)}_{et,k},r'_{\textsf{PN}},ck_{\textsf{L}_1})\big\}.$
Note that they are Pedersen-like commitments.
We make use of the previously proposed CaP zk-SNARK $\textsf{CP}_{\textsf{link}}$~\cite{campanelli2019legosnark} to composite two separate proofs by proving the validity of the relation $\mathtt{R}_{\textsf{PN}}^{\textsf{link}}$.
Also, the similar methodology can be applied into other-layer relations for linking the proofs regarding any two subsequent layers.
The key idea of the methodology is to reshape the computation of Pedersen commitment into the linear subspace computation.
Concretely, the relation of proving $C_{\text{w},k}^{1}=h^{r_{w,k}}g_{0,i}^{\sum_{i=0}^{l^2_{\text{new}}-1}w_i}$ (line 21 of Fig.~\ref{fig:PNS2-1}) and $d_1=g^{\frac{v\eta}{\gamma}}\cdot g^{\frac{\beta}{\gamma}\sum_{i=0}^{l^2_{\text{new}}-1}w_i}$ (line 9 of Fig.~\ref{fig:PNS2}) committing to $\textbf{W}^{r(1)}_{ed,k}$ is transformed into the linear subspace relation $\mathtt{R}_{\textbf{W}}: \textbf{C}= \textbf{ck} \cdot \textbf{wt}=\overbrace{
{\begin{bmatrix}
 g^{\frac{\beta}{\gamma}\text{w}_0}\cdot g^{v\eta/\gamma} \\ 
 g^{\frac{\beta}{\gamma}\text{w}_1}\cdot g^{v\eta/\gamma}  \\ 
 ...\\
 h^{r_{\text{w},k}}g_{0,i}^{\sum_{i=0}^{l^2_{\text{new}}-1}w_i}
\end{bmatrix}}}^{\textbf{C} \in \mathtt{G}^{l_{new}^2}}$}
\noindent $=
\overbrace{
{\begin{bmatrix}
0 & g^{\frac{\eta}{\gamma}} & g^{\frac{\beta}{\gamma}} & 0 & ... & 0 \\
0 & g^{\frac{\eta}{\gamma}} & 0 &  g^{\frac{\beta}{\gamma}} & ... & 0 \\
... & ... & ... &  ... & ... & ... \\
0 & g^{\frac{\eta}{\gamma}} & 0 &  0 & ... & g^{\frac{\beta}{\gamma}} \\
h & 0 & g_{0,0} &  ... & ... & g_{0,(l_{new}^2-1)} 
\end{bmatrix}}}^{\textbf{ck} \in \mathtt{G}^{(l_{new}^2)\times (l_{new}^2+2)}}$
$\times\overbrace{
{\begin{bmatrix}
r_{\text{w},k}\\
v\\
\text{w}_0\\
\text{w}_1\\
...\\
\text{w}_{(l_{new}^2-1)}\\
\end{bmatrix}_.}}^{\textbf{wt} \in \mathtt{Z}_p^{(l_{new}^2+2)}}$.

\wjs{Similarly, we can deduce relations $\mathtt{R}_{\textbf{X}}$ and $\mathtt{R}_{\textbf{Y}}$ for proving $C_{\text{x},k}^{1}$ and $d_2$ committing to $\textbf{X}^{r(1)}_{et,k}$, and $C_{\text{y},k}^{1}$ and $d_3$ committing to $\textbf{Y}^{r(1)}_{et,k}$, respectively.
As a result, we represent the former $\mathtt{R}_{\textsf{PN}}^{\textsf{link}}$ as  $\mathtt{R}_{\textsf{PN}}^{\textsf{link}}:=\big\{(\textsf{st}_{\textsf{PN}}^{\textsf{link}},\textsf{wt}_{\textsf{PN}}^{\textsf{link}})\mid \mathtt{R}_\textbf{W} \wedge \mathtt{R}_\textbf{X} \wedge \mathtt{R}_\textbf{Y}\big\}$.
Then, as demonstrated in Fig.~\ref{fig:link}, the proof $\pi^{\textsf{link}}$ for such a relation is generated by leveraging $\textsf{CP}_{\textsf{link}}$ that calls a scheme for proving linear subspace relations 
$\mathtt{R}_{\textbf{W}}$, $\mathtt{R}_{\textbf{X}}$ and $\mathtt{R}_{\textbf{Y}}$.}

\vspace{2pt}
%\noindent\textbf{\emph{(iii)~Aggregating multiple proofs}}
\subsection{Aggregating multiple proofs}~\label{sec:aggregation}
\vspace{-5pt}

Due to that a CNN model is tested with the test data from multiple testers, there are multiple proofs with respect to proving the correctness of each inference process.
For ease of storage overhead and verification cost, it is desirable to aggregate multiple proofs into a single proof.
%
%Specifically, storing the multiple proofs on the public platform is naturally not satisfactory due to high cost, \emph{e.g,} gas cost.
%
Besides, ensuring the verification computation as simple and low as possible can make our public platform more easily accessible to a later-coming user.
To the end, this section proceeds to introduce the algorithm of aggregating multiple proofs.

\vspace{3pt}
\textbf{Relation Definition.}
Suppose there are $n_t$ CaP proofs $\{\pi^{\textsf{m},i}=(A_i,B_i,C_i,D_i)\}_{i\in [n_t]}$ w.r.t a CNN $\textsf{m}$ which is tested by $n_t$ testers.
We define the aggregation relation that the multiple generated proofs are valid at the same time:
$\mathtt{R}_{\text{agg}}=\{(\textsf{st}_{\text{agg}},\textsf{wt}_{\text{agg}})\mid \{\textsf{AC-}\prod.\textsf{Verify}(\textsf{vk}^{\textsf{m}},\pi^{\textsf{m},i}, \textsf{st}^{\textsf{m},i})\rightarrow 1\}_{i\in [n_t]} \}.$
Here, $\textsf{vk}^{\textsf{m}}$ is the verification key, extracted from the common randomness string in the \textsf{Setup} algorithm, and 
$\textsf{vk}^{\textsf{m}}=(g^{\alpha},h^{\beta},\{\textsf{OP}_i\}_{i=0}^t, h^{\gamma},h^{\delta})$.
We note that $\textsf{vk}^{\textsf{m}}$ is the same for the different statements $\{\textsf{st}^{\textsf{m},i}\}_{i\in [n_t]}$, each of which contains $\textsf{st}^{\textsf{m},i}=(a_{i,0},...,a_{i,t})$.
$\pi^{\textsf{m},i}$ is the proof for the statement $\textsf{st}^{\textsf{m},i}$ based on the different test data and the different intermediates during the computation of $\textsf{m}$.

\vspace{3pt}
\textbf{Proof Aggregation.}
With the defined relation, we are ready to provide a proof for it.
A crucial tool to generate such a proof is recently proposed SnarkPack~\cite{gailly2021snarkpack}.
The external effort for us is to extend this work targeted at the original \textsf{Groth16} scheme~\cite{groth2016size} without a commit-and-prove component to our case, that is, the proofs based on the CaP \textsf{Groth16} scheme~\cite{campanelli2019legosnark} as well as our QMP-based proofs (see Fig.~\ref{fig:PNS2}).
As shown in line $18$ of Fig.~\ref{fig:agg}, we additionally compute a multi-exponentiation inner product for $\{D_i\}_{i\in [n_t]}$ of the multiple proofs, which is similar to computing the original $\{C_i\}_{i\in [n_t]}$.
As a result, we enable a single verification on the aggregated proof $\pi_{\textsf{agg}}$ for a later-coming user who is interested in the model \textsf{m} that is tested by $n_t$ testers.
\begin{figure}[!h]
\fbox{
\begin{minipage}[h]{0.98\linewidth}
%\small
\scriptsize
\hspace{60pt} \textbf{Aggregating multiple CaP proofs}

\vspace{3pt}

\hspace{-3pt}\text{1:}
\underline{\textsf{SnarkPack}.\textsf{Setup}($1^{\lambda},\mathtt{R}_{\text{agg}})\rightarrow \textsf{crs}_{\textsf{agg}}$:}

\hspace{-3pt}\text{2:}
obtain $g$ and $h$ from \textsf{crs}; $a,b \overset{\$}{\leftarrow} \mathtt{Z}_p$

\hspace{-3pt}\text{3:}
generate commitment keys $ck_{\textsf{two}}$ and $ck_{\textsf{one}}$:

\hspace{-3pt}\text{4:}
~$\vec{v_1}=(h,h^a,...,h^{a^{n-1}})$, ~$\vec{w_1}=(g^{a^n},...,g^{a^{2n-1}})$,

\hspace{-3pt}\text{5:}
~$\vec{v_2}=(h,h^b,...,h^{b^{n-1}})$, ~$\vec{w_2}=(g^{b^n},...,g^{b^{2n-1}})$

\hspace{-3pt}\text{6:}
$ck_{\textsf{two}}=(\vec{v_1},\vec{v_2},\vec{w_1},\vec{w_2})$,
$ck_{\textsf{one}}=(\vec{v_1},\vec{v_2})$

%\hspace{-3pt}\text{7:}
\textcolor[rgb]{0.1,0.1,0.8}{~//\textsf{call} a generalized inner product argument \textsf{MT\_IPP} (see Section 5.2 of \cite{gailly2021snarkpack})}

%\hspace{1pt}
%\textcolor[rgb]{0.1,0.1,0.8}{~//for a merging relation of multi-exponentiation inner product and target inner pairing product}

\hspace{-3pt}\text{7:}
\textsf{MT\_IPP}.Setup($1^{\lambda}$,$\mathtt{R}_{\textsf{MT}}$)$\rightarrow \textsf{crs}_{\textsf{MT}}$

\hspace{-3pt}\text{8:}
$\textsf{crs}_{\textsf{agg}}=(vk^{\textsf{m}},ck_{\textsf{two}},ck_{\textsf{one}},\textsf{crs}_{\textsf{MT}})$

\vspace{3pt}
\hspace{-3pt}\text{9:}
\underline{$\textsf{SnarkPack}.\textsf{Prove}(\textsf{crs}_{\textsf{agg}},\{\textsf{st}^{\textsf{m},i}\}_{i\in [n_t]},\{\pi^{\textsf{m},i}\}_{i\in [n_t]})$:}

\hspace{-5pt}\text{10:}
\textcolor[rgb]{0.1,0.1,0.8}{~//commit to $\{A_i\}_{i\in[n_t]}$ and $\{B_i\}_{i\in[n_t]}$ using $ck_{\textsf{two}}$}

\hspace{-5pt}\text{11:}
$C_{1-AB}=e(A_0,h)...e(A_{n_t-1},h^{a^{n-1}})...e(g^{a^n},B_0)...e(g^{a^{2n-1}},B_{n_t-1})$

\hspace{-5pt}\text{12:}
$C_{2-AB}=e(A_0,h)...e(B_{n_t-1},h^{b^{n-1}})...e(g^{b^n},B_0)...e(g^{b^{2n-1}},B_{n_t-1})$

\hspace{-5pt}\text{13:}
\textcolor[rgb]{0.1,0.1,0.8}{~//commit to $\{C_i\}_{i\in[n_t]}$ and $\{D_i\}_{i\in[n_t]}$ separately using $ck_{\textsf{one}}$}

\hspace{-5pt}\text{14:}
$C_{C}=e(C_0,h)...e(C_{n_t-1},h^{a^{n-1}})$, $C_{C}=e(C_0,h)...e(C_{n_t-1},h^{b^{n-1}})$

\hspace{-5pt}\text{15:}
$C_{D}=e(D_0,h)...e(D_{n_t-1},h^{a^{n-1}})$, $C_{D}=e(D_0,h)...e(D_{n_t-1},h^{b^{n-1}})$

\hspace{-5pt}\text{16:}
\textcolor[rgb]{0.1,0.1,0.8}{~//generate a challenge}

\hspace{-5pt}\text{17:}
$r=\textsf{HASH}(\{\textsf{st}^{\textsf{m},i}\}_{i\in [n_t]},C_{1-AB},C_{2-AB},C_{C},C_{D})$, $\vec{r}=(r^0,...,r^{n_t-1})$

\hspace{-5pt}\text{18:}
$I_{AB}=\prod_{i=0}^{n_t-1}e(A_i,B_i)^{r^i}$, 
$I_C=\prod_{i=0}^{n_t-1}(C_i)^{r^i}$, $I_D=\prod_{i=0}^{n_t-1}(D_i)^{r^i}$

\vspace{2pt}
\hspace{-5pt}\text{19:}
\textsf{MT\_IPP}.\textsf{Prove}($\textsf{crs}_{\textsf{MT}},C_{1-AB},C_{2-AB},C_{C},C_{D},I_{AB},I_C,I_D$,

\hspace{5pt}
$(A_i,B_i,C_i,D_i)_{i\in [n_t]},\vec{r})\rightarrow \pi_{\textsf{MT}}$

\hspace{-5pt}\text{21:}
$\pi_{\textsf{agg}}=(C_{1-AB},C_{2-AB},C_{C},C_{D},I_{AB},I_C,I_D,\pi_{\textsf{MT}})$

\vspace{3pt}
\hspace{-5pt}\text{22:}
\underline{\textsf{SnarkPack}.\textsf{Verify}($vk^{\textsf{m}},\textsf{crs}_{\textsf{MT}},\{\textsf{st}^{\textsf{m},i}\}_{i\in [n_t]},\pi_{\textsf{agg}})\rightarrow b_1\wedge b_2$}

\vspace{3pt}
\hspace{-5pt}\text{23:}\textsf{MT\_IPP}.\textsf{Verify}($\textsf{crs}_{\textsf{MT}},C_{1-AB},C_{2-AB},C_{C},C_{D},I_{AB},I_C,I_D,\vec{r},\pi_{\textsf{MT}})\rightarrow b_1$

\hspace{-5pt}\text{24:}
($I_{AB}\overset{?}{=}e(g^{\alpha\sum_{i=0}^{n_t-1}r^i},h^{\beta})e(I_D^{\sum_{i,j=0}^{n_t-1,t}a_{i,j}},h^{\gamma})e(I_C,h^{\delta}))\rightarrow b_2$

\end{minipage}
}
\caption{Generating zk-SNARK proofs for relation $\mathtt{R}_{\textsf{agg}}$.}\label{fig:agg}
\vspace{-20pt}
\end{figure}

\section{SECURITY ANALYSIS}\label{sec:analysis}
%
%This section mainly analyses that our designs based on the CaP zk-SNARK and the Leveled FHE satisfy correctness, security and privacy.
%
\begin{theorem}\label{the:VC}
If the underlying CaP zk-SNARK schemes are secure commit-and-prove arguments of knowledge $\prod_s$, the used leveled FHE scheme \textsf{L-FHE} is semantically secure and commitments \textsf{Com} are secure, and moreover a model is partitioned into \textsf{PriorNet} and \textsf{LaterNet} with an optimal splitting strategy, then our publicly verifiable model evaluation satisfies the following three properties:

\noindent\textbf{Correctness.} 
If \textsf{PriorNet}~$F_{\textbf{M}_1}$ runs on correctly encrypted inputs $\textbf{X}_{et}$ and outputs $\textbf{Y}_{et}^{(3)}$, and meanwhile, \textsf{LaterNet}~$F_{\textbf{M}_2}$ runs on the correctly decrypted $\textbf{Y}^{(3)}$ and returns $l_{\textsf{test}}$, then $l_{\textsf{test}}=F_{\textbf{M}_2}(\textbf{Y}^{(3)})$ and $\textbf{Y}_{et}^{(3)}=F_{\textbf{M}_1}(\textbf{X}_{et})$ are verified;

\noindent\textbf{Security.} If a PPT $\mathcal{A}$ knows all public parameters and public outputs of $\prod_s$, \textsf{L-FHE} and \textsf{Com}, as well as, the computation of $F_{\textbf{M}_1}$ and $F_{\textbf{M}_2}$, but has no access to the model parameters of $F_{\textbf{M}_1}$, it cannot generate $l_{\textsf{test}}^*$ which passes verification but $l_{\textsf{test}}^* \ne F_{\textbf{M}_2}(\textbf{Y}^{(3)})$;

\noindent\textbf{Privacy.} If a PPT $\mathcal{A}$ knows all public parameters and public outputs of $\prod_s$, \textsf{L-FHE} and \textsf{Com}, as well as, the computation of $F_{\textbf{M}_1}$ and $F_{\textbf{M}_2}$, but has no access to the model parameters of $F_{\textbf{M}_1}$, it cannot obtain any information about the clear $\textbf{X}$.\looseness=-1
\end{theorem}

We recall that a secure CaP zk-SNARK scheme should satisfy the properties of completeness, knowledge soundness and zero-knowledge; a secure commitment should be correct, hiding and binding, and a secure \textsf{L-FHE} scheme satisfies semantic security and correctness. %Due to the page limitation, analyses are presented in Appendix \ref{app:analysis}.
Now we are ready to analyze how the three properties of correctness, security and privacy can be supported by the underlying used cryptographic components of our publicly verifiable model evaluation method.
\wjs{Notice, as we mentioned in Section~\ref{subsec:cnn}, the proof generation for $\mathtt{R}_{\textsf{PN}}$ needs an additional step (namely \textsf{Step~1} in Fig.~\ref{fig:PNS2-1}), since $\mathtt{R}_{\textsf{PN}}$ relates to encrypted computation. $\mathtt{R}_{\textsf{LN}}$ only needs \textsf{Step~2}, relying on the completeness, knowledge soundness and zero-knowledge of the underlying CaP zk-SNARK scheme~\cite{campanelli2019legosnark}.
Our following analysis is exactly for $\mathtt{R}_{\textsf{PN}}$.}

%\wjs{//CaP zk-SNARK on encrypted data}
%
%\wjs{//security proof of CaP composition}
%

%\noindent\emph{A.~Correctness}
%

The correctness property relies on the correctness of \textsf{L-FHE} and commitments, and the completeness of the underlying CaP zk-SNARK schemes.
We employ a widely-adopted \textsf{L-FHE} scheme~\cite{fansomewhat} which has been proved correct.
For CaP zk-SNARK schemes, our designs use (\emph{a}) Fiore et al.'s CaP zk-SNARK for simultaneous evaluation of multiple ciphertexts generated by the \textsf{L-FHE} scheme~\cite{fiore2020boosting} (see Fig.~\ref{fig:PNS2-1}), (\emph{b}) the CaP \textsf{Groth16} scheme~\cite{groth2016size, campanelli2019legosnark}, (\emph{c}) our QMP-based zk-SNARK derived from the CaP \textsf{Groth16} (see Fig.~\ref{fig:PNS2}), and (\emph{d}) the CaP zk-SNARK $\textsf{CP}_{\textsf{link}}$ for compositing separate proofs~\cite{campanelli2019legosnark} (see Fig.~\ref{fig:link}).
%
%The schemes (\textsf{S1}), (\textsf{S2}) and (\textsf{S4}) satisfy completeness as proved in their papers.
%

Based on \emph{Theorem~10} of scheme (\emph{a})~\cite{fiore2020boosting}, we deduce the correctness of zk-SNARK proofs for relation $\mathtt{R}_{\textsf{PN}}^{S1}$ by direct verification (see Fig.~\ref{fig:PNS2-1}).
Specifically, if $C_{\textsf{L}_1},C_{\textbf{X}}$ are correct commitments, and for $\textsf{L}_1(x,y)=\textbf{W}(x,y)+ \textbf{X}(x,y)+\textbf{Y}(x,y)=\sum_{i=0,j=0}^{d_c, l_{\text{new}}^2-1}l_{i,j}x^iy^j$ and random point $k\in \mathtt{Z}_p$, $s,t\in \mathtt{Z}_p$, the following formula holds,
\begin{small}
\begin{align*}
&\textbf{T}(s,t) =\frac{\textsf{L}_1(s,t)-\textsf{L}_1(k,t)}{(s-k)}\\
&=\underline{[(\textsf{L}_{1,0}(s)\cdot t^0 + ... +\textsf{L}_{1,l_{\text{new}}^2-1}(s)\cdot t^{l_{\text{new}}^2-1}) -}\\
&~~~~~\frac{(\textsf{L}_{1,0}(k)\cdot t^0 + ... +\textsf{L}_{1,l_{\text{new}}^2-1}(k)\cdot t^{l_{\text{new}}^2-1})]}{(s-k)}\\
&=\underline{[(l_{0,0}(s-k)+...+l_{d_c,0}(s^{d_c}-k^{d_c}))\cdot t^{0}+...+}\\
&~~~~~\frac{(l_{0,l_{\text{new}}^2-1}(s-k)+...+l_{d_c,l_{\text{new}}^2-1}(s^{d_c}-k^{d_c}))\cdot t^{l_{\text{new}}^2-1}]}{(s-k)}\\
%
%&=\underline{[(l_{0,0}(x-k)+...+l_{d_c,0}(x-k)(x^{d_c-1}k^0+...+x^0k^{d_c-1}))}\\
%
%&\underline{\cdot y^{0}+...+}\\
%
%&\frac{(l_{0,0}(x-k)+...+l_{d_c,0}(x-k)(x^{d_c-1}k^0+...+x^0k^{d_c-1}))\cdot y^{l_{\text{new}}^2-1}]}{(x-k)}\\
%
&=\sum_{j=0}^{l_{\text{new}}^2-1}\sum_{t=0}^{d_c-1}\sum_{i=0}^{d_c-t}l_{i+t,j}k^{i-1}s^i t^j.
\end{align*}
\end{small} 
Note that $C_{\textbf{T}}$ correctly commits to $\textbf{T}(s,t)$, then given $\pi^{S1}=(C_{\textbf{T}}, \mathtt{e}, \sigma, \tau)$, a verifier can compute $U$ which is equal to $e(h^a\bar{g}^b, g^{*})$ (see line~$28$ in Fig.~\ref{fig:PNS2-1}) by
\begin{small}
\begin{align*}
U&=e(h^{\sigma}\bar{g}^{\tau},g^{*})\cdot e(C_{\textbf{T}}^1, g^{*}_1/{g^{*}}^k)^{\textsf{e}}\cdot e(C_{\textbf{L}}^1/C_{\textbf{L},k}^1, g^{*})^{-\textsf{e}}\\
&=e(h^{\sigma}\bar{g}^{\tau},g^{*})\cdot e(h^{r_{\textsf{T}}} \bar{g}^{\textsf{T}(s,t)},{g^*}^{(s-k)})^{\textsf{e}}\cdot e(\frac{h^{r_{\textsf{T}}}}{h^{r_{\textsf{T},k}}} \frac{g^{\textsf{L}_1(s,t)}}{g^{\textsf{L}_1(k,t)}},g^*)^{-\textsf{e}}\\
&=e(h^{\sigma}\bar{g}^{\tau},g^{*})\cdot e(h,g^*)^{\textsf{e} r_{\textsf{T}} (s-k)}\cdot e(g,g^*)^{\textsf{e}\textsf{T}(s,t)(s-k)}\cdot \\
&~~~~~~~e(h,g^*)^{-\textsf{e}(r_{\textsf{T}}-r_{\textsf{T},k})}\cdot e(g,g^*)^{-\textsf{e}(\textsf{L}_1(s,t)-\textsf{L}_1(k,t))}\\
&=e(h, g^{*})^{\sigma}\cdot e(h,g^{*})^{\textsf{e}(r_{\textsf{T},k}-r_{\textsf{T}})}\cdot
e(\bar{g},g^{*})^{\tau+\textsf{e} r_{\textsf{T}}}\cdot\\
&~~~~~~~e(g,g^*)^{\textsf{e}\textsf{T}(s,t)(s-k)}\cdot e(g,g^*)^{-\textsf{e}(\textsf{L}_1(s,t)-\textsf{L}_1(k,t))}\\
&=e(h, g^{*})^a\cdot e(\bar{g},g^{*})^b \cdot 1 \\
&=e(h^a\bar{g}^b, g^{*}).
\end{align*}
\end{small}

\noindent Based on \emph{Theorem H.1.} of scheme (\emph{b}) \cite{campanelli2019legosnark}, we proceed to derive the correctness of our QMP-based zk-SNARK proofs for relation $\mathtt{R}_{\textsf{PN}}^{S2}[\textsf{L}_1]$ (\emph{i.e.}, scheme (\emph{c})) by verifying $\pi^{S2}_{\textsf{L}_1}:=(A,B,C,D,C_{\textsf{L}_1,k}^{'})$, where $C_{\textsf{L}_1,k}^{'}$ will be used in $\textsf{CP}_{\textsf{Link}}$.
Specifically, a verifier needs to verify that (\textbf{\textsf{$\star$}}) $\textsf{tr}\{e(A,B)\}\overset{?}{==}\textsf{tr}\{e(g^{\alpha},h^{\beta})\cdot e(D, h^{\gamma})\cdot e(C,h^{\delta})\}$, where
\begin{small}
\begin{align*}
&\textsf{tr}\{e(A,B)\}=\textsf{tr}\{e(g^{\alpha}\cdot g^{Z^TL(z)}\cdot g^{\delta t},h^{\beta}\cdot h^{R(z)}\cdot h^{\delta s})\}~~~~~~~~\\
&=e(g,h)^{\textsf{tr}\{(\alpha+Z^TL(z)+\delta t)(\beta+R(z)+\delta s)\}}\\
&=e(g,h)^{(\alpha+\delta t)(\beta+R(z)+\delta s)+\textsf{tr}\{Z^TL(z)(\beta+R(z)+\delta s)\}}\\
&=e(g,h)^{(\alpha+\delta t)(\beta+R(z)+\delta s)+\beta Z^TL(z)+\textsf{tr}\{Z^TL(z)R(z)\}+\delta s Z^TL(z)},
\end{align*} 
\end{small}
\begin{small}
\vspace{-10pt}
\begin{align*}
&\textsf{tr}\{e(g^{\alpha},h^{\beta})\cdot e(D, h^{\gamma})\cdot e(C,h^{\delta})\}\\
&=\textsf{tr}\{e(g,h)^{\alpha \beta}\cdot e(g^{\frac{\beta Z^TL(z)+\alpha R(z)+Z^TO(z)+v\eta}{\gamma}},h^{\gamma})\\
&\cdot e(g^{h(z, Z)t(z)/\delta+(\alpha+Z^TL(z)+\delta t)s+(\beta+R(z)+\delta s)t-ts\delta-v\eta/\delta},h^{\delta})\}\\
&=e(g,h)^{\alpha \beta+\beta Z^TL(z)+\alpha R(z)+\textsf{tr}\{Z^TO(z)\}+v\eta}\\
&\cdot e(g,h)^{h(z, Z)t(z)+(\alpha+Z^TL(z)+\delta t)s\gamma+(\beta+R(z)+\delta s)t\gamma-ts\gamma^2-v\eta}.
\vspace{-30pt}
\end{align*}
\end{small}
Due to $p(x,Z)=\textsf{tr}\{Z^T\cdot L(x)\cdot R(x)\}-\textsf{tr}\{Z^T\cdot O(x)\}=h(x,Z)t(x)$, 
formula~(\textbf{\textsf{$\star$}}) holds.

The presentation of the correctness of schemes (\emph{b}) and (\emph{d}) is omitted, as they can be directly found in \cite{campanelli2019legosnark}.
%
%In addition, the correctness of ($\textsf{S}_4$) refers to Section $4$ in \cite{campanelli2019legosnark}.
%

%\vspace{5pt}
%\noindent\emph{B.~Security}
%

The security property relies on the correctness of \textsf{L-FHE}, and the knowledge soundness of CaP zk-SNARK we leveraged.
The correctness of \textsf{L-FHE} ensures that
$\textbf{Y}^{r(3)}_{et}=f^{\text{avg}}(f^{\textsf{act}}(f^{\text{conv}}(\textbf{X}_{et}^{r(1)})))$ decrypts to  $\textbf{Y}^{r(3)}=f^{\text{avg}}(f^{\textsf{act}}(f^{\text{conv}}(\textbf{X}^{r(1)})$.
For the knowledge soundness of our two-step zk-SNARK proofs (Fig.~\ref{fig:PNS2-1} and Fig.~\ref{fig:PNS2}), we rely on the knowledge soundness of the two schemes (\emph{a}) and (\emph{c}), and assume an adversary can black-box query a random oracle \textsf{HASH}.
We next analyze it specifically for the first layer $\textsf{L}_1$ from two aspects.
%
%We emphasize again scheme (\emph{c}) deduced from scheme (\emph{b}).
%

\wjs{On the one hand, based on the knowledge soundness of schemes (\emph{a}) and (\emph{c}), if any adversary~$\mathcal{A}$ can provide valid proofs with regard to layer-by-layer computations, \emph{e.g.,} $\pi^{\textsf{S1}}$ and $\pi^{\textsf{S2}}$ for $\textsf{L}_1$, there exists an extractor who is able to output the witnesses satisfying the corresponding defined relations~$\mathtt{R}_{\textsf{PN}}^{S_1}, \mathtt{R}_{\textsf{PN}}^{S_2}[\textsf{L}_1]$, with all but negligible probability.
Particularly, the knowledge soundness of scheme (\emph{a}) gives us that $\textbf{T}(s,t)=\sum_{j=0}^{l_{\text{new}}^2-1}\sum_{t=0}^{d_c-1}\sum_{i=0}^{d_c-t}l_{i+t,j}k^{i-1}s^i t^j$ is correct evaluation value in the random point $k$ of the polynomial $\textsf{L}_1(x,t)$; the knowledge soundness of scheme (\emph{c}) gives us that $\textsf{tr}\{Z^T\cdot L(z)\cdot R(z)\}-\textsf{tr}\{Z^T\cdot O(z)\}=h(z,Z)t(z)$.}

\wjs{On the other hand, the remaining probability $\mathcal{A}$ can cheat is that $\textbf{W}(x,t)\odot \textbf{X}(x,t)-\textbf{Y}^{*}(x,t)$ is a non-zero polynomial while $\textbf{W}(k,t)\odot \textbf{X}(k,t)-\textbf{Y}^{*}(k,t)=0$ (meaning $\textbf{Y}^{*r(1)}_{et} =\textbf{W}^{r(1)}_{ed}\odot \textbf{X}^{r(1)}_{et}$), which is negligible in $\textsf{L}_1$.
This depends on that we ensure parameters $q\gg d_c$ ($q=2^{109}$ and $d_c=4096$) and the point $k$ is randomly generated by the random oracle \textsf{HASH}.
The probability that $k$ is the root of the non-zero polynomial $\textbf{W}(x,t) \odot \textbf{X}(x,t)-\textbf{Y}^{*}(x,t)$ thus is negligible, \emph{i.e.}, $d_c$/$q$=$4096/2^{109}$.
Last, to analyze the knowledge soundness of the proofs for other layers of \textsf{PriorNet} is similar to the above two-aspect analysis for $\textsf{L}_1$, due to the nature of layer-wise computation.}
%

\iffalse
\wjs{we majorly analyse that (1)~if any adversary~$\mathcal{A}$ can provide valid proofs with regard to layer-by-layer computations, \emph{e.g.,} $\pi^{\textsf{S1}}$ and $\pi^{\textsf{S2}}$ for the first layer $\textsf{L}_1$, there exists an extractor outputting the witnesses, \emph{e.g.}, $\textbf{W}_{ed}^{r(1)}$ plus $\textbf{X}_{et}^{r(1)}$, and $\textbf{W}_{ed,k}^{r(1)}$ plus $\textbf{X}_{et,k}^{r(1)}$, with all but negligible probability, for satisfying our defined relations;
%
(2) the probability that $\mathcal{A}$ generates $l_{\textsf{test}}^* = F_{\textbf{M}_2}(F_{\textbf{M}_1}(\textbf{X}_{et}^{r(1)}))$ is negligible, that is, $d_c$/$q$=$4096/2^{109}$. 
%
In our case, it can be partitioned into the probability of generating $\textbf{Y}_{et}^{r(3)^*}=f^{\text{avg}}(f^{\text{s}}(f^{\text{conv}}(\textbf{X}_{et}^{r(1)})$ is negligible, and meanwhile, the probability of generating $l_{\textsf{test}}^*=f^{o}(f^{fc}(f^{fc}(f^{avg}(f^{s}(f^{conv}(\textbf{Y}^{r(3)^*}))))))$ is negligible.}
%
\fi

%\wjs{to be added}

%\vspace{2pt} 
%\noindent\emph{C.~Privacy}

The privacy property relies on the semantic security of \textsf{L-FHE}, the hiding property of comments and the zero-knowledge of the leveraged CaP zk-SNARK schemes.
In terms of semantic security, we derive from the security of a previously proposed \textsf{L-FHE} (applied to \textsf{PriorNet}), such that for any PPT adversary $\mathcal{A}$ who has acess to the resulting proofs, the probability of the following experiment $\textsf{Exp}_{\mathcal{A}}^{Privacy}[\textsf{L-FHE},f^{\text{avg}}(f^{\text{act}}(f^{\text{conv}}(\cdot), \lambda]$ outputting 1 is not larger than $1/2+\text{negl}(\lambda)$:

\begin{small}
\begin{align*}
   \vspace{-20pt}
    &\textsf{Exp}_{\mathcal{A}}^{Privacy}[\textsf{L-FHE},f^{\text{avg}}(f^{\text{act}}(f^{\text{conv}}(\cdot), \lambda]:\\
    &~~~b \leftarrow {0,1};\\
    &~~~(pk_u, sk_u)\leftarrow \textsf{L-FHE}.\textsf{KeyGen}(1^{\lambda});\\
    &~~~(\textbf{X}^{r(1)}_0,\textbf{X}^{r(1)}_1) \leftarrow \mathcal{A}(pk_u);\\
    &~~~(\textbf{X}^{r(1)}_{et,b}) \leftarrow \textsf{L-FHE}.\textsf{Enc}(\textbf{X}^{r(1)}_{b});\\
    &~~~b^* \leftarrow \mathcal{A}(pk_u, \textbf{X}^{r(1)}_{et,b}, \pi^{\textsf{S1}}, \pi^{\textsf{S2}}),\\
    &~~~\text{If}~b^*=b, \text{output}~1, \text{else}~0.
    \vspace{-20pt}
\end{align*}
\end{small}

For commitments, we leverage the \textsf{MPoly.Com} scheme which is perfect hiding as proven in \emph{Theorem}~8 of~\cite{fiore2020boosting}. 

\wjs{We now analyse zero-knowledge based on the zero-knowledge simulators of the underlying zk-SNARKs, including $(\textsf{Sim}^{\textsf{MUniEv}-\Pi.\textsf{Setup}}, \textsf{Sim}^{\textsf{MUniEv}-\Pi.\textsf{Prove}})$ and $(\textsf{Sim}^{\textsf{AC}-\Pi.\textsf{Setup}}, \textsf{Sim}^{\textsf{AC}-\Pi.\textsf{Prove}})$, respectively.
$\textsf{Sim}^{\textsf{MUniEv}-\Pi.\textsf{Setup}}$ and $\textsf{Sim}^{\textsf{AC}-\Pi.\textsf{Setup}}$ are the same as the algorithms of $\textsf{MUniEv}-\Pi.\textsf{Setup}$ and $\textsf{AC}-\Pi.\textsf{Setup}$ respectively, and generate specific common random strings for the two zk-SNARKs by running $\textsf{Sim}^{\textsf{MUniEv}-\Pi.\textsf{Setup}}(1^{\lambda})\rightarrow (\textsf{crs}^{S1}, \textsf{td}^{S1})$ and $\textsf{Sim}^{\textsf{AC}-\Pi.\textsf{Setup}} (1^{\lambda})\rightarrow (\textsf{crs}^{S2}, \textsf{td}^{S2})$. 
Then, we use the simulations $\textsf{Sim}^{\textsf{MUniEv}-\Pi.\textsf{Prove}}$ and $\textsf{Sim}^{\textsf{AC}-\Pi.\textsf{Prove}}$ to generate simulated proofs with the \textsf{crs} for the corresponding statements.
With the simulators, we proceed with the following games, in which any PPT distinguisher $\mathcal{D}$ successfully distinguishes the simulation with a negligible probability.}

\wjs{\textsf{Hybrid~0.} This starts with the real algorithms in Fig.~\ref{fig:PNS2-1} and Fig.~\ref{fig:PNS2}, where the proofs $\pi^{S1}$ and $\pi^{S2}$ are generated by $\textsf{MUniEv}-\Pi.\textsf{Prove}$ and $\textsf{AC}-\Pi.\textsf{Prove}$, respectively.}

\wjs{\textsf{Hybrid~1.} This remains unchanged in the usage of witnesses to be proven and we generate commitments as the line 11-12, 19-23 and 27 presented of Fig.~\ref{fig:PNS2-1}. But we adopt the zero-knowledge simulators including $(\textsf{Sim}^{\textsf{MUniEv}-\Pi.\textsf{Setup}}, \textsf{Sim}^{\textsf{MUniEv}-\Pi.\textsf{Prove}})$ and $(\textsf{Sim}^{\textsf{AC}-\Pi.\textsf{Setup}}, \textsf{Sim}^{\textsf{AC}-\Pi.\textsf{Prove}})$ to generate proofs.
Based on that the real $\textsf{MUniEv}-\Pi.\textsf{Prove}$ and $\textsf{AC}-\Pi.\textsf{Prove}$ are zero-knowledge algorithms, \textsf{Hybrid~0} and \textsf{Hybrid~1} are indistinguishable.}\looseness=-1

\wjs{\textsf{Hybrid~2.} This runs the same algorithms as \textsf{Hybrid~1} except that we replace the values in the above commitments with zeros. Based on the hiding property of the leveraged \textsf{MPoly.Com} scheme, \textsf{Hybrid~1} and \textsf{Hybrid~2} are indistinguishable.}
%

\iffalse
\begin{small}
\begin{align*}
    \textsf{Pr}&\{\textsf{MUniEv}-\Pi.\textsf{Setup}(1^{\lambda}, d_{c}, n_{c})\rightarrow (ck^{S1},\textsf{td}), \\
    %
    &\textsf{MUniEv}-\Pi.\textsf{Prove}(ck^{S1}, \textbf{W}^{r(1)}_{ed},\textbf{X}^{r(1)}_{et})\rightarrow \pi^{S1},\\
    &\textsf{Sim}^{\textsf{MUniEv}-\Pi}(\textsf{td},C_{\textbf{W}_1}, C_{\textbf{X}}) \rightarrow \pi^{S1^*};\\
    %
    &\textsf{AC}-\Pi.\textsf{Setup}(1^{\lambda}, QMP)\rightarrow (\textsf{crs},\textsf{td})\\
    %
    &\textsf{AC}-\Pi.\textsf{Prove}(\textsf{crs}, \textsf{st},\textsf{wt})\rightarrow \pi^{S2},\\
    &\textsf{Sim}^{\textsf{AC}-\Pi}(\textsf{td},\textsf{st}) \rightarrow \pi^{S2^*}\\
    &|\mathcal{D}(\pi^{S1},\pi^{S1^*}=1\wedge\mathcal{D}(\pi^{S2},\pi^{S2^*}=1)\}\leq \text{negl}(\lambda).
\end{align*}
\end{small}
\fi
% 

%----VC is secure-----

\section{Experiments}\label{sec: experiments}

\noindent\textbf{Implementation.}
The main components of our implementation include zk-SNARK systems, FHE and polynomial commitment. 
We firstly use the libsnark library in C++ to implement proving/verification based on \textsf{Groth16} of QAP and QMP-based circuits during the process of CNN prediction.
We then use the Microsoft SEAL library to encrypt test inputs and evaluate \textsf{PriorNet} on ciphered inputs, with necessary parameters setting (\emph{i.e.}, $d_c=4096, q=2^{109}, t=1032193\approx 2^{20}$) as recommended.
%\footnote{https://github.com/microsoft/SEAL/} 
%\footnote{https://github.com/scipr-lab/libsnark}
We proceed to implement the component of polynomial commitment using the \textsf{libff} library of libsnark, for generating commitments to encrypted/unencrypted data.

In the aspect of machine learning models, we use well-trained CNN models to run the prediction process over the single-channel MNIST and \wjs{three-channel CIFAR-10 datasets.} 
Specifically, we start from a toy CNN with a convolution layer, a ReLU layer plus an average pooling layer, and generate proofs w.r.t the toy CNN running on the MNIST dataset.
We then extend it to \textsf{LeNet-5} over the MNIST dataset.
\wjsRR{The \textsf{LeNet-5} architecture is $32\times 32\times 1$ $\xrightarrow[\text{filter~size}=5\times 5\times 6]{conv_1 ~\text{and}~act_1}$~$28\times 28\times 6$~$\xrightarrow[\text{size}=2\times 2]{pool_1}$~$14\times 14\times 6$~$\xrightarrow[\text{filter~size}=5\times 5\times 16]{conv_2~\text{and}~act_2}$~$10\times 10\times 16$~$\xrightarrow[\text{size}=2\times 2]{pool_2}$~$5\times 5\times 16$~$\xrightarrow[\text{ }]{fc_3~\text{and}~act_3}$~$120\times 1\times 1$~$\xrightarrow[\text{ }]{fc_4~\text{and}~act_4}$~$84\times 1\times 1$~$\xrightarrow[\text{ }]{fc_o}$~$10\times 1\times 1$.} 
We additionally run the model on the CIFAR-10 dataset. 
In this part, we spend major efforts on approximately handling convolution operations over the two datasets with different-scale matrix multiplication, before generating proofs.
Due to the parameters and pixel values of data are floating-point numbers, not adapting to zk-SNARK systems, we use a generic $8$-bit unsigned quantization technique to transform them into integers in [$0,255$].
We conduct our MNIST experiments on a Ubuntu $18.04.6$ server with a 6-core Ryzen5 processor, $7.8$~GB RAM and $97.2$~GB Disk space, and execute CIFAR-10 experiments on a docker container with Ubuntu18.04, Quad-core Intel i5-7500 processor running at 3.4 GHz and 48 GB RAM.

\noindent\textbf{Evaluation.}
\wjsSec{We evaluate the running time and the storage overhead of performing \textsf{PriorNet} over a single encrypted image, by taking \textsf{LeNet-5} as an example and selecting different split points.
The evaluation results are shown in TABLE~\ref{tab:fhe-1}. 
We also evaluate the overhead with an increasing number of encrypted images, when a split point is selected at the activation layer after the first full connection layer, see TABLE~\ref{tab:fhe-2}.}

\begin{table}[h]
\vspace{-8pt}
\scriptsize
  \centering
  \caption{\small Running time (s) and storage overhead (MB) of performing \textsf{PriorNet}. If the split point is at $fc_3$, the \textsf{PriorNet} $F_{\textbf{M}_1}$ can be represented by $f^{fc_3}(f^{pool_2}(f^{act_2}(f^{conv_2}(f^{pool_1}(f^{act_1}(f^{conv_1}(\cdot)))))))$.}
    \begin{tabular}{cccccc}
    \toprule
    \textbf{Split point}  & $conv_1$ & $conv_2$ & $fc_3$ & $fc_4$ & $fc_5$ \\
    \midrule
    \textbf{Running time} & $12.95$ & $46.08$ & $53.70$ & $55.50$ & $56.26$\\
    \textbf{Storage} & $5.11$ &	$84.31$ & $1702.23$ & $2035.23$ & $2063.13$\\
    \bottomrule
    \end{tabular}
  \label{tab:fhe-1}
\vspace{-5pt}
\end{table}
\begin{table}[h]
\vspace{-10pt}
\scriptsize
  \centering
  \caption{\small Running time (s) with an increasing number of encrypted images.}
    \begin{tabular}{cccccc}
    \toprule
    \textbf{Dataset size}  & $10$ & $30$ & $50$ & $80$ & $100$ \\
    \midrule
    \textbf{Running time} & $1272.68$ & $3807.05$ & $6343.02$ & $10261.01$ & $12687.47$\\ 
    \bottomrule
    \end{tabular}
  \label{tab:fhe-2}
\vspace{-5pt}
\end{table}

\begin{figure}[ht]
\vspace{-10pt}
\scriptsize
	\centering
        \hspace{-50pt}
	\subfigure{
		\begin{minipage}[b]{0.135\textwidth}
			\includegraphics[width=1.4\textwidth,height=3.0cm]{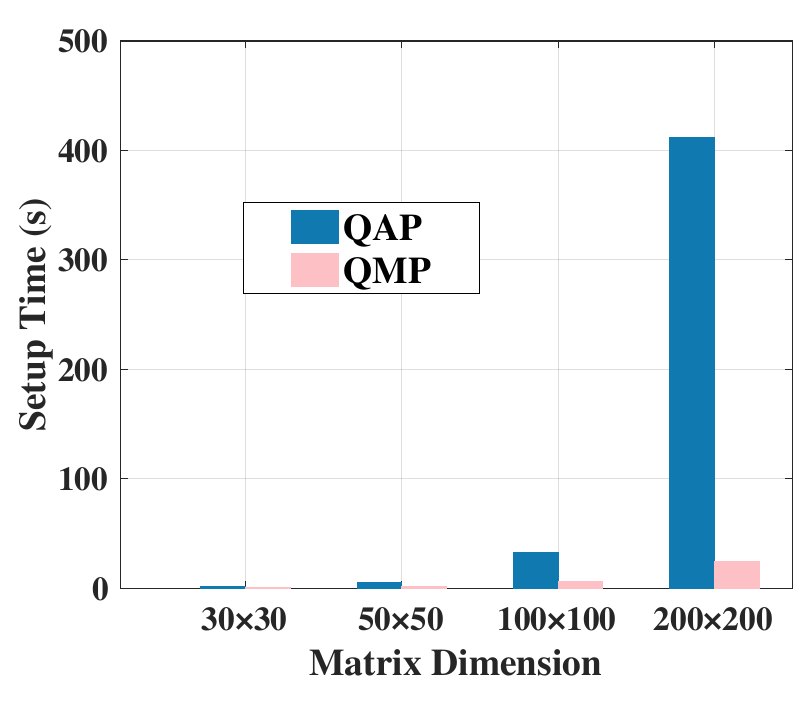}
		\end{minipage}
		
	} 
	\hspace{30pt}    
    	\subfigure{
    		\begin{minipage}[b]{0.135\textwidth}
   		 	\includegraphics[width=1.4\textwidth,height=3.0cm]{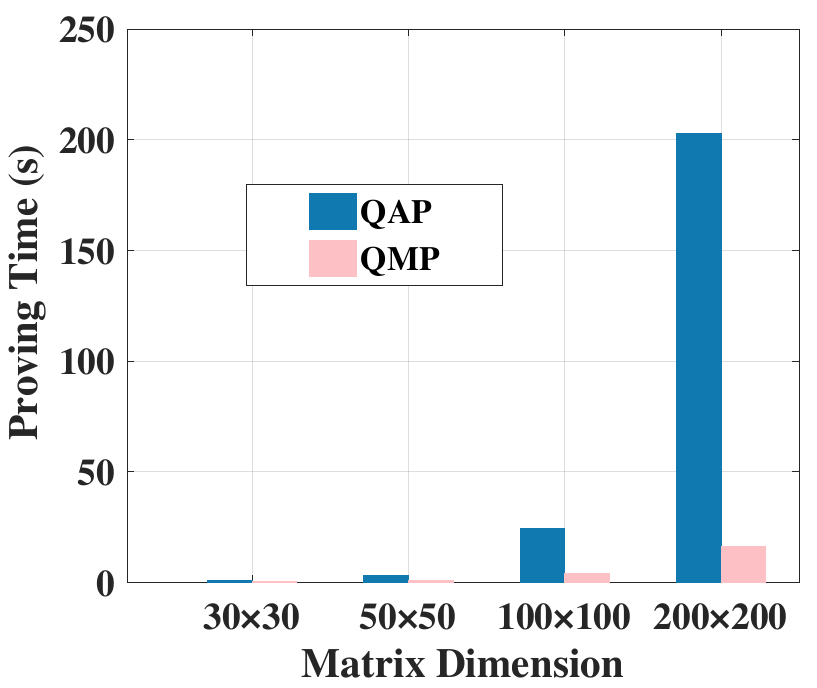}		
       \end{minipage}
    	}
	\caption{Comparison on \textsf{Setup} time and proving time.}
	\label{fig:crs}
	\vspace{-5pt} 
\end{figure}

Next, we compare QMP-based zk-SNARK and QAP-based zk-SNARK for matrix multiplication in terms of the proving time, \textsf{Setup} time and CRS size, as shown Fig.\ref{fig:crs} and TABLE~\ref{tab:crsSize}.
%
%Evaluation results with regard to verification time and proof size are discussed in Appendix~\ref{app:res}.
%  
Concretely, we simulate the operations of matrix multiplication in increasing dimensions up to $200\times 200$ padding with random integers in [$0,10$].
Then, we generate proofs for the matrix multiplication operations using the QAP-based and QMP-based zk-SNARK.
Fig.\ref{fig:crs} shows that the QMP-based zk-SNARK is efficient than the QAP-based one in generating CRS and proof, as the matrix dimension increases.
%
%In terms of proving time, QMP-based zk-SNARK enables more efficient proof generation than QAP-based zk-SNARK on matrix multiplication.
%
For the $200\times 200$ matrix multiplication, the QMP-based zk-SNARK is $17.6\times$ and $13.9\times$ faster than the QAP-based one in \textsf{Setup} time and proving time, respectively.
Besides, the QMP-based zk-SNARK obviously produces smaller CRS size than the QAP-based zk-SNARK, see  TABLE~\ref{tab:crsSize}. 

\begin{table}[h]
\vspace{-7pt}
\scriptsize
  \centering
  \caption{CRS size comparison (KB).}
    \begin{tabular}{ccccc}
    \toprule
    \textbf{Matrix dimension} & $\mathbf{30\times 30}$ & $\mathbf{50\times 50}$ & $\mathbf{100\times 100}$ & $\mathbf{200\times 200}$ \\
    \midrule
    \textbf{QAP} & 2,911.22 &	12,445.73 & 97,230.09 &	768,503.09 \\
    \textbf{QMP} & 84.42 &233.83 & 934.21& 3,735.72  \\
    \bottomrule
    \bottomrule
    \textbf{Matrix dimension} & $\mathbf{500\times }$ & $\mathbf{1000\times }$ & $\mathbf{2000\times }$ & $\mathbf{3000\times }$ \\
    \midrule
    \textbf{QMP} &  23,346.32 &	93,384.16 & 	373,535.53	& 840,454.47  \\
    \bottomrule
    \end{tabular}
  \label{tab:crsSize}
\vspace{-10pt}
\end{table}
\begin{figure}[h]
\vspace{-5pt}
\scriptsize
	\centering
         \hspace{-50pt}
	\subfigure{
		\begin{minipage}[b]{0.135\textwidth}
			\includegraphics[width=1.2\textwidth,height=3.0cm]{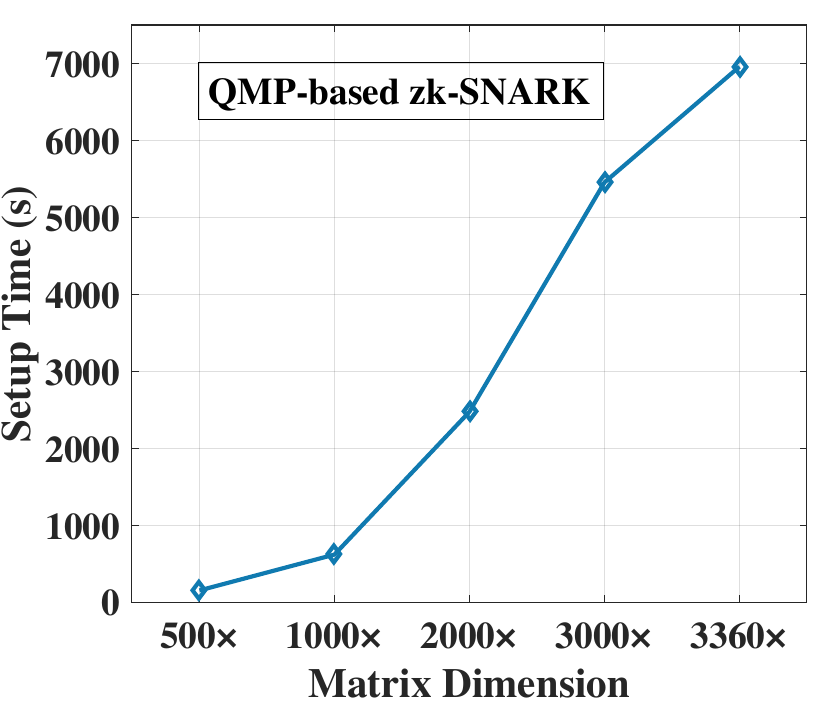}
		\end{minipage}
		
	}
	\hspace{30pt}
    	\subfigure{
    		\begin{minipage}[b]{0.135\textwidth}
   		 	\includegraphics[width=1.2\textwidth,height=3.0cm]{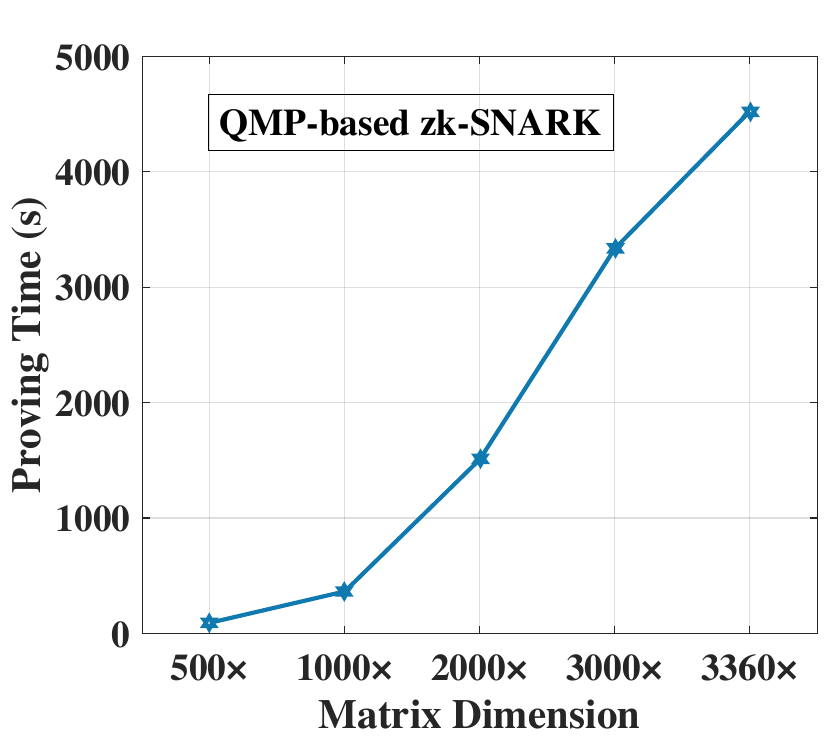}		
   		 	\end{minipage}
    	}
	\caption{Performance of our QMP-based zk-SNARK handling matrix multiplication in dimensions greater than $220\times 220$.}
	\label{fig:proving}
	\vspace{-10pt}
\end{figure}  
    
\begin{figure}[htbp]
\vspace{-12pt}
\scriptsize
	\centering
         \hspace{-50pt}
	\subfigure{
		\begin{minipage}[b]{0.135\textwidth}
			\includegraphics[width=1.4\textwidth,height=3.0cm]{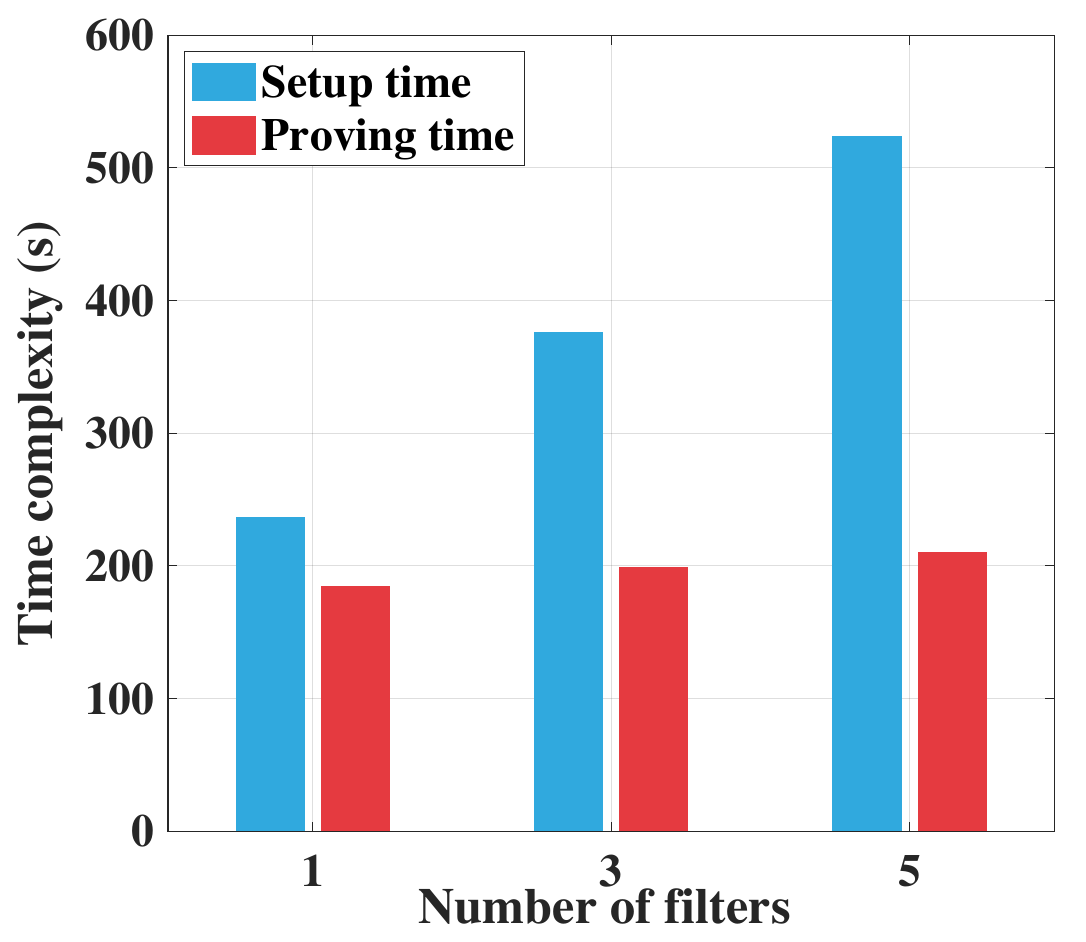}
		\end{minipage}
	} 
	\hspace{30pt}
    	\subfigure{
    		\begin{minipage}[b]{0.135\textwidth}
   		 	\includegraphics[width=1.4\textwidth,height=3.0cm]{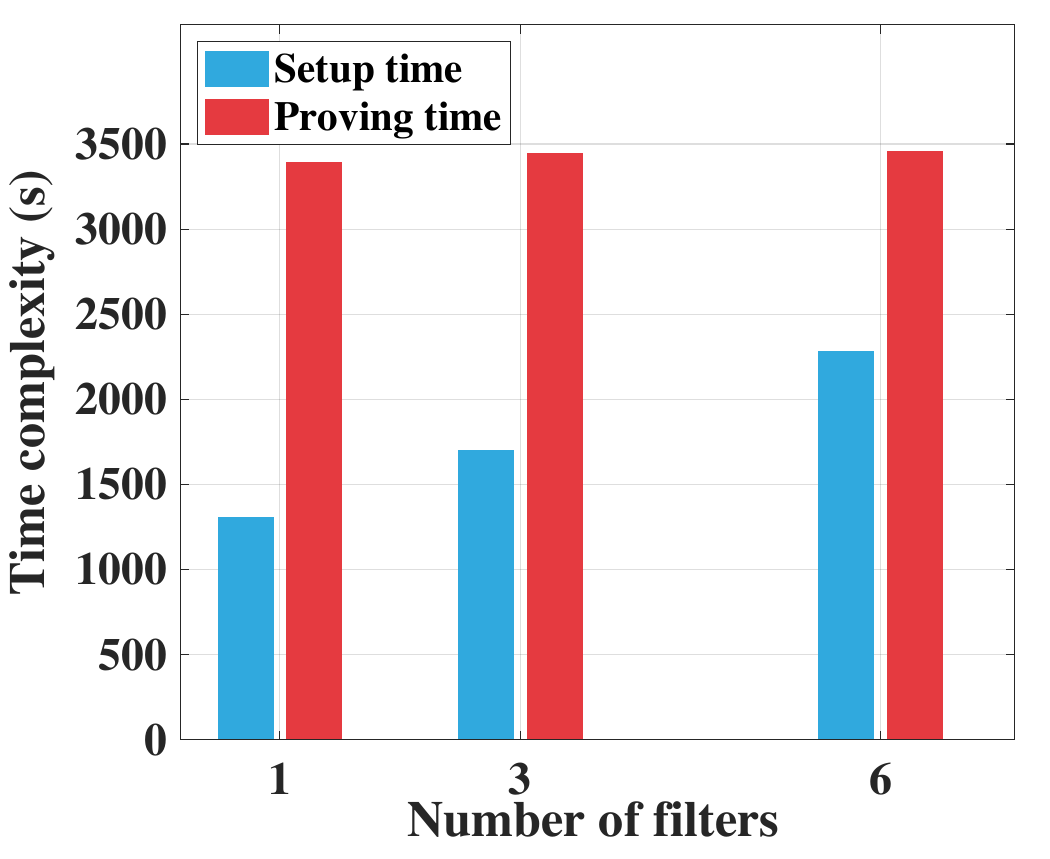}		\end{minipage}
    	}
	\caption{\textsf{Setup} time and proving time for a conv. layer with 1000 images from MNIST and CIFAR-10.}
	\label{fig:mnist_data}
	\vspace{-10pt} 
\end{figure}
\begin{table}
\vspace{-10pt}
\scriptsize
  \centering
  \caption{Other performance metrics.}
    \begin{tabular}{lcccc}
    \toprule
    \multirow{2}*{\textbf{Datasets}}&\multirow{2}*{\textbf{\#Filter}}&\multirow{2}*{\textbf{Verification time (s)}}&\multicolumn{2}{c}{\textbf{Size (KB)}}\\
    \cline{4-5}
    & &  & \textbf{CRS}& \textbf{Proof}\\
    \midrule
    \multirow{3}*{\textbf{MNIST}} & 1 & 54559&1,054,266.00&351,421.97\\
    & 3 & 54002&1,054,266.00&351,421.97\\
    & 5 & 56011&1,054,266.00&351,421.97\\
    \midrule
    \multirow{3}*{\textbf{CIFAR-10}} & 1 &375942 &15353859056
 &351,421.97\\
    & 3 & 376144 & 15353859056 &351,421.97\\
    & 6 & 408911 & 16927629296 &351,421.97\\
    \bottomrule
    \end{tabular}%
  \label{tab:other_metrics} %
  \vspace{-10pt}
\end{table}
We can see in Fig.~\ref{fig:proving} that the QMP-based zk-SNARK can handle the matrix multiplication in increasing dimensions up to $3360\times3360$, which is its merit, compared to the QAP-based zk-SNARK supporting the maximum number of multiplication bound $10^7$~\cite{keuffer2018efficient}.
Also, the proving time almost increases linearly by the dimension of matrices.
%

\iffalse
In terms of verification,  the verification
time of QMP-based zk-SNARK grows faster than that of QAP-based one, as shown in TABLE~\ref{tab:verTime}.   
%
In our future work, we may introduce a random sampling strategy in proof generation to reduce verification cost, \emph{e.g.}, randomly sampling a bounded number of values in two matrices to be multiplied, and resetting the non-chosen values as zeros.
%
In such a way, only the sampled values in the two matrices are multiplied and only their multiplication correctness need to be proved and verified.
%
The proof size also becomes lager when the matrix dimension turns lager as shown in TABLE~\ref{tab:proofSize}, while the proof size in the QAP-based zk-SNARK keeps $1019~$bits regardless of the matrix dimension.
%
\fi

% 
\begin{figure}[t]
%\vspace{-15pt}
	\centering
	\hspace{-30pt}
	\subfigure{
		\begin{minipage}{0.135\textwidth}
	\includegraphics[width=1.2\textwidth,height=2.8cm]{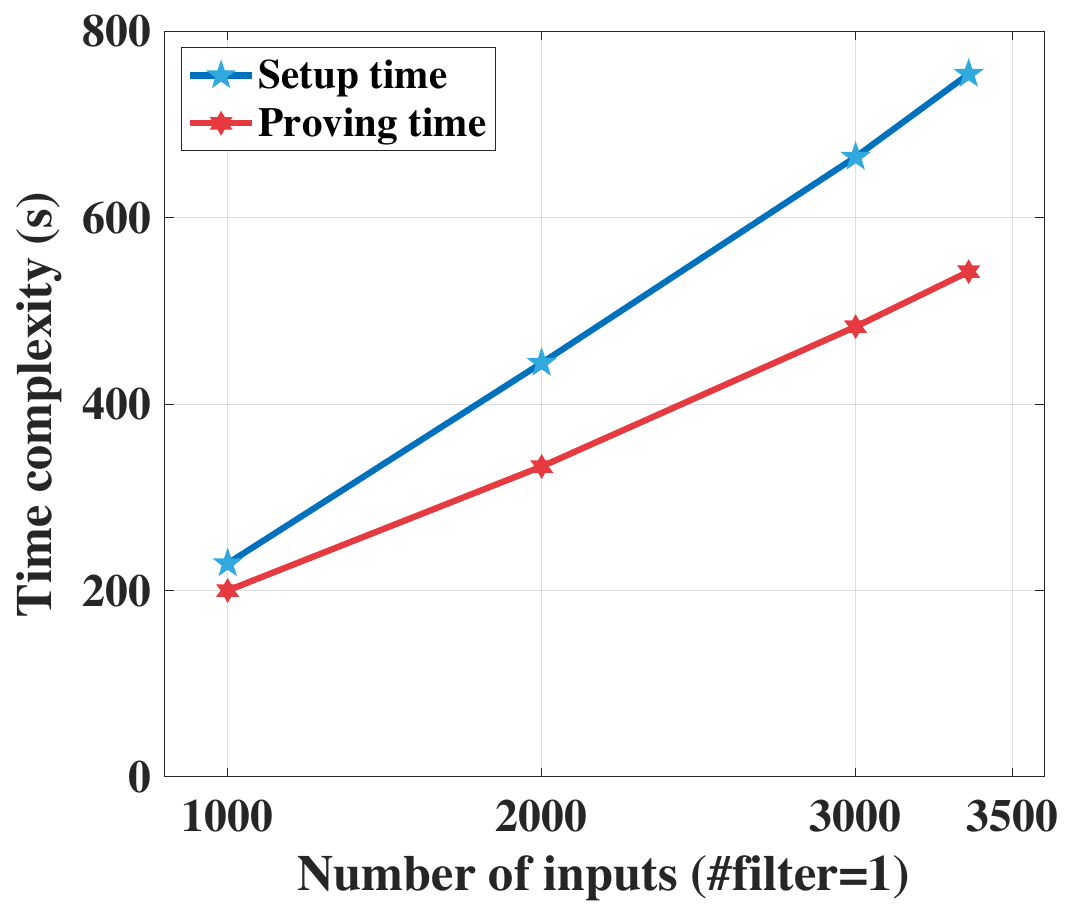}
		\end{minipage} 
			\label{fig:filter:one}
	}
	\hspace{2pt}
    	\subfigure{
    		\begin{minipage}{0.135\textwidth}
   	 \includegraphics[width=1.2\textwidth,height=2.8cm]{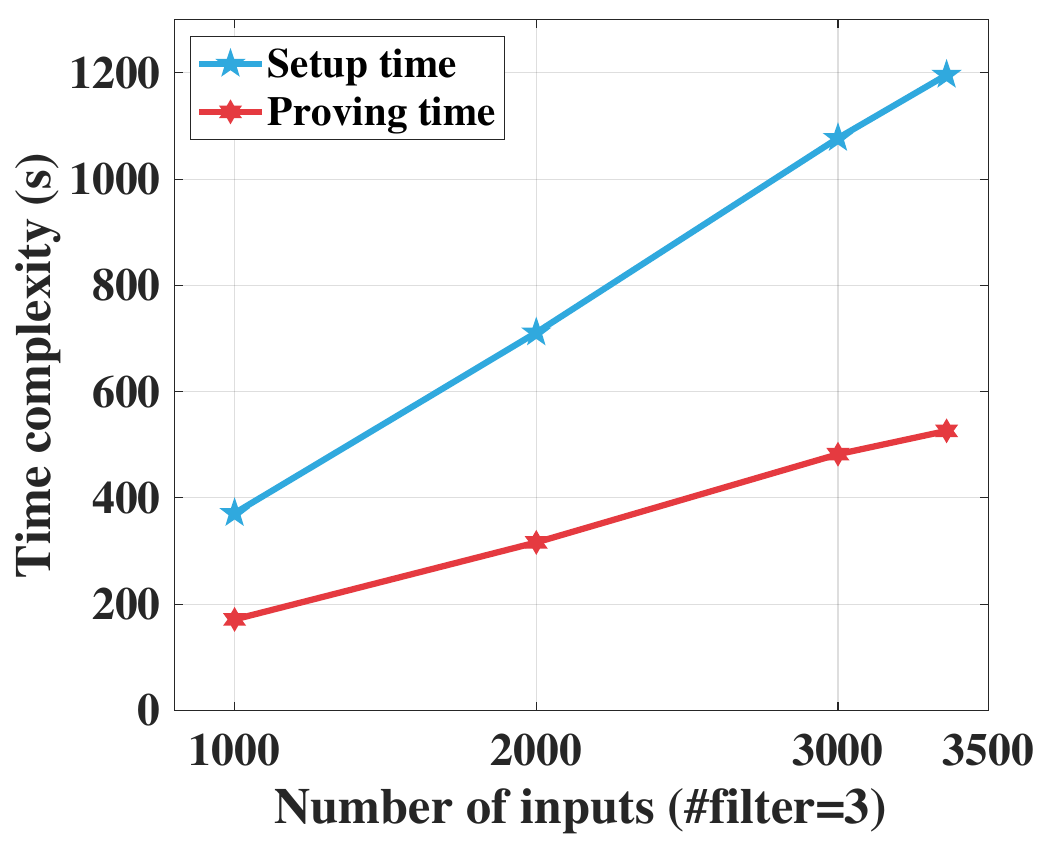}
    		\end{minipage}
		\label{fig:filter:two}
    	}
    \hspace{2pt}
    	\subfigure{
    		\begin{minipage}{0.135\textwidth}
   	 \includegraphics[width=1.2\textwidth,height=2.8cm]{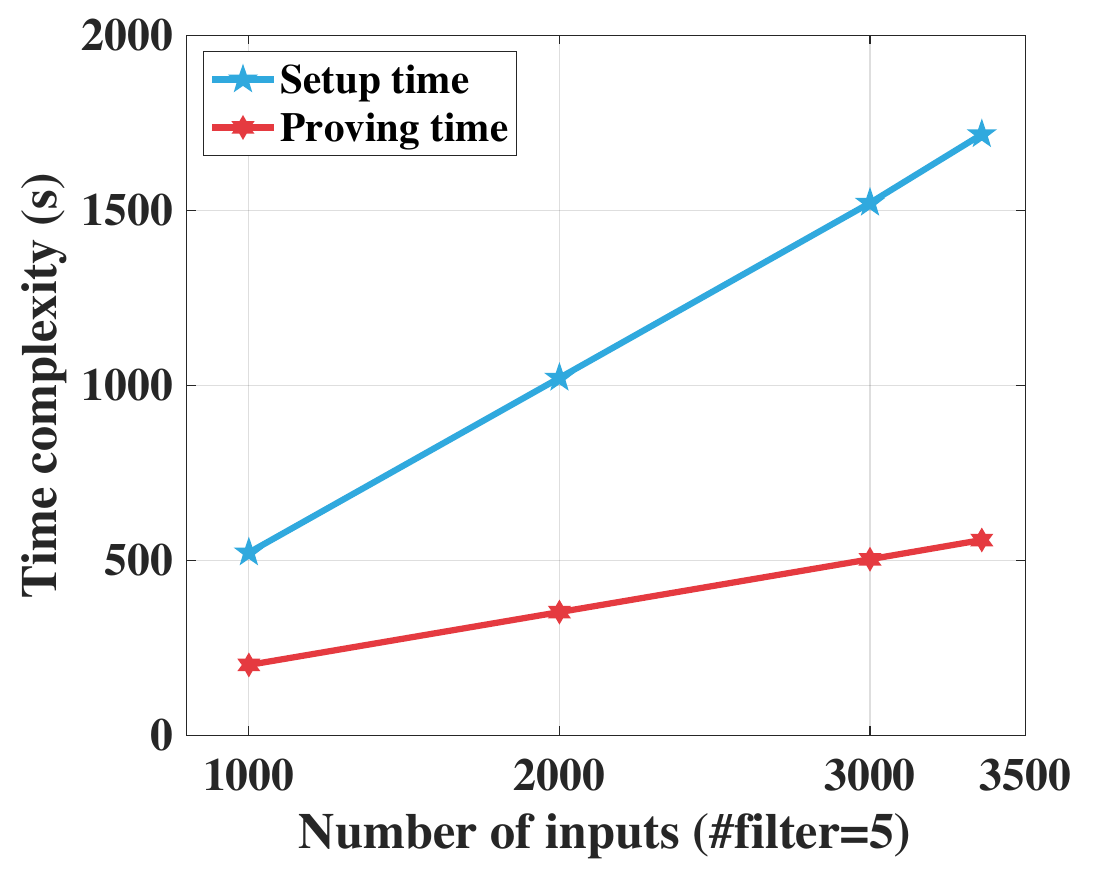}
    		\end{minipage}
		\label{fig:filter:three}
    	}
	\caption{Time complexity on MNIST dataset.}\label{fig:MNIST-time}
\end{figure}

\begin{figure}[t]
\vspace{-10pt}
	\centering
	\hspace{-30pt}
	\subfigure{
		\begin{minipage}{0.135\textwidth}
	\includegraphics[width=1.2\textwidth,height=2.8cm]{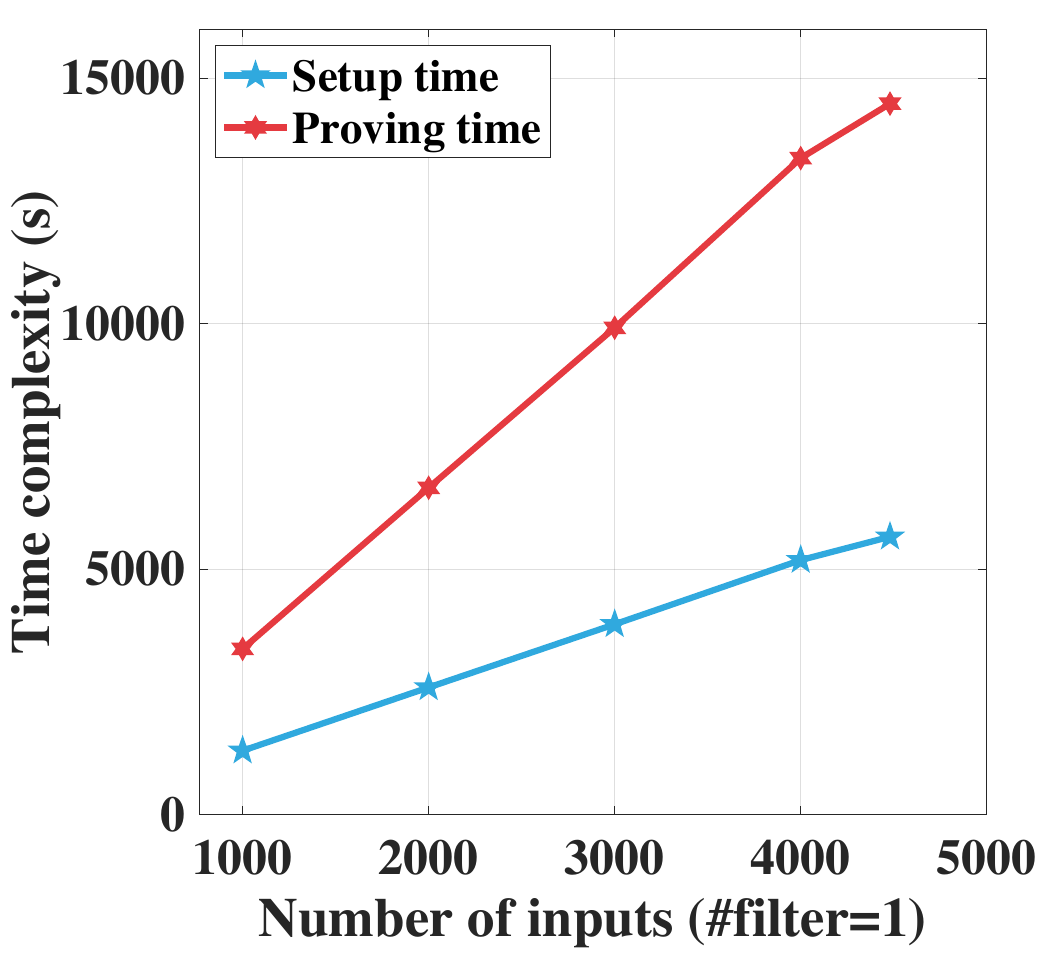}
		\end{minipage}
			\label{fig:filter-2:one}
	}
	\hspace{3pt}
    	\subfigure{
    		\begin{minipage}{0.135\textwidth}
   	\includegraphics[width=1.2\textwidth,height=2.6cm]{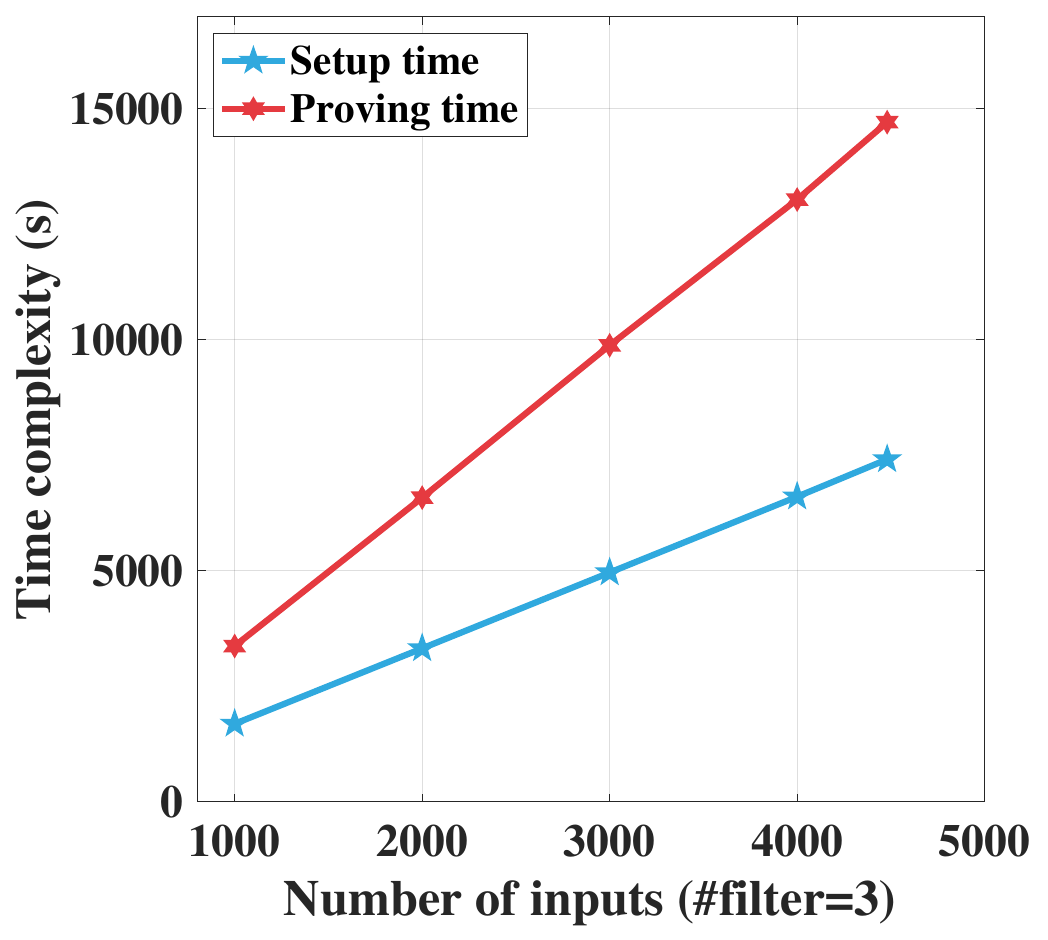}
    		\end{minipage}
		\label{fig:filter-2:three}
    	}
    \hspace{3pt} 
    	\subfigure{
    		\begin{minipage}{0.135\textwidth}
   	\includegraphics[width=1.2\textwidth,height=2.6cm]{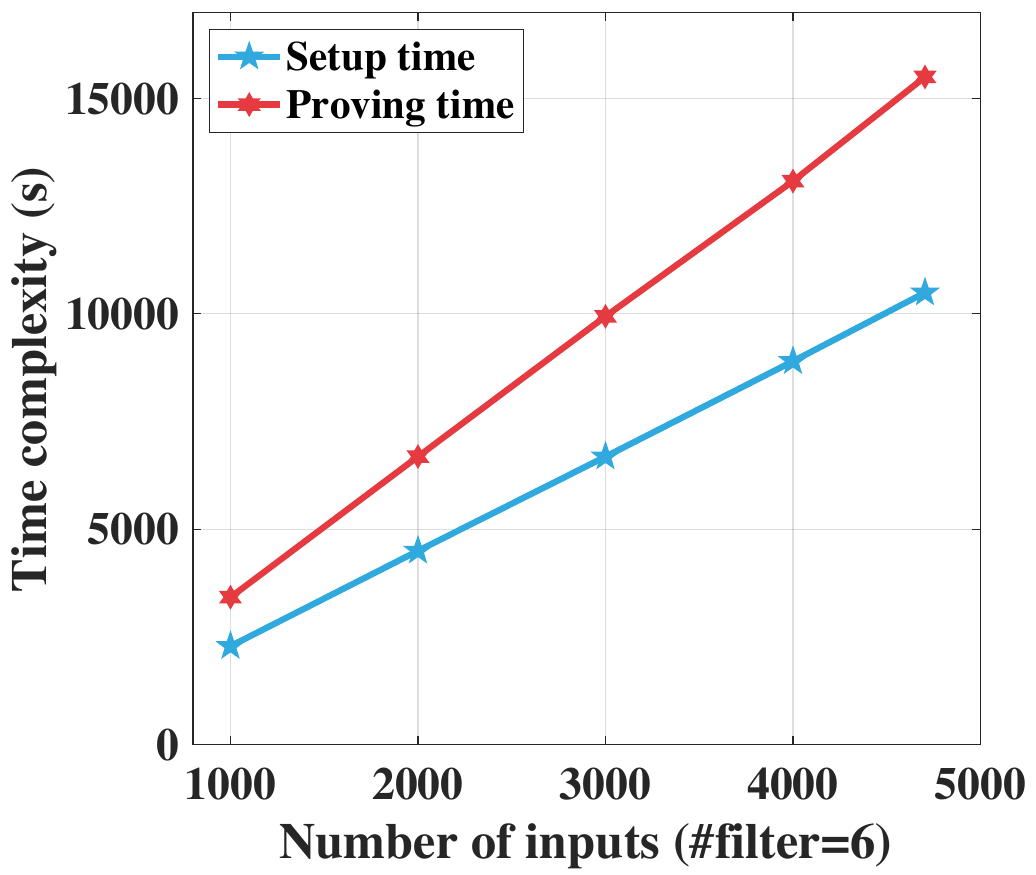}
    		\end{minipage}
		\label{fig:filter-2:six}
    	}
	\caption{Time complexity on CIFAR-10 dataset.}\label{fig:CIFAR-time} 
	\vspace{-18pt}
\end{figure}

We proceed to apply the QMP-based zk-SNARK in a convolution layer with stride $(1,1)$, taking as inputs $1000$ single-channel images of $28\times 28$ from MNIST and  $1000$ three-channel images of $32\times 32$ from CIFAR-10.
For the former dataset, we use $5$ filters of $5\times 5$ while for the latter one, we use $6$ filters.
We transform the convolution operations into a matrix multiplication between a weight matrix and an input matrix both in dimension $3360\times 3360$ for MNIST (similarly, $4704\times 4704$ for CIFAR-10). We note $3360=28\times(28-5+1)\times5 \approx mn^2$, where $m, n$ mean the dimension of a filter and an image, respectively. Here, $m\ge M$ which is the number of filters.
Different from the aforementioned experiments where the values of matrices are random integers, the weight matrix of $3360\times 3360$ here is strategically assigned with the weights of filters plus some zero elements as padding, for ensuring computation correctness (recall it in Fig.\ref{fig:conv}), and similarly, the input matrix of $3360\times 3360$ is assigned with the pixel values of $1000$ images, padding with $3360\times (3360-1000)$ zeros.
As a result, the average performance results of 5 runs are shown in Fig.~\ref{fig:mnist_data}.
We discover that compared to the aforementioned experiments as presented in Fig.\ref{fig:proving}, the \textsf{Setup} and proving time here are relatively smaller.
The main reason can be the padding zero elements inside the weight and input matrices cancel a lot of multiplications.
Besides the \textsf{Setup} and proving time, TABLE~\ref{tab:other_metrics} demonstrates the corresponding verification time, as well as constant CRS size and proof size.

We further conduct additional experiments (named  \textsf{Exp~($\star$)}) on different number of filters and an increasing number of inputs from MNIST and CIFAR-10 datasets, as elaborated in Fig.~\ref{fig:MNIST-time} and Fig.~\ref{fig:CIFAR-time}.
%
%Note that the resulting CRS size and proof size in the above experiments are the same.
%
We can see from the figures that the proving time basically stays stable regardless of the number of filters, but the \textsf{Setup} time increases linearly by the number of filters $M$.
We note that the complexity of the proving time is $O(mn^2 \cdot mn^2), m\ge M$ (resp. $O(Mn^2\cdot Mn^2)$ if $m<M$), where $mn^2$ (resp. $Mn^2$) is the number of inputs, and the proving time is independent of the number of filters $M$.

We proceed to generate QAP-based proofs for the ReLU and average pooling operations on the $3360\times 3360$ matrix, named  \textsf{Exp~($\star\star$)}.
Note that a ReLU operation needs $20$ constraints and an average pooling operation needs $144$ constraints.
The evaluated performance is elaborated in TABLE~\ref{tab:act_pool}.
\begin{table}[htpp]
\vspace{-5pt} 
\scriptsize
  \centering
  \caption{Performance for \textsf{Exp~($\star\star$)} (s).}
    \begin{tabular}{cccc}
    \toprule 
    \textbf{Layer} & \textbf{Setup Time} & \textbf{Proving Time} & \textbf{Verification Time}\\
    \midrule
    ReLU & 5520.61 & 1448.83 & 14.78 \\
    Pooling & 196.43 & 49.93 &  14.78\\
    \bottomrule
    \end{tabular}%
  \label{tab:act_pool}
  \vspace{-10pt}
\end{table}

\section{Conclusion}
The paper discusses using zk-SNARK systems for verifiable CNN testing on encrypted test data. The authors optimize matrix multiplication relations by representing convolution operations with a single MM computation and using a new QMP. This reduces the multiplication gate and proof generation overhead. They also aggregate multiple proofs into a single proof for the same CNN but different test datasets. They provide a proof-of-concept implementation and share their implementation code publicly.
%
\iffalse
We study verifiable CNN testing over ciphered test data by leveraging the state-of-the-art zk-SNARK systems.
%
Following existing related works, we make efforts to optimize matrix multiplication relations.
% 
Specifically, we represent convolution operations applying to multiple filters and inputs with a single MM computation.
%
We then use a new QMP for expressing the MM computation, so as to reduce the multiplication gate of the finally generated circuit for the MM computation, and thereby reducing proof generation overhead. 
%
%
We also aggregate multiple proofs with respect to a same CNN but different test datasets into a single proof.
%
Besides, we present a proof-of-concept implementation and conduct experiments to demonstrate the validity of our approach.
%
For ease of following, we share our implementation code publicly.
\fi
%Our experimental results demonstrate that our QMP-based zk-SNARK performs nearly $13.9\times$ faster than the existing QMP-based zk-SNARK in proving time, and $17.6\times$ faster in \textsf{Setup} time.
%
%Besides, we can overcome the QAP-based zk-SNARK's limitation of failing to handle a number of multiplication beyond $10^7$.

\section*{Acknowledgment}
We appreciate all reviewers for their constructive comments and suggestions.
Jian Weng was supported by National Natural Science Foundation of China (No. 61825203), National Key Research and Development Plan of China (No. 2020YFB1005600), Major Program of Guangdong Basic and Applied Research Project (No. 2019B030302008), Guangdong Provincial Science and Technology Project under Grant (No. 2021A0505030033), National Joint Engineering Research Center of Network Security Detection and Protection Technology, and Guangdong Key Laboratory of Data Security and Privacy Preserving.
Anjia Yang was supported by the Key-Area R\&D Program of Guangdong Province (No. 2020B0101090004, 2020B0101360001), the National Key R\&D Program of China (No. 2021ZD0112802), the National Natural Science Foundation of China (No. 62072215).
Ming Li was supported by the National Natural Science Foundation of China (No. 62102166, 62032025), the Guangdong Provincial Science and Technology Project (No. 2020A1515111175), and the Science and Technology Major Project of Tibet Autonomous Region (No. XZ202201ZD0006G).
Jia-Nan Liu was supported by the National Natural Science Foundation of China (No. 62102165).
%

%-------------------------------------------------------------------------------
\bibliographystyle{IEEEtran}
%\biboptions{sort&compress}
\bibliography{references}

\begin{thebibliography}{10}
\providecommand{\url}[1]{#1}
\csname url@samestyle\endcsname
\providecommand{\newblock}{\relax}
\providecommand{\bibinfo}[2]{#2}
\providecommand{\BIBentrySTDinterwordspacing}{\spaceskip=0pt\relax}
\providecommand{\BIBentryALTinterwordstretchfactor}{4}
\providecommand{\BIBentryALTinterwordspacing}{\spaceskip=\fontdimen2\font plus
\BIBentryALTinterwordstretchfactor\fontdimen3\font minus
  \fontdimen4\font\relax}
\providecommand{\BIBforeignlanguage}[2]{{%
\expandafter\ifx\csname l@#1\endcsname\relax
\typeout{** WARNING: IEEEtran.bst: No hyphenation pattern has been}%
\typeout{** loaded for the language `#1'. Using the pattern for}%
\typeout{** the default language instead.}%
\else
\language=\csname l@#1\endcsname
\fi
#2}}
\providecommand{\BIBdecl}{\relax}
\BIBdecl

\bibitem{lecun1989backpropagation}
Y.~LeCun, B.~Boser \emph{et~al.}, ``Backpropagation applied to handwritten zip
  code recognition,'' \emph{Neural computation}, vol.~1, no.~4, pp. 541--551,
  1989.

\bibitem{das2018convolutional}
R.~Das, E.~Piciucco \emph{et~al.}, ``Convolutional neural network for
  finger-vein-based biometric identification,'' \emph{IEEE Transactions on
  Information Forensics and Security}, vol.~14, no.~2, pp. 360--373, 2018.

\bibitem{google16driving}
G.~Accident, ``A google self-driving car caused a crash for the first time.''
  \url{https://www.theverge.com/2016/2/29/11134344/
  google-self-driving-car-crash-report}, 2016.

\bibitem{amazon20face}
``The two-year fight to stop amazon from selling face recognition to the
  police.''
  \url{https://www.technologyreview.com/2020/06/12/1003482/amazon-stopped-selling-police-face-recognition-fight/},
  2020.

\bibitem{wicker2018feature}
M.~Wicker, X.~Huang \emph{et~al.}, ``Feature-guided black-box safety testing of
  deep neural networks,'' in \emph{Proc. of TACAS}, 2018.

\bibitem{aggarwal2021testing}
A.~Aggarwal, S.~Shaikh, S.~Hans, S.~Haldar, R.~Ananthanarayanan, and D.~Saha,
  ``Testing framework for black-box ai models,'' in \emph{Proc. of IEEE/ACM
  ICSE-Companion}, 2021.

\bibitem{ma2018deepgauge}
L.~Ma, F.~Juefei-Xu \emph{et~al.}, ``Deepgauge: Multi-granularity testing
  criteria for deep learning systems,'' in \emph{Proc. of ACM/IEEE ICASE},
  2018.

\bibitem{hendrycks2018benchmarking}
D.~Hendrycks and T.~Dietterich, ``Benchmarking neural network robustness to
  common corruptions and perturbations,'' in \emph{ICLR}, 2018.

\bibitem{cats4ML}
L.~Aroyo and P.~Paritosh, ``Uncovering unknown unknowns in machine learning,''
  \url{https://ai.googleblog.com/2021/02/uncovering-unknown-unknowns-in-machine.html},
  2021.

\bibitem{ghodsi2017safetynets}
Z.~Ghodsi, T.~Gu, and S.~Garg, ``Safetynets: Verifiable execution of deep
  neural networks on an untrusted cloud,'' in \emph{Proc. of NIPS}, 2017.

\bibitem{collantes2020safetpu}
M.~I.~M. Collantes \emph{et~al.}, ``Safetpu: A verifiably secure hardware
  accelerator for deep neural networks,'' in \emph{Proc. Of IEEE VTS}, 2020.

\bibitem{lee2020vcnn}
S.~Lee, H.~Ko, J.~Kim, and H.~Oh, ``vcnn: Verifiable convolutional neural
  network.'' \url{https://eprint.iacr.org/2020/584.pdf}, 2020.

\bibitem{fengzen}
B.~Feng, L.~Qin, Z.~Zhang, Y.~Ding, and S.~Chu, ``Zen: An optimizing compiler
  for verifiable, zero-knowledge neural network inferences,''
  \url{https://eprint.iacr.org/2021/087.pdf}, 2021.

\bibitem{keuffer2018efficient}
J.~Keuffer, R.~Molva, and H.~Chabanne, ``Efficient proof composition for
  verifiable computation,'' in \emph{Proc. of ESORICS}, 2018.

\bibitem{zhao2021veriml}
L.~Zhao, Q.~Wang \emph{et~al.}, ``Veriml: Enabling integrity assurances and
  fair payments for machine learning as a service,'' \emph{IEEE Transactions on
  Parallel and Distributed Systems}, vol.~32, no.~10, pp. 2524--2540, 2021.

\bibitem{madi2020computing}
A.~Madi, R.~Sirdey \emph{et~al.}, ``Computing neural networks with homomorphic
  encryption and verifiable computing,'' in \emph{ACNS Workshops}, 2020.

\bibitem{liuzkcnn}
T.~Liu, X.~Xie \emph{et~al.}, ``zkcnn: Zero knowledge proofs for convolutional
  neural network predictions and accuracy,'' in \emph{Proc. of ACM CCS}, 2021.

\bibitem{thaler2013time}
J.~Thaler, ``Time-optimal interactive proofs for circuit evaluation,'' in
  \emph{Annual Cryptology Conference}, 2013, pp. 71--89.

\bibitem{groth2016size}
J.~Groth, ``On the size of pairing-based non-interactive arguments,'' in
  \emph{EUROCRYPT}, 2016, pp. 305--326.

\bibitem{agrawal2018non}
S.~Agrawal, C.~Ganesh \emph{et~al.}, ``Non-interactive zero-knowledge proofs
  for composite statements,'' in \emph{CRYPTO}, 2018, pp. 643--673.

\bibitem{campanelli2019legosnark}
M.~Campanelli \emph{et~al.}, ``Legosnark: Modular design and composition of
  succinct zero-knowledge proofs,'' in \emph{Proc. of ACM CCS}, 2019.

\bibitem{fiore2020boosting}
D.~Fiore, A.~Nitulescu, and D.~Pointcheval, ``Boosting verifiable computation
  on encrypted data,'' in \emph{PKC}, no. 12111, 2020, pp. 124--154.

\bibitem{mouris2021zilch}
D.~Mouris and N.~G. Tsoutsos, ``Zilch: A framework for deploying transparent
  zero-knowledge proofs,'' \emph{IEEE Transactions on Information Forensics and
  Security}, 2021.

\bibitem{caffe2019zoo}
``Bvlc/caffe,'' \url{https://github.com/BVLC/caffe/wiki/Model-Zoo}, 2019.

\bibitem{kaggle2010data}
Kaggle, ``Data science competition platform.'' \url{https://www.kaggle.com/}.

\bibitem{awsModel}
A.~Marketplace, ``Machine learning solutions.''
  \url{https://aws.amazon.com/marketplace/solutions/machine-learning}.

\bibitem{xu2019verifynet}
G.~Xu \emph{et~al.}, ``Verifynet: Secure and verifiable federated learning,''
  \emph{IEEE Transactions on Information Forensics and Security}, vol.~15, pp.
  911--926, 2019.

\bibitem{he2019model}
Z.~He, T.~Zhang, and R.~B. Lee, ``Model inversion attacks against collaborative
  inference,'' in \emph{Proc. of ACM ACSAC}, 2019.

\bibitem{ryffel2019partially}
T.~Ryffel, E.~Dufour-Sans \emph{et~al.}, ``Partially encrypted machine learning
  using functional encryption,'' in \emph{NIPS}, 2019.

\bibitem{aono2017privacy}
Y.~Aono, T.~Hayashi \emph{et~al.}, ``Privacy-preserving deep learning via
  additively homomorphic encryption,'' \emph{IEEE Transactions on Information
  Forensics and Security}, vol.~13, no.~5, pp. 1333--1345, 2017.

\bibitem{alesiani2021method}
S.~G. Francesco~Alesiani, ``Method for verifying information,''
  \url{https://patentimages.storage.googleapis.com/53/3c/62/0ed0b3f9bb163f/US20210091953A1.pdf},
  2021.

\bibitem{gailly2021snarkpack}
N.~Gailly, M.~Maller, and A.~Nitulescu, ``Snarkpack: Practical snark
  aggregation.'' in \emph{Proc. of RWC}, 2022.

\bibitem{zhang2020doubly}
J.~Zhang, T.~Liu \emph{et~al.}, ``Doubly efficient interactive proofs for
  general arithmetic circuits with linear prover time,'' in \emph{Proc. of ACM
  CCS}, 2021.

\bibitem{goldwasser2015delegating}
S.~Goldwasser \emph{et~al.}, ``Delegating computation: interactive proofs for
  muggles,'' \emph{Journal of the ACM}, vol.~62, no.~4, pp. 1--64, 2015.

\bibitem{niu2020toward}
C.~Niu, F.~Wu, S.~Tang, S.~Ma, and G.~Chen, ``Toward verifiable and privacy
  preserving machine learning prediction,'' \emph{IEEE Transactions on
  Dependable and Secure Computing}, 2020.

\bibitem{zhang2020zero}
J.~Zhang, Z.~Fang \emph{et~al.}, ``Zero knowledge proofs for decision tree
  predictions and accuracy,'' in \emph{Proc. of ACM CCS}, 2020.

\bibitem{Weng21Mystique}
C.~Weng, K.~Yang, X.~Xie, J.~Katz, and X.~Wang, ``Mystique: Efficient
  conversions for zero-knowledge proofs with applications to machine
  learning,'' in \emph{Proc. of USENIX Security}, 2021, pp. 501--518.

\bibitem{dong2023fusion}
C.~Dong, J.~Weng \emph{et~al.}, ``Fusion: Efficient and secure inference
  resilient to malicious server and curious clients,'' in \emph{Proc. of NDSS},
  2023.

\bibitem{gilad2016cryptonets}
R.~Gilad-Bachrach, N.~Dowlin \emph{et~al.}, ``Cryptonets: Applying neural
  networks to encrypted data with high throughput and accuracy,'' in
  \emph{ICML}, 2016, pp. 201--210.

\bibitem{mohassel2017secureml}
P.~Mohassel and Y.~Zhang, ``Secureml: A system for scalable privacy-preserving
  machine learning,'' in \emph{Proc. Of IEEE S\&P}, 2017.

\bibitem{mishra2020delphi}
P.~Mishra, R.~Lehmkuhl \emph{et~al.}, ``Delphi: A cryptographic inference
  service for neural networks,'' in \emph{USENIX Security}, 2020, pp.
  2505--2522.

\bibitem{juvekar2018gazelle}
C.~Juvekar, V.~Vaikuntanathan, and A.~Chandrakasan, ``$\{$GAZELLE$\}$: A low
  latency framework for secure neural network inference,'' in \emph{Proc. of
  USENIX Security}, 2018.

\bibitem{kumar2020cryptflow}
N.~Kumar, M.~Rathee \emph{et~al.}, ``Cryptflow: Secure tensorflow inference,''
  in \emph{IEEE S\&P}, 2020, pp. 336--353.

\bibitem{kai2019lightweight}
K.~Huang, X.~Liu \emph{et~al.}, ``A lightweight privacy-preserving cnn feature
  extraction framework for mobile sensing,'' \emph{IEEE Transactions on
  Dependable and Secure Computing}, vol.~18, no.~3, pp. 1441--1455, 2019.

\bibitem{hesamifard17cryptoDL}
E.~Hesamifard, H.~Takabi, and M.~Ghasemi, ``Cryptodl: Deep neural networks over
  encrypted data,'' \url{https://arxiv.org/abs/1711.05189}, 2017.

\bibitem{liu2016efficient}
X.~Liu, R.~H. Deng \emph{et~al.}, ``An efficient privacy-preserving outsourced
  calculation toolkit with multiple keys,'' \emph{IEEE Transactions on
  Information Forensics and Security}, vol.~11, no.~11, pp. 2401--2414, 2016.

\bibitem{liu2016privacy}
X.~Liu \emph{et~al.}, ``Efficient and privacy-preserving outsourced calculation
  of rational numbers,'' \emph{IEEE Transactions on Dependable and Secure
  Computing}, vol.~15, no.~1, pp. 27--39, 2016.

\bibitem{liu2018privacy}
X.~Liu, R.~H. Deng \emph{et~al.}, ``Privacy-preserving outsourced calculation
  toolkit in the cloud,'' \emph{IEEE Transactions on Dependable and Secure
  Computing}, vol.~17, no.~5, pp. 898--911, 2018.

\bibitem{gennaro2012quadratic}
R.~Gennaro, C.~Gentry \emph{et~al.}, ``Quadratic span programs and succinct
  nizks without pcps,'' \url{https://eprint.iacr.org/2012/215.pdf}, 2012.

\bibitem{fansomewhat}
J.~Fan and F.~Vercauteren, ``Somewhat practical fully homomorphic encryption,''
  \url{https://eprint.iacr.org/2012/144.pdf}, 2012.

\bibitem{chen2017targeted}
X.~Chen, C.~Liu \emph{et~al.}, ``Targeted backdoor attacks on deep learning
  systems using data poisoning,'' \url{https://arxiv.org/abs/1712.05526}, 2017.

\bibitem{xu2020secure}
G.~Xu, H.~Li \emph{et~al.}, ``Secure and verifiable inference in deep neural
  networks,'' in \emph{Proc. of ACM ACSAC}, 2020.

\bibitem{he2018verideep}
Z.~He, T.~Zhang \emph{et~al.}, ``Sensitive-sample fingerprinting of deep neural
  networks,'' in \emph{CVPR}, 2019.

\bibitem{zhang2021stealing}
Y.~Zhang \emph{et~al.}, ``Stealing neural network structure through remote fpga
  side-channel analysis,'' \emph{IEEE Transactions on Information Forensics and
  Security}, vol.~16, pp. 4377--4388, 2021.

\bibitem{yan2020cache}
M.~Yan \emph{et~al.}, ``Cache telepathy: Leveraging shared resource attacks to
  learn dnn architectures,'' in \emph{Proc. of USENIX Security}, 2020.

\bibitem{saileshwar2021mirage}
G.~Saileshwar \emph{et~al.}, ``Mirage: Mitigating conflict-based cache attacks
  with a practical fully-associative design.'' in \emph{Proc. of USENIX
  Security}, 2021.

\bibitem{KosbaPPSST14TRUESET}
A.~E. Kosba, D.~Papadopoulos \emph{et~al.}, ``Trueset: Faster verifiable set
  computations,'' in \emph{Proc. of USENIX Security}, 2014.

\end{thebibliography}

\ifCLASSOPTIONcaptionsoff
  \newpage
\fi

\subsection{Discussion}\label{app:res}
In terms of verification time and proof size, the QMP-based zk-SNARK has a higher overhead than the QAP-zk-SNARK. 
Its verification time grows faster than that of QAP-based zk-SNARK, see TABLE~\ref{tab:verTime}. 
The proof size also becomes lager when the matrix dimension turns lager as shown in TABLE~\ref{tab:magnitude}, while the proof size in the QAP-based zk-SNARK keeps $1019~$bits regardless of the matrix dimension. 
We next see how the proof size becomes longer with the matrix dimension increasing.
We note that the bounded number of multiplications the QAP zk-SNARK can handle is $220*220*220=10,648,000$, and then the QAP zk-SNARK would be called multiple times when handling the multiplication operations more than $10,648,000$, which results in multiple $1019$-bit proofs.
Suppose that the QAP-based zk-SNARK proofs for the $3000\times 3000$ matrix multiplication are totally $\frac{3000*3000*3000}{220*220*220}$
$\times1019$ bits.
Also, the QMP-based zk-SNARK proofs for the same matrix multiplication are $2,295,000,764$ bits.
The proof size is nearly $888$ times larger than that of the above QAP-based zk-SNARK proof.
We observe that the times of magnitude become smaller as the matrix dimension increases, see TABLE~\ref{tab:magnitude}, which may mean that the QMP-based zk-SNARK is more suitable to handle large matrix multiplication.
%. 
%Thus, we suggest that the QMP-based zk-SNARK can be more suitable to handle the matrix multiplication in a larger dimension, and in the case, the times of magnitude might approximate to zero.
% 

\begin{table}[h]
\vspace{-5pt}
\scriptsize
  \centering
  \caption{Times of magnitude in proof size.}
    \begin{tabular}{cccccc}
    \toprule
    \textbf{Matrix dimension} &  $\mathbf{1000\times }$ & $\mathbf{2000\times }$ & $\mathbf{3000\times }$& $\mathbf{3360\times }$&
    $\mathbf{4000\times }$\\
    \midrule
    \textbf{Times} & 2665 & 1332 & 888& 793&666\\
    \bottomrule
    \end{tabular}%
  \label{tab:magnitude}
\end{table}

\begin{table}[h]
\vspace{-10pt}
\scriptsize
  \centering
  \caption{Verification time comparison (s).}
    \begin{tabular}{ccccc}
    \toprule
    \textbf{Matrix dimension} & $\mathbf{30\times 30}$ & $\mathbf{50\times 50}$ & $\mathbf{100\times 100}$ & $\mathbf{200\times 200}$ \\
    \midrule
    \textbf{QAP} & 0.005 &	0.007 & 0.014 &	0.044 \\
    \textbf{QMP} &  4.896 &	13.580 & 54.630 & 215.800 \\
    \bottomrule
    \end{tabular}%
  \label{tab:verTime} %
\vspace{-8pt}
\end{table}
% 

\iffalse

\begin{table}[h]
\scriptsize
  \centering\caption{Proof size of the QMP-based zk-SNARK for matrix multiplication (KB).}
    \begin{tabular}{ccccc}
    \toprule
    \textbf{Matrix dimension} & $\mathbf{30\times 30}$ & $\mathbf{50\times 50}$ & $\mathbf{100\times 100}$ & $\mathbf{200\times 200}$ \\
    \midrule
    \textbf{Size} &  28.10 &	77.91 & 311.37	& 1,245.21 \\
    \bottomrule
    %
    \bottomrule
    \textbf{Matrix dimension} & $\mathbf{500\times }$ & $\mathbf{1000\times }$ & $\mathbf{2000\times }$ & $\mathbf{3000\times }$ \\
    \midrule
    %
    \textbf{Size} &  7,782.08 &	31,128.02 & 	124,511.81	& 280,151.46 \\
    \bottomrule
    \end{tabular}%
  \label{tab:proofSize} %
\end{table}
\fi
%

In our future work, we would introduce a random sampling strategy in the phase of proof generation, aiming to reduce the verification time. 
A straightforward idea can be adopted by randomly sampling a bounded number of values in two matrices to be multiplied, and resetting the non-chosen values as zeros.
In such a way, only the sampled values in the two matrices are multiplied and only their multiplication correctness need to be proved.
But here two noteworthy issues should be considered: (1) how to generate the randomness used for sampling against a distrusted prover; (2) how to determine the bound of the number of sampled values for ensuring computational soundness in zk-SNARK.

\iffalse
\noindent\textbf{Complementary explanation for Fig.~\ref{fig:conv}.}
% 
In Section~\ref{sec:overview}, for ease of presentation, we assume the number of filters $M$ is more than the dimension of a filter matrix $m$ within a convolutional layer, and as a result, both of the transformed matrices for the filter matrices and input matrices are in dimension $Mn^2\times Mn^2$.
%
Herein, we add the complementary explanation for the case where the number of filters $M$ is not more than the dimension of a filter matrix $m$.
%
In this case, the transformed matrices are in dimension $mn^2\times mn^2$, as elaborated in Fig.~\ref{fig:mn2}.
% 
Note that the number of inputs is $mn^2$ here.
% 
For the convolutional layers during a CNN testing, we would take into account the number of filters, the dimension of filter and input, as well as the number of inputs, and then strategically reshape the filters and inputs into two square matrices in the same dimension.
%
\begin{figure}[h]
    \centering
    \includegraphics[width=1.0\linewidth]{image/mn2.png}
    \caption{Matrix multiplication for convolution operations ($M\leq m$).}
    \label{fig:mn2} 
\end{figure}
% 
\fi
 
\end{document}